\documentclass{article}
\usepackage[english]{babel}
\usepackage[utf8]{inputenc}
\usepackage{amsmath}
\usepackage{amsfonts}
\usepackage{amssymb}
\usepackage{graphicx}
\usepackage[backend=biber,maxcitenames=2,citetracker=true,uniquelist=false,uniquename=false,style=authortitle,citestyle=authoryear,maxbibnames=99,doi=false,isbn=false,url=false,eprint=false]{biblatex}
\usepackage{setspace}
\usepackage{wrapfig}
\usepackage{fancyhdr}
\usepackage{tabularx}
\usepackage[a4paper,margin=2.5cm]{geometry}
\usepackage[right]{eurosym}
\usepackage[printonlyused]{acronym}
\usepackage{subcaption}
\usepackage{floatflt}
\usepackage[usenames,dvipsnames]{color}
\usepackage{colortbl}
\usepackage{paralist}
\usepackage{array}
\usepackage{indentfirst}
\usepackage{titlesec}
\usepackage{longtable}
\usepackage[colorlinks=true,linkcolor=black,urlcolor=black,citecolor=black,pdfpagelabels=true]{hyperref}
\usepackage{tablefootnote}
\usepackage{csquotes}
\usepackage{mathtools}
\usepackage{booktabs,caption}
\usepackage[flushleft]{threeparttable}
\usepackage{hyperref}
\usepackage[bottom,flushmargin]{footmisc}
\usepackage{multicol}
\usepackage{pdflscape}
\usepackage{titlesec}
\usepackage{titling}
\usepackage{authblk}
\usepackage[inline]{enumitem}
\usepackage{xcolor, soul}
\sethlcolor{white}
\setulcolor{white}
\usepackage{pifont}
\usepackage{threeparttable}
\usepackage{threeparttablex}

\usepackage{xpatch}
\xpatchbibmacro{cite}{\printnames{labelname}}%
{\ifciteseen{\printnames{labelname}}{\printnames[][-\value{listtotal}]{labelname}}}
{}{}

\xpatchbibmacro{textcite}{\printnames{labelname}}%
{\ifciteseen{\printnames{labelname}}{\printnames[][-\value{listtotal}]{labelname}}}
{}{}

\setul{0.3ex}{.5ex}

\definecolor{babyblueeyes}{rgb}{0.63, 0.79, 0.95}

\newcommand{\cmark}{\ding{51}}%
\newcommand{\xmark}{\ding{55}}%

\makeatletter
\AtBeginDocument{\let\hl\@firstofone}
\makeatother

\newcolumntype{H}{>{\setbox0=\hbox\bgroup}c<{\egroup}@{}}

\providecommand{\keywords}[1]
{
  \noindent
  \small	
  \textbf{Keywords:} #1
}

\providecommand{\jelclass}[1]
{
  \noindent
  \small	
  \textbf{JEL classification:} #1
}

\numberwithin{equation}{section}

\title{\huge Random Forest Estimation\\of the Ordered Choice Model\thanks{\noindent A previous version of the paper was presented at research seminars of the University of St.Gallen, at the German Statistical Week in Trier and the Statistics of Machine Learning Conference in Prague. We thank participants, in particular Francesco Audrino, Martin Biewen, Daniel Goller, Michael Knaus and David Preinerstorfer for helpful comments and suggestions. The usual disclaimer applies.}\\
~\\
}

\author[1]{Michael Lechner\thanks{\noindent Michael Lechner is also affiliated with CEPR, London, CESifo, Munich, IAB, Nuremberg, and IZA, Bonn.\\
Email: \url{michael.lechner@unisg.ch}}}
\author[2]{Gabriel Okasa\thanks{\noindent Email: \url{gabriel.okasa@epfl.ch}}}

\affil[1]{Swiss Institute for Empirical Economic Research\\ University of St.Gallen}
\affil[2]{Swiss Federal Institute of Technology in Lausanne\\ EPFL}

\begin{document}

\maketitle
\vspace{1cm}
\begin{abstract}
In this paper we develop a new machine learning estimator for ordered choice models based on the random forest. The proposed \textit{Ordered Forest} flexibly estimates the conditional choice probabilities while taking the ordering information explicitly into account. In addition to common machine learning estimators, it enables the estimation of marginal effects as well as conducting inference and thus provides the same output as classical econometric estimators. An extensive simulation study reveals a good predictive performance, particularly in settings with non-linearities and near-multicollinearity. An empirical application contrasts the estimation of marginal effects and their standard errors with an ordered logit model. A software implementation of the \textit{Ordered Forest} is provided both in \textsf{R} and \textsf{Python} in the package \textsf{orf} available on \textsf{CRAN} and \textsf{PyPI}, respectively.\\

\\

\jelclass{C14, C25, C40.}
\end{abstract}
\vfill

\thispagestyle{empty}

\pagebreak
\onehalfspacing
\setcounter{page}{1}

\section{Introduction}
Many empirical models deal with categorical dependent variables which have an inherent ordering. In such cases the outcome variable is measured on an ordered scale such as level of education defined by primary, secondary and tertiary education or income coded into low, middle and high income level. Further examples include survey outcomes on self-assessed health status \parencites(bad, good, very good, see e.g.)(){Case2002}[or][]{Murasko2008}, \ul{level of life satisfaction and happiness} \parencites()(){Boes2010}[and][]{Boes2010a} or political opinions \parencites(do not agree, agree, strongly agree, see e.g.)(){Jackson2005}[or][]{Jackman2009} as well as grades, scores and various ratings and valuations \parencites(see)(){Butler1998, Hamermesh2005, Afonso2009}[or][for some further examples]{Gogas2014}. Moreover, even sports outcomes resulting in loss, draw and win are part of such modelling framework \parencite[e.g.][]{Okasa2018}. So far, the ordered probit or ordered logit model represent workhorse models in such cases. The main advantage of these models is the ease of estimation, usually done by maximum likelihood. However, the major disadvantage are the strong parametric assumptions which are imposed for convenience rather than derived from any substantive knowledge about the application. Unfortunately, the desired marginal effects are sensitive to these assumptions. Although there is a large literature on how to generalize these assumptions in case of binary choice models \parencites{Matzkin1992, Ichimura1993, Klein1993}, or multinomial (unordered) choice models \parencites{Lee1995, Fox2007}, limited work has been done for ordered choice models \parencites()(){Lewbel2000, Klein2002}[also see][for an overview]{Stewart2005}.

In this paper, we exploit recent advances in the machine learning literature to \ul{develop} an estimator for \ul{conditional} choice probabilities as well as marginal effects together with inference procedures when the outcome variable has an ordered categorical nature. The proposed \ul{\textit{Ordered Forest}} estimator is based on the \ul{regression} random forest algorithm \ul{as introduced by} \textcite{Breiman2001} and makes use of cumulative probability predictions based on binary indicators of respective ordered categories to flexibly estimate the single choice probabilities of the particular ordered category, conditional on covariates. Furthermore, to analyze the relationship of the ordered choice probabilities with the covariates, the \textit{Ordered Forest} exploits numerical derivative approximations for estimation of the mean marginal effects and marginal effects at mean as the typical quantities of interest in the field of discrete choice models \parencite[see e.g.][]{Greene2010}. Finally, in order to quantify the estimation uncertainty of the above parameters, the \textit{Ordered Forest} adapts the weight-based inference proposed by \textcite{Lechner2019} \ul{using the asymptotic results of }\textcite{Wager2018} \ul{for the consistency and normality of random forest predictions for the case of ordered categorical outcomes. Thus \textit{Ordered Forest} estimator provides not only the point estimate for the conditional choice probabilities and the corresponding marginal effects, but also an estimate for the respective standard errors.}

We investigate the predictive performance of the estimator by comparing it to classical and other competing methods via a large-scale Monte Carlo simulation study as well as using real datasets. The results from the synthetic simulation reveal good performance of the \textit{Ordered Forest} in finite samples throughout all simulation designs, including high-dimensional settings. In particular, the superior performance of the estimator over the parametric ordered logit becomes apparent when dealing with nonlinear functional forms and near-multicollinearity among covariates. Furthermore, the \textit{Ordered Forest} outperforms the competing forest-based estimators in the most complex simulation designs. Additionally, the results from the empirical evaluation further confirm the good predictive performance of the estimator in real datasets. Lastly, an empirical application demonstrates the estimation of the marginal effects and the associated inference procedure. The empirical results highlight the value of the additional flexibility in the effect estimation of relevant economic parameters. Moreover, to enable the usage of the method by applied researchers a free software implementation of the \textit{Ordered Forest} estimator has been developed in \textsf{R} \parencite{rstats} as well as in \textsf{Python} \parencite{python} and is provided in the package \textsf{orf} \ul{available on the official} \href{https://CRAN.R-project.org/package=orf}{\textsf{CRAN}} \parencite{orf2019} and \href{https://pypi.org/project/orf/}{\textsf{PyPI}} \parencite{orf2022} \ul{repositories.}\footnote{\hl{Source codes of both \textsf{R} and \textsf{Python} versions of the estimator are available on} \href{https://github.com/orf-lab}{\textsf{GitHub}}. \ul{Additionally, an implementation of the estimator in \textsf{GAUSS} is available} \href{https://www.michael-lechner.eu/statistical-software/}{\textsf{online}} \ul{and on} \href{https://www.researchgate.net/publication/334230505_GAUSS_Code_for_the_Ordered_Forest_Estimator_Version_33}{\textsf{ResearchGate}}.}

\ul{This paper contributes to the econometric as well as machine learning literature in several ways. In terms of econometrics, this paper develops a new estimator of the ordered choice models based on a machine learning algorithm.} The proposed \textit{Ordered Forest} estimator improves on the classical \ul{parametric} models such as ordered logit and ordered probit models by allowing \textit{ex-ante} flexible functional forms as well as allowing for a large\ul{r} covariate space. The latter is a feature of many machine learning methods, but is typically absent from standard econometrics. \ul{In terms of machine learning, this paper develops a new type of random forest estimator adapted to ordered categorical outcomes. As such, the proposed \textit{Ordered Forest} extends the classical regression forests as developed by} \textcite{Breiman2001} and \textcite{Wager2018} \ul{specifically for estimation of ordered choice models and thus expands the forest-based estimators for particular econometric models such as for example the survival forest} \parencite{Hothorn2004} \ul{designed for estimation of survival models or the quantile regression forest} \parencite{Meinshausen2006} \ul{for estimation of conditional quantiles.} Additionally \ul{to the above forest-based estimators}, the \textit{Ordered Forest} \ul{further} advances machine learning methods with the estimation of marginal effects and the inference thereof, a feature of many parametric models, but generally missing in the machine learning literature. Hence, \ul{our} contribution is twofold. First, with respect to the literature on parametric estimation of the ordered choice models, the \textit{Ordered Forest} represents a flexible estimator without any parametric assumptions, while providing essentially the same information as an ordered parametric model. Second, with respect to the machine learning literature, the \textit{Ordered Forest} achieves more precise estimation of ordered choice probabilities, while adding estimation of marginal effects as well as statistical inference thereof.

This paper is organized as follows. Section \ref{sec:lit} discusses the related literature concerning \ul{parametric and }machine learning methods for the estimation of ordered choice models. Section \ref{sec:RF} reviews the random forest algorithm and its theoretical properties. In Section \ref{sec:ORF} the \textit{Ordered Forest} estimator is introduced including the estimation of the conditional choice probabilities, marginal effects and the inference procedure. The Monte Carlo simulation is presented in Section \ref{sec:MC}. Section \ref{sec:emp} shows an empirical application. Section \ref{sec:conclusion} concludes. Further details regarding estimation methods, the simulation study and the empirical application are provided in Appendices \ref{Appendix:methods}, \ref{Appendix:simulation} and \ref{Appendix:applications}, respectively.

\section{Literature}\label{sec:lit}
In econometrics, the ordered probit and ordered logit models are widely used when there are ordered response variables \parencite{McCullagh1980}. These models build on the latent regression model assuming an underlying continuous outcome $Y_i^*$ as a linear function of regressors $X_i$ with unknown coefficients $\beta$, while assuming that the latent error term $u_i$ follows the standard normal or the logistic distribution. Furthermore, the ordered discrete outcome $Y_i$ represents categories that cover a certain range of the latent continuous $Y_i^*$ and is determined by unknown threshold parameters $\alpha_m$. Formally, in the case of the ordered logit the latent model is defined as:
\begin{align}\label{eq:ologit}
Y_i^* &= X_i'\beta +u_i \text{, \phantom{space}} (u_i \mid X_i) \sim Logistic(0,\pi^2/3)
\end{align}
with unknown threshold parameters \ul{$\alpha_0<\alpha_1<...<\alpha_M$} such that:
\begin{align}\label{eq:thresholds}
Y_i &= m \quad \text{ if } \quad \alpha_{m-1} < Y_i^* \leq \alpha_m \quad \text{ for } \quad m=1,...,M ,
\end{align}
where the coefficients and the thresholds are commonly estimated via maximum likelihood with the delta method or bootstrapping used for inference. \ul{Notice, that the outer thresholds are $\alpha_0=-\infty$ and $\alpha_M=\infty$.} The above latent model is also often motivated by the quantity of interest, i.e. the conditional choice probabilities which are given by: 
\begin{align}\label{eq:ologitprobs}
P[Y_i=m \mid X_i=x]= \Lambda\big(\alpha_m-X_i'\beta\big)-\Lambda\big(\alpha_{m-1}-X_i'\beta\big) ,
\end{align}
where the link function $\Lambda(\cdot)$ is the logistic cdf mapping the real line onto the unit interval. Thus, the estimated probabilities are bounded between $0$ and $1$. The marginal effects are further given as partial derivative of the probabilities in \eqref{eq:ologitprobs} \ul{as}:
\begin{align}\label{eq:ologitme}
\frac{\partial P[Y_i=m \mid X_i=x]}{\partial x^k}= \bigg[\lambda\big(\alpha_{m-1}-X_i'\beta\big)-\lambda\big(\alpha_{m}-X_i'\beta\big)\bigg]\beta_k ,
\end{align}
where $x^k$ is the $k$-th element of $X_i$ and $\beta_k$ is the corresponding coefficient, while $\lambda(\cdot)$ being the logistic pdf.

Although such models are relatively easy to estimate, they impose strong parametric assumptions which hinder the flexibility of these models. Apart from the assumptions about the distribution of the error term, further functional form assumptions are being imposed. As is clear from \eqref{eq:ologit}, the coefficients $\beta$ are constant across the outcome classes which is often labelled as the parallel regression assumption \parencite{Williams2016}. This inflexibility affects both the estimation of the choice probabilities as well as the estimation of marginal effects. For these reasons, generalizations of these models have been proposed in the literature in order to relax some of the assumptions. An example of such models is the generalized ordered logit model \parencite{McCullagh1989}, where the parallel regression assumption is abandoned. \textcite{Boes2006} provide an excellent overview of several other generalized parametric models. However, all of these models retain some of the distributional assumptions which limit their modelling flexibility.

Besides the standard econometric literature on parametric specifications of ordered choice models \parencites(for an overview see)(){Agresti2002}[or][]{Boes2006}, a new strand of literature devoted to relaxing the parametric assumptions by using novel machine learning methods is emerging. Particularly, the tree-based methods have gained considerable attention. Although the classical CART algorithms introduced by \textcite{Breiman1984} are very powerful in both regression as well as in classification \parencite[see][for a review]{Loh2011}, there is a need for adjustment when predicting ordered response. In the case of regression, the \textit{discrete} nature of the outcome is not being taken into account and in the case of classification, the \textit{ordered} nature of the outcome is not being taken into account. For these reasons, a strand of the literature focused particularly on adjustments towards ordered classification rather than regression which excludes the estimation of the conditional probabilities as is the case in the parametric ordered choice models. For example, \textcite{Kramer2000} propose a simple procedure for constructing a distance-sensitive classification learner \ul{using post-processing classification rules}. Another approach suggested in the literature is to modify the splitting criterion directly. In particular, the usage of alternative impurity measures as opposed to the Gini coefficient in case of classification trees have been suggested, namely the generalized Gini criterion \parencite{Breiman1984} or the ordinal impurity function \parencite{Piccarreta2008}. Both of these measures put higher penalty on misclassification the more distant the predicted category is from the true one. It follows that the above methods focus on estimating ordered classes rather than estimating ordered class probabilities, as is the focus of this paper.

The above ideas, however, have not been much used in practice. The reason might be the well-known drawbacks of single trees which suffer from unstable splits and a lack of smoothness \parencite{Hastie2009}. A natural extension of the CART algorithms is the random forest first introduced by \textcite{Breiman2001}. However, the random forest algorithm as well as CART is primarily suitable for either regression or classification exercises. As such, appropriate modifications of the standard random forest algorithm are desired in order to predict conditional probabilities of discrete outcomes while taking the ordering nature into account. \textcite{Hothorn2006} propose a random forest algorithm building on their conditional inference framework for recursive partitioning which can also deal with ordered outcomes. The difference to standard regression forests lies in a different splitting criterion using a test statistic where the conditional distribution at each split is based on permutation tests \parencites(for details see)(){Strasser1999}[and][]{Hothorn2006}. Their proposed ordinal forest regression assumes an underlying latent continuous response $Y_i^*$ as is the case in standard ordered choice models. \textcite{Hothorn2006} define a score vector $s(m) \in \mathbb{R}^M$, with $m=1,...,M$ observed ordered classes. This scores reflect the distances between the classes. The authors suggest to set the scores as midpoints of the intervals of $Y_i^*$ which define the classes. As the underlying $Y_i^*$ is unobserved, such a suggestion results in $s(m)=m$ and ordinal forest regression collapses to a standard forest regression as pointed out by \textcite{Janitza2016}.\footnote{\textcite{Janitza2016} \ul{perform also a simulation study to test the robustness of the suggested score values by setting $s(m)=m^2$, but do not find any significant differences to simple $s(m)=m$.}} However, although the tree building step coincides, the prediction step differs as the estimates are the choice probabilities calculated as the proportions of the respective outcome classes falling into the same leaf instead of averages of the outcomes. \ul{As such, for each leaf within a tree, the prediction is computed for each value of the ordered categorical outcome as its share within the leaf, resulting in a probability predictions between 0 and 1. This is in contrast to standard prediction procedures, which would compute an average of all values of the ordered categorical outcome. Nevertheless, after computing the single-tree predictions as the relative frequencies of the ordered outcomes, the forest estimates of} the conditional choice probabilities $\hat{P}[Y_i=m \mid X_i=x]$ are computed by taking the averages of the choice probabilities produced by each tree, i.e. the same aggregation scheme as in a regression forest. \textcite{Hornung2017} \ul{points out that setting $s(m)=m$ implies inherently assuming} that the class widths, i.e. the adjacent intervals of the continuous outcome variable $Y_i^*$ determining the descrete outcome $Y_i$ are of the same length. This, however, does not have to hold in general and these intervals might not follow any particular pattern.\footnote{\ul{Recently,} \textcite{Buri2020} \ul{and} \textcite{Tutz2021} \ul{proposed score-free methods based on random forests that do not rely on the underlying continuous intervals of the observed ordered classes.}} In order to address this issue, \textcite{Hornung2017} proposes an ordinal forest method, which optimizes these interval widths by maximizing the out-of-bag (OOB) prediction performance of the forests.\footnote{This approach could be regarded as semiparametric as it uses the nonparametric structure of the trees and assumes a particular parametric distribution (standard normal) within its optimization procedure.} However, on the contrary to the approach of \textcite{Hothorn2006}, the forest algorithm used is based on the forest as developed by \textcite{Breiman2001}, while the primary target is to predict the ordinal class and the choice probabilities are obtained as relative frequencies of trees predicting the particular class. \ul{As such, each tree predicts the most probable value of the ordered categorical outcome. Thereupon, the forest prediction for the conditional choice probability is computed as the share of trees predicting the particular categorical value of the ordered outcome. This is in contrast to the estimation scheme by} \textcite{Hothorn2006}, \ul{where the probability prediction step occurs at the level of trees, instead of at the level of forest as is the case here.} \textcite{Hornung2017} shows better prediction performance of such ordinal forests which optimize the class widths of $Y_i^*$ in comparison to the conditional forests. Without the optimization step, the author denotes such forest as the naive ordinal forest.\footnote{A more detailed description of the conditional as well as the ordinal forest is provided in Appendix \ref{Appendix:hornik} and \ref{Appendix:hornung}, respectively.}

While both of the discussed approaches take the ordering information of the outcomes into account, they focus mainly on prediction and variable importance without considering estimation of the marginal effects or the associated inference for the effects which are a fundamental part of the classical econometric ordered choice models. In addition, although both of these methods demonstrate good predictive performance, none of them provides theoretical guarantees with regards to the distribution of these predictions. Further, it is worth to mention that in practice both methods suffer from considerable computational costs. In case of the conditional forest, the additional permutation tests that need to be performed to evaluate the test statistic at each split result in a considerably longer computation time. For the ordinal forest, the additional optimization step for the class widths requires a prior estimation of a large number of forests (1000 by default) which also leads to a substantially longer computation time (see Tables \ref{tab:abs_time} and \ref{tab:rel_time} in Appendix \ref{Appendix:tuning} for further details).

There is also a strand of literature which is concerned with the estimation of ordered outcome models in high-dimensional settings based on regularization methods. Examples of this approach include penalized ordered outcome models by \textcite{Wurm2017} who make use of a standard ordered logit/probit regression while introducing an elastic net penalization term. \textcite{Harrell2015} describes a cumulative logit model with a ridge type of penalty. \textcite{Archer2014} implement the GMIFS (generalized monotone incremental forward stagewise) algorithm for penalized ordered outcome models which is similar to the Lasso type penalty. However, although the penalized models can deal with high dimensions, when the true model is relatively "sparse", they \ul{nevertheless belong to a specific parametric class such as the ordered logit/probit} \parencite[see][]{Hastie2009}. \ul{Such models can only become more flexible, and thus partially relax the parametric assumptions when generating} a large number of polynomials and interactions of available covariates prior to estimation. \ul{It follows that these models use a global approximation of the functional form and cannot learn it adaptively as tree-based approaches do. In contrast, random forests do not impose parametric assumptions and can learn any arbitrary relationship in a nonparametric way by locally adaptive estimation in small neighbourhoods of the data. It follows that random forests use a local approximation of the functional form, without any need for prior pre-processing of the data. As such, random forests are nonlinear in covariates and although there are no specific statistical tests to find such a random forest structure, essentially random forests can approximate any structure, including a global linear structure, if a sufficient amount of data is provided.} \ul{For these reasons, the remainder of this paper focuses on the forest-based methods.}

\section{Random Forests}\label{sec:RF}
Random forests as introduced by \textcite{Breiman2001} became quickly a very popular prediction method thanks to its good prediction accuracy, while being relatively simple to tune. Further advantages of random forests as a nonparametric technique are the high degree of flexibility and ability to deal with large number of predictors, while coping better with the curse of dimensionality problem in comparison to classical nonparametric methods such as kernel or local linear regression \parencite[see for example][]{Racine2008}. Random forests are based on bootstrap aggregation, i.e. the so-called bagging of single regression (or classification) trees where the covariates considered for each next split within a tree are selected at random. More precisely, the random forest algorithm draws a bootstrap sample \ul{$Z_i^*(X_i, Y_i)$} of size $N$ from the available training data for $b=1,...,B$ bootstrap replications. For each bootstrapped sample, a random-forest tree $\hat{T}_b$ is grown by recursive partitioning until the minimum leaf size is reached. At each of the splits, $m$ out of $p$ covariates chosen at random are considered. After all $B$ trees are grown in this fashion, the regression random forest estimate of the conditional mean $E[Y_i\mid X_i=x]$ is the ensemble of the trees:
\begin{align}\label{RF1}
\hat{RF}^B(x) = \frac{1}{B}\sum^{B}_{b=1}\hat{T}_b(x) \quad \text{ with } \quad \hat{T}_b(x)=\frac{1}{\mid \{ i:X_i \in L_b(x) \} \mid} \sum_{\{ i: X_i \in L_b(x) \}} Y_i,
\end{align}
where $L_b(x)$ denotes a leaf containing $x$. Single trees, if grown sufficiently deep, have a low bias, but fairly high variance. By averaging over many single trees with randomly choosing the set of observations and split covariates, the variance of the estimator is being reduced substantially. First, the variance reduction is achieved through bagging. The higher the number of bootstrap replications, the lower the variance. Second, the variance is further reduced through the random selection of covariates. The lower is the number of considered covariates for a split, the more is the correlation between the trees reduced and consequently, the bigger is the variance reduction of the average \parencite{Hastie2009}.

Another attractive feature of random forests is the weighted average representation of the final estimate of the conditional mean $E[Y_i\mid X_i=x]$. As such we can rewrite the random forest prediction as follows:
\begin{align}\label{RF2}
\hat{RF}^B(x) = \sum^N_{i=1}\hat{w}_i(x)Y_i,
\end{align}
where the weights are defined as:
\begin{align}
\hat{w}_{b,i}(x)=\frac{\mathbf{1}(\{X_i \in L_b(x) \})}{\mid \{ i:X_i \in L_b(x) \} \mid} \quad \text{ with } \quad \hat{w}_i(x)=\frac{1}{B}\sum^B_{b=1}\hat{w}_{b,i}(x).
\end{align}
As such the forest weights $\hat{w}_i(x)$ are again an average over all single tree weights. These tree weights capture if the training example $X_i$ falls into the leaf $L_b(x)$ scaled by the size of that leaf. Notice, that the weights are locally adaptive. Intuitively, random forests resemble the classical nonparametric kernel regression with an adaptive, data-driven bandwidth and with limited curse of dimensionality. One can show that in the regression case, the random forest estimate as defined in \eqref{RF1} is equivalent to the weighting estimate defined in \eqref{RF2}. This weighting perspective of random forests has been firstly suggested by \textcite{Hothorn2004} and \textcite{Meinshausen2006} in the scope of survival and quantile regression, respectively. Recently, \textcite{Athey2019} point out the usefulness of the random forest weights in various estimation tasks. In this spirit, we will later on in Section \ref{sec:inf} use the forest induced weights explicitly for inference as has been recently suggested by \textcite{Lechner2019}.

\ul{Besides the huge popularity of random forests for prediction, the statistical literature focused} on establishing asymptotic properties of random forests \ul{as well} \parencites{Meinshausen2006, Biau2012, Scornet2015, Mentch2016}. A major step towards formally valid inference has been done in a recent work by \textcite{Wager2014a} and \textcite{Wager2018} who prove consistency and asymptotic normality of random forest predictions, under some modifications of the standard random forest algorithm. These modifications concern both the tree-building procedure as well as the tree-aggregation scheme. First, the tree aggregation is now done using subsampling without replacement instead of bootstrapping. Second, the tree building procedure introduces the major and crucial condition of so-called honesty as first suggested by \textcite{Athey2016}. A tree is honest, if it does not use the same responses for both, placing splits and estimating the within-leaf \ul{predictions}. This can be achieved by \ul{the so-called} double-sample trees, which split the random subsample of training data \ul{$Z_i^*(X_i, Y_i)$} into two disjoint sets of the same size, while the one is used for placing splits and the other one for estimating the \ul{predictions}. Furthermore, for the consistency it is essential that the size of the leaves $L$ of the trees becomes small relative to the sample size as $N$ gets large.\footnote{\textcite{Wager2018} point out that the leaves need to be relatively small in all dimensions of the covariate space. This implies that the high-dimensional settings are not considered and hence the theoretical asymptotic results might not hold in such settings.} This is achieved by introducing some randomness in choosing the splitting variables. Particularly, each covariate receives a minimum amount of positive chance of a split. Such constructed tree is then said to be a random-split tree. Additionally, the trees are required to be $\alpha$-regular, meaning that after each split, both of the child nodes contain at least a fraction $\alpha$ of the training data (specifically, $\alpha\leq 0.2$ is required). Lastly, trees have to be symmetric in a sense that the order of the training data is independent of the predictor output. Overall, apart from subsampling and honesty the above conditions are not particularly binding and do not fundamentally deviate from the standard \ul{regression} random forest. \ul{Lastly, some additional regularity conditions need to be satisfied for the asymptotic arguments to hold. In particular, the data $Z_i(X_i, Y_i) \in [0,1]^p \times \mathbb{R}$ comes from \textit{i.i.d.} sampling, the $p$-dimensional covariates $X_i \sim \mathcal{U}([0,1]^p)$ are independently and uniformly distributed}, \ul{the conditional means $E[Y_i\mid X_i=x]$ and $E[Y_i^2\mid X_i=x]$ are Lipschitz-continuous, the variance is bounded away from zero, $Var[Y_i\mid X_i=x]>0$, and the number of subsampling replications is large enough to eliminate the Monte Carlo effects, while an appropriate scaling of the subsample size $s_N$ is ensured.\footnote{\ul{The condition of uniformity of covariates is due to simplicity and is not particularly binding as the result holds also with a density bounded away from zero and infinity as argued by} \textcite{Wager2018}. \ul{Furthermore, the Lipschitz-continuity of the conditional mean appears not too restrictive as the random forest estimates have in general smooth response surfaces} \hl{when $B\to\infty$, i.e. the number of bootstrap or subsampling iterations goes to infinity} \parencite{Buhlmann2002}. \ul{Lastly, the appropriate scaling of the subsample size $s_N$ does not affect the asymptotic normality, but violations might lead to asymptotic bias as pointed out by} \textcite{Wager2018}. For a detailed description of the conditions as well as of the proof, see \textcite{Wager2018}.}} \ul{Then, under the above assumptions,} the random forest predictions can be shown to be (pointwise) asymptotically Gaussian and unbiased. \ul{We use this result to provide an inference procedure for the marginal effects of the \textit{Ordered Forest} discussed in Section} \ref{sec:inf}.

\section{Ordered Forest Estimator}\label{sec:ORF}
The general idea of the \textit{Ordered Forest} estimator is to provide a flexible alternative for estimation of ordered choice models that can deal with a large-dimensional covariate space. As such, the main goal is the estimation of conditional ordered choice probabilities, i.e. $P[Y_i=m \mid X_i=x]$ as well as marginal effects, i.e. the changes in the estimated probabilities in association with changes in covariates. Correspondingly, the variability of the estimated effects is of interest and therefore a method for conducting statistical inference is provided as well. The latter two features go beyond the traditional machine learning estimators which focus solely on the prediction exercise, and complement the prediction with the same econometric output as the traditional parametric estimators.

\subsection{Conditional Choice Probabilities}
The main idea of the estimation of the ordered choice probabilities by a random forest algorithm lies in the estimation of cumulative, i.e. nested probabilities based on binary indicators. As such, for an $i.i.d$ random sample of size $N(i=1,...,N)$, consider an ordered outcome variable $Y_i \in \{1,...,M \}$ with ordered classes $m$. Then the binary indicators are given as $Y_{m,i}=\mathbf{1}(Y_i \leq m)$ for outcome classes $m=1,...,M-1$. First, the ordered model is transformed into multiple overlapping binary models which are estimated by random forests yielding the predictions for the cumulative probabilities, i.e $\hat{Y}_{m,i}=\hat{P}[Y_{m,i}=1 \mid X_i=x]$. Second, the estimated cumulative probabilities are differenced to isolate the respective class probabilities $P_{m,i}=P[Y_i=m \mid X_i=x]$. Hence the estimate for the conditional probability of the $m$-th ordered class is given by subtracting two adjacent cumulative probabilities as $\hat{P}_{m,i}=\hat{Y}_{m,i}-\hat{Y}_{m-1,i}$. Formally, the proposed estimation procedure can be described as follows:

\begin{enumerate}
\item Create $M-1$ binary indicator variables such as
\begin{align}
Y_{m,i}=\mathbf{1}(Y_i \leq m) \text{\phantom{kkt} for \phantom{kkt}} m=1,...,M-1, \label{indicator_ordered}
\end{align}
\vspace{-0.2cm}
\ul{where $m$ is known and given by the definition of the dependent variable.}
\item Estimate regression random forest for each of the $M-1$ indicators \ul{as}
\begin{align}
P[Y_{m,i}=1 \mid X_i=x] = \sum^N_{i=1}w_{m,i}(x)Y_{m,i} \text{\phantom{kkt} for \phantom{kkt}} m=1,...,M-1,
\end{align}
\vspace{-0.2cm}
where the forest weights are defined as $w_{m,i}(x)=\frac{1}{B}\sum^B_{b=1}w_{m,b,i}(x)$ with trees weights given by\\

\vspace{-0.4cm}
$w_{m,b,i}(x)=\frac{\mathbf{1}(\{X_i \in L_{b,m}(x) \})}{\mid \{ i:X_i \in L_{b,m}(x) \} \mid}$ with leaves $L_{b,m}(x)$ for a total of $B$ trees.
\item Obtain \ul{forest} predictions \ul{for each of the $M-1$ indicators as}
\begin{align}
\hat{Y}_{m,i}=\hat{P}[Y_{m,i}=1 \mid X_i=x] = \sum^N_{i=1}\hat{w}_{m,i}(x)Y_{m,i} \text{\phantom{kkt} for \phantom{kkt}} m=1,...,M-1,
\end{align}
\vspace{-0.2cm}
\ul{where $\hat{Y}_{m,i}$ are estimated cumulative probabilities.}
\item Compute \ul{ordered} probabilities for each \ul{distinct} class as
\begin{align}
\hat{P}_{m,i}&=\hat{Y}_{m,i}-\hat{Y}_{m-1,i} \quad \text{\phantom{.....} for \phantom{.}} \quad m=2,...,M \label{eq3}
\end{align}
\vspace{-0.2cm}
\ul{with}
\vspace{-0.2cm}
\begin{align}
\hspace{0.4cm}\hat{Y}_{M,i}&=1 \quad \text{\phantom{.................} and } \quad \hat{P}_{1,i}=\hat{Y}_{1,i} \label{eqq1}
\end{align}
\vspace{-0.2cm}
and
\vspace{-0.2cm}
\begin{align}
\hspace{0.7cm}\hat{P}_{m,i}&=0 \quad \text{\phantom{..................} if } \quad \text{\phantom{..}}\hat{P}_{m,i} < 0 \label{eq4} \\[5pt] 
\hspace{0.4cm}\hat{P}_{m,i}&=\frac{\hat{P}_{m,i}}{\sum^{M}_{m=1}\hat{P}_{m,i}} \quad \text{\phantom{.} for } \quad m=1,...,M, \label{eq5}
\end{align}
\end{enumerate}
where equation \eqref{eq3} makes use of the cumulative (nested) probability feature. As such, the predicted values of two subsequent binary indicator variables $Y_{m,i}$ are subtracted from each other to isolate the probability of the higher order class.\footnote{Similar transformations of an ordered model into multiple binary models have been proposed in the classification literature. \textcite{Kwon1997} introduce the so-called ordinal pairwise partitioning method in the context of neural networks. Yet the closest to our work is the approach by \textcite{Frank2001} who make use of the cumulative model explicitly.} In equation \eqref{eqq1} the first part is given by construction as follows from the indicator function \eqref{indicator_ordered} that all values of $Y_i$ fullfil the condition for $m=M$ and from the fact that cumulative probabilities must add up to 1. The second part defines the probability of the lowest value of the ordered outcome variable. This follows directly from the random forest estimation as the created indicator variable $Y_{1,i}$ describes the very lowest value of the ordered outcome classes and as such, no modification of its predicted value is necessary to obtain a valid probability prediction. Line \eqref{eq4} ensures that the computed probabilities from \eqref{eq3} do not become negative. This might occasionally happen especially if the respective outcome classes comprise of very few observations. This issue is well-known also from the generalized ordered logit model where the parallel regression assumption is relaxed \parencite[see][p. 155]{McCullagh1989}. However, even though it is possible in theory, growing honest trees seems to largely prevent this from happening in practice. Lastly, in case if negative predictions should occur and thus being set to zero, \eqref{eq5} defines a normalization step to ensure that all class probabilities sum up to 1. Notice, that such an approach requires estimation of $M-1$ forests in the training data, which might appear to be computationally expensive. However, given that most empirical problems involve a rather limited number of outcome classes (usually not exceeding 10 distinct classes) and the relatively fast estimation of standard regression forest\footnote{The computational speed of the regression forests depends on many tuning parameters, of which the number of bootstrap replications, i.e. grown trees is the most decisive one.} without any additional permutation test nor optimization steps needed as is the case for the conditional or the ordinal forests, respectively, the here proposed procedure shall be computationally advantageous (see Tables \ref{tab:abs_time} and \ref{tab:rel_time} in Appendix \ref{Appendix:tuning}).

\subsection{Marginal Effects}
After estimating the conditional \ul{ordered} choice probabilities, it is of interest to investigate how the estimated probabilities are associated with covariates, i.e. how the changes in the covariates translate into changes in the probabilities. Typical measures for such relationships in standard nonlinear econometrics are the marginal, or, partial effects. Thus, for nonlinear models, including ordered choice models, two fundamental measures are of common interest, mean marginal effects and marginal effects at the mean of the covariates.\footnote{One can evaluate the marginal effect at any arbitrarily chosen value. The default option is usually the mean or the median.} These quantities are feasible also in the case of the \textit{Ordered Forest} estimator. Due to the character of the ordered choice model, the marginal effects on all probabilities of different values of the ordered outcome classes are estimated, i.e. $P[Y_i=m \mid X_i=x]$. In the following, let us define the marginal effect for an element $x^k$ of $X_i$ as follows:
\begin{align}\label{eq:me1}
ME_i^{k,m}(x)=\frac{\partial P[Y_i=m \mid X_i^k=x^k, X_i^{-k}=x^{-k}]}{\partial x^k} ,
\end{align}
with $X_i^k$ and $X_i^{-k}$ denoting the elements of $X_i$ with and without the $k$-th element, respectively.\footnote{As a matter of notation, capitals denote random variables, whereas small letters refer to the particular realizations of the random variable.}
Next, let us define the marginal effect for categorical variables as a discrete change in the following way:
\begin{align}\label{eq:me3}
ME_i^{k,m}(x)=P[Y_i=m \mid X_i^k=\left\lceil{x^k}\right\rceil, X_i^{-k}=x^{-k}]-P[Y_i=m \mid X_i^k=\left\lfloor{x^k}\right\rfloor, X_i^{-k}=x^{-k}] ,
\end{align}
where $\left\lceil{\cdot}\right\rceil$ and $\left\lfloor{\cdot}\right\rfloor$ denote upper and lower integer values, respectively, such that a difference of one unit is respected. Notice, that in the case of a binary variable this leads to the respective probabilities being evaluated at $\left\lceil{x^k}\right\rceil = 1$ and $\left\lfloor{x^k}\right\rfloor =0$ as is usual for ordered choice models. From the above definitions of marginal effects, we obtain the desired quantity of interest, i.e. the marginal effect at mean by evaluating $ME_i^{k,m}(x)$ at the population mean of $X_i$, for which the sample mean is a natural proxy. The mean marginal effect is obtained by taking sample averages of $ME_i^{k,m}(x)$, i.e. $\frac{1}{N}\sum^{N}_{i=1}ME_i^{k,m}(x)$.

Having formally defined the desired marginal effects, the next issue is the estimation of these effects. For the case of binary and categorical covariates $X^k$, this appears straightforward as the estimated \textit{Ordered Forest} model provides predicted values for all probabilities at all values $x^k$. As such, the estimate $\hat{ME}_i^{k,m}(x)$ of marginal effects defined in equation \eqref{eq:me3} remains as a difference of the two conditional probabilities estimated by the \textit{Ordered Forest}. However, it is less obvious for continuous variables, where derivatives are needed. As the estimates of the choice probabilities are averaged leaf means, the marginal effect is not explicit and not differentiable. In the nonparametric literature \textcite{Stoker1996} and \textcite{Powell1996}, among others, are directly concerned with estimating average derivatives. However, these methods lack convenience of estimation and have thus not been widely adopted by empirical researchers.\footnote{\ul{The issues range from estimation difficulty, possibly non-standard distribution of the estimator, to ambiguous choices of nuisance parameters.}} Therefore, we approximate the derivative by a discrete analogue based on the definition of a derivative as follows:
\begin{align}\label{eq:me4}
\hat{ME}_i^{k,m}(x)&=\frac{\hat{P}[Y_i=m \mid X_i^k=x^{kU}, X_i^{-k}=x^{-k}]-\hat{P}[Y_i=m \mid X_i^k=x^{kL}, X_i^{-k}=x^{-k}]}{x^{kU}-x^{kL}}\\
&=\frac{\hat{P}_{m,i}(x^{kU}) - \hat{P}_{m,i}(x^{kL})}{x^{kU}-x^{kL}} ,
\end{align}
with $x^{kU},x^{kL}$ \ul{defined as $x^{kU}=x^{k}+h \cdot \sigma(x^{k})$ and $x^{kL}=x^{k}-h \cdot \sigma(x^{k})$, while ensuring that the support of $x^k$ is respected, and where $\sigma(\cdot)$ denotes standard deviation and $h$ controls the window size for evaluating the marginal effect. We recommend to set $h=0.1$ to achieve accurate evaluation at the margin.}\footnote{\ul{We have additionally experimented with $h=0.5$ and $h=1$ which resulted in incrementally larger effect sizes. Generally, the lower the window size $h$, the more local the effect and the higher the window size $h$, the more global the effect becomes. As} \textcite{Burden2011} \ul{point out, the window size $h$ should not be chosen too small due to the instability of the numerical derivative approximations. In the software implementation in the \textsf{R} package \textsf{orf}, users can control this parameter by changing the argument \textsf{window}. See} \textcite{orf2019} \ul{for more details.}} Hence, the approximation targets the marginal change in the value of the covariate $X_i^k$. Notice, that such an estimation of marginal effects is much more demanding exercise than solely predicting the choice probabilities. Therefore, it is expected that considerably more subsampling iterations are needed for a good performance.

\subsection{Inference}\label{sec:inf}
\ul{The building block of the \textit{Ordered Forest} are the estimates of conditional probabilities
such as $P[Y_{m,i}=1\mid X_i=x]$. Particularly, the \textit{Ordered Forest} makes use of linear combinations of such probability estimates made by the random forest for both the conditional ordered choice probabilities as well as for the corresponding marginal effects. Therefore, for conducting inference on these quantities, it is sufficient to ensure that the underlying estimates of conditional probabilities are asymptotically normally distributed. Here, we combine the results of} \textcite{Wager2018} and \textcite{Lechner2019}. \ul{First, we use the asymptotic results of} \textcite{Wager2018} \ul{who show that the consistency and normality of random forest predictions hold also when dealing with binary outcomes, and thus also hold for probability predictions of type $P[Y_{m,i}=1\mid X_i=x]$. Hence, the final \textit{Ordered Forest} estimates for the conditional ordered choice probabilities and the marginal effects, based on a forest algorithm respecting the conditions discussed in Section} \ref{sec:RF}, \ul{inherit the consistency and normality properties. Second, we adapt the inference procedure for random forests as developed by} \textcite{Lechner2019} \ul{to estimate the variance of the conditional ordered choice probabilities and the corresponding marginal effects.}

The here proposed method for conducting approximate inference of the estimated marginal effects utilizes the weight-based representation of random forest predictions and adapts the weight-based inference proposed by \textcite{Lechner2019} for the case of the \textit{Ordered Forest} estimator.\footnote{See also \textcite{Lechner2002} and \textcite{Imbens2006} \ul{for related approaches.}} The main condition for conducting weight-based inference is to ensure that the weights and the outcomes are independent. In general, the weights are functions of the covariates for the observation $i$ and the training data. In order to estimate the variance of the marginal effects successfully, the conditioning set of the weights must be reduced. Therefore, if the observation $i$ is not part of the training data and there is $i.i.d.$ sampling, then the weights depend only on the observation $i$ and are furthermore independent of the outcomes \parencite[for a formal analysis, see][]{Lechner2019}. This is achieved through sample splitting where one half of the sample is used to build the forest, and thus to determine the weights, and the other half to estimate the effects using the respective outcomes. Notice that this condition goes beyond honesty as defined in \textcite{Wager2018} as this requires not only estimating honest trees but estimating honest forest as a whole. \ul{The reason for this is the fact that the weights are not based on the estimated trees, but on the estimated forest. Therefore, to ensure independence between the weights and outcomes, the honesty condition must be w.r.t. to the forest and it is not sufficient to build honest trees only.} This comes, however, at the expense of the efficiency of the estimator as less data are effectively used. Nevertheless, the simulation evidence in \textcite{Lechner2019} suggests that this efficiency loss is small, if present at all.\footnote{The so-called cross-fitting to avoid the efficiency loss as suggested by \textcite{Chernozhukov2018c} does not appear to be applicable here as the independence of the weights and the outcomes would not be ensured.}

Since the \textit{Ordered Forest} estimator is based on differences of random forest predictions for adjacent outcome categories, also the covariance term enters the variance formula of the final estimator\footnote{One could avoid the covariance term with an additional sample split, which might, however, further lead to a decreased efficiency of the estimator.} as opposed to the Modified Causal Forests developed in \textcite{Lechner2019}. Further, the estimation of marginal effects is based on differences of single \textit{Ordered Forest} predictions which also needs to be taken into account.\footnote{Notice, that for outcome classes $m=1$ and $m=M$, the variance formula simplifies substantially.} Let us first rewrite the marginal effects in terms of weighted means of the outcomes as follows:
\scalebox{.9}{\parbox{.99\linewidth}{%
\begin{align*}
&\hat{ME}_i^{k,m}(x)=\frac{\hat{P}_{m,i}(x^{kU}) - \hat{P}_{m,i}(x^{kL})}{x^{kU}-x^{kL}}\\[5pt]
&= \frac{1}{x^{kU}-x^{kL}}\cdot\Bigg(\Bigg[\sum^N_{i=1}\hat{w}_{i,m}(x^{kU})Y_{i,m} - \sum^N_{i=1}\hat{w}_{i,m-1}(x^{kU})Y_{i,m-1}\Bigg]-\Bigg[\sum^N_{i=1}\hat{w}_{i,m}(x^{kL})Y_{i,m} - \sum^N_{i=1}\hat{w}_{i,m-1}(x^{kL})Y_{i,m-1}\Bigg]\Bigg)\\[2pt]
&=  \frac{1}{x^{kU}-x^{kL}}\cdot\Bigg(\Bigg[\sum^N_{i=1}\hat{w}_{i,m}(x^{kU})Y_{i,m} - \sum^N_{i=1}\hat{w}_{i,m}(x^{kL})Y_{i,m}\Bigg] - \Bigg[\sum^N_{i=1}\hat{w}_{i,m-1}(x^{kU})Y_{i,m-1} - \sum^N_{i=1}\hat{w}_{i,m-1}(x^{kL})Y_{i,m-1}\Bigg]\Bigg)\\[2pt]
&=  \frac{1}{x^{kU}-x^{kL}}\cdot\Bigg(\sum^N_{i=1}\tilde{w}_{i,m}(x^{kU}x^{kL})Y_{i,m} - \sum^N_{i=1}\tilde{w}_{i,m-1}(x^{kU}x^{kL})Y_{i,m-1}\Bigg) ,
\end{align*}
}}\\
where $\tilde{w}_{i,m}(x^{kU}x^{kL})=\hat{w}_{i,m}(x^{kU})-\hat{w}_{i,m}(x^{kL})$, and $\tilde{w}_{i,m-1}(x^{kU}x^{kL})=\hat{w}_{i,m-1}(x^{kU})-\hat{w}_{i,m-1}(x^{kL})$ are the new weights defining the marginal effect. As such the quantity of interest 
for inference becomes the variance of the above expression given as:\\
\scalebox{.9}{\parbox{.99\linewidth}{%
\begin{align*}
Var\bigg(\hat{ME}_i^{k,m}(x)\bigg)&= Var\Bigg(\frac{1}{x^{kU}-x^{kL}}\cdot\Bigg(\sum^N_{i=1}\tilde{w}_{i,m}(x^{kU}x^{kL})Y_{i,m} - \sum^N_{i=1}\tilde{w}_{i,m-1}(x^{kU}x^{kL})Y_{i,m-1}\Bigg)\Bigg)\\[2pt]
&= Var\Bigg( \frac{\sum^N_{i=1}\tilde{w}_{i,m}(x^{kU}x^{kL})Y_{i,m}}{x^{kU}-x^{kL}} \Bigg) + Var\Bigg( \frac{\sum^N_{i=1}\tilde{w}_{i,m-1}(x^{kU}x^{kL})Y_{i,m-1}}{x^{kU}-x^{kL}} \Bigg)\\[2pt]
&- 2\cdot Cov\Bigg( \frac{\sum^N_{i=1}\tilde{w}_{i,m}(x^{kU}x^{kL})Y_{i,m}}{x^{kU}-x^{kL}}; \frac{\sum^N_{i=1}\tilde{w}_{i,m-1}(x^{kU}x^{kL})Y_{i,m-1}}{x^{kU}-x^{kL}} \Bigg) ,
\end{align*}
}}\\
which suggests the following estimator for the variance:\footnote{Here, we estimate the variance with sample counterparts. An alternative approach, as in \textcite{Lechner2019}, would be to first apply the law of total variance and, second, estimate the conditional moments by nonparametric methods. However, due to the presence of the covariance term the conditioning set contains 2 variables which causes the convergence rate to decrease and hence such variance estimation might even result in less precise estimates, depending on the sample size.}\\
\scalebox{.84}{\parbox{.99\linewidth}{%
\begin{align*}
&\hat{Var}\bigg(\hat{ME}_i^{k,m}(x)\bigg)=\frac{N}{N-1} \cdot \frac{1}{(x^{kU}-x^{kL})^2} \cdot \\[2pt]
& \cdot \Bigg( \sum^N_{i=1}\bigg(\tilde{w}_{i,m}(x^{kU}x^{kL})Y_{i,m} -\frac{1}{N}\sum^N_{i=1}\tilde{w}_{i,m}(x^{kU}x^{kL})Y_{i,m}\bigg)^2 + \sum^N_{i=1}\bigg(\tilde{w}_{i,m-1}(x^{kU}x^{kL})Y_{i,m-1} -\frac{1}{N}\sum^N_{i=1}\tilde{w}_{i,m-1}(x^{kU}x^{kL})Y_{i,m-1}\bigg)^2\\[2pt]
&- 2 \cdot \sum^N_{i=1}\bigg(\tilde{w}_{i,m}(x^{kU}x^{kL})Y_{i,m} -\frac{1}{N}\sum^N_{i=1}\tilde{w}_{i,m}(x^{kU}x^{kL})Y_{i,m}\bigg) \cdot \bigg(\tilde{w}_{i,m-1}(x^{kU}x^{kL})Y_{i,m-1} -\frac{1}{N}\sum^N_{i=1}\tilde{w}_{i,m-1}(x^{kU}x^{kL})Y_{i,m-1}\bigg) \Bigg) ,
\end{align*}
}}\\
where for the marginal effects at the mean of the covariates the weights $\tilde{w}_{i,m}(x^{kU}x^{kL})$ and the scaling factor $1/(x^{kU}-x^{kL})^2$ are evaluated at the respective sample means, whereas for the mean marginal effects the average of the weights $\frac{1}{N} \sum^{N}_{i=1} \tilde{w}_{i,m}(x^{kU}x^{kL})$ and of the scaling factor $1/\big(\frac{1}{N} \sum^{N}_{i=1}(x^{kU}-x^{kL})\big)^2$ is used. Notice also the fact that the scaling factor drops out in the case of categorical covariates. According to the simulation study in \textcite{Lechner2019} the weight-based inference in case of the Modified Causal Forests tends to be rather conservative for the individual effects and rather accurate for aggregate effects. The results from the here conducted empirical application resemble this pattern where inference for the marginal effects at the mean of the covariates is more conservative in comparison to inference for the mean marginal effects (see \ul{also} Appendix \ref{AppendixME} \ul{for a comparison}).

\section{Monte Carlo Simulation}\label{sec:MC}
In order to investigate the finite sample \ul{performance} of the proposed \textit{Ordered Forest} estimator, we perform a Monte Carlo simulation study comparing competing estimators for ordered choice models based on the random forest algorithm. As a parametric benchmark, we take the ordered logistic regression. The considered models are specifically the following:
\begin{enumerate*}[label=(\roman*)]
\item ordered logit \parencite{McCullagh1980},
\item naive ordinal forest \parencite{Hornung2017},
\item ordinal forest \parencite{Hornung2017},
\item conditional forest \parencite{Hothorn2006}, and
\item \textit{Ordered Forest} \ul{as developed in Section} \ref{sec:ORF}
\end{enumerate*}.
Within the simulation study the \textit{Ordered Forest} estimator is analyzed more closely to study the finite sample \ul{performance} of the estimator depending on the particular forest building schemes and the way the ordering information is being taken into account. Regarding the former we study the \textit{Ordered Forest} based on the standard random forest as in \textcite{Breiman2001}, i.e. with boostrapping and \textit{without} honesty as well as based on the \ul{adjusted} random forest as in \textcite{Wager2018}, i.e. with subsampling and \textit{with} honesty. Regarding the latter we study an alternative approach for estimating the conditional choice probabilities which could be labelled as a 'multinomial' forest. In that case, the ordering information is not being taken into account and the probabilities of each category are estimated directly. The details of this approach are provided in Appendix \ref{AppendixMRF}. Given this, the \textit{Ordered Forest} estimator should perform better than the multinomial forest in terms of the prediction accuracy thanks to the incorporation of additional information from the ordering of the outcome classes. 

\begin{table}[h!]
\centering
\caption{General Settings of the Simulation}
\label{tab:sim}
\begin{tabular}{lr}
\toprule
\multicolumn{2}{c}{Monte Carlo} \\
\midrule
observations in training set & 200 (800) \\
observations in testing set & 10000 \\
replications & 100 \\
covariates with effect & 15 \\
trees in a forest & 1000 \\
randomly chosen covariates & $\sqrt{p}$ \\
minimum leaf size\tablefootnote{Due to the conceptual differences of the conditional forests, an alternative stopping rule ensuring growing deep trees is chosen. See details in Appendix \ref{Appendix:tuning}.} & 5 \\
\bottomrule
\end{tabular}
\end{table}

General settings regarding the sample size, the number of replications, as well as forest-specific tuning parameters for the Monte Carlo simulation are depicted in Table \ref{tab:sim}. Furthermore, a detailed description of the software implementation of the respective estimators as well as the software specific tuning parameters are discussed in Appendix \ref{Appendix:tuning}.

\subsection{Data Generating Process}
In terms of the data generating process, we built upon an ordered logit model as defined in \eqref{eq:ologit} and \eqref{eq:thresholds}. As such we simulate the underlying continuous latent variable $Y_i^*$ as a linear function of regressors $X_i$, while drawing the error term $u_i$ from the logistic distribution. Then, the continuous outcome $Y_i^*$ is discretized into an ordered categorical outcome $Y_i$ based on the threshold parameters $\alpha_m$.\footnote{\ul{The thresholds are determined beforehand according to fixed threshold quantiles $\alpha_m^q$ of a large sample of $N=1'000'000$ observations of the latent $Y_i^*$ from the very same DGP to reflect the realized outcome distribution and then used afterwards in the simulations as a part of the deterministic component.}} Furthermore, the intercept term is fixed to zero, i.e. $\beta_0=0$ and thus the thresholds are relative to this value of the intercept. As a result, such DGP captures the probability of the latent variable $Y_i^*$ falling into a particular class given the location defined by the deterministic component of the model together with its stochastic component \parencite{Harden2013}.

In simulations of the data generating process, different numbers of possible discrete \ul{ordered} classes are considered, particularly $M = \{3,6,9\}$ which corresponds to the simulation set-up used in \textcite{Janitza2016} and \textcite{Hornung2017}. Further, both equal class widths, i.e. equally spaced threshold parameters $\alpha_m$, as well as randomly spaced thresholds, while still preserving the monotonicity of the discrete outcome $Y_i$, are considered. For the latter, the threshold quantiles are drawn from the uniform distribution, i.e. $\alpha_m^q \sim U(0,1)$ and ordered afterwards. For the former, the threshold quantiles are equally spaced between 0 and 1 depending on the number of classes. The $\beta$ coefficients are specified as having fixed coefficient size, namely $\beta_1,...,\beta_5 = 1$, $\beta_6,...,\beta_{10} = 0.75$ and $\beta_{11},...,\beta_{15} = 0.5$ as is also the case in \textcite{Janitza2016} and \textcite{Hornung2017}. Moreover, an option for nonlinear effects is introduced, too. As such, the \ul{covariates do not enter the functional form linearly,} but are given by a sine function $sin(2X_i)$ as for example in \textcite{Lin2014}, which is hard to model as opposed to other nonlinearities such as polynomials or interactions. The set of covariates $X_i$ is drawn from the multivariate normal distribution with zero mean and a pre-specified variance-covariance matrix $\Sigma$, i.e. $X_i \sim \mathcal{N}(0,\Sigma)$, where $\Sigma$ is specified either as an identity matrix and as such implying zero correlation between regressors, or it is specified to have a specific correlation structure between regressors\footnote{Note that with a too high multicollinearity, the ordered logit model breaks down. With restricting the level of multicollinearity, the logit model can be still reasonably compared to the other competing methods.} as follows:
\begin{align*}
\rho_{i,j} = \begin{cases}
1 \quad &\text{ for } \quad i=j \\
0.8 \quad &\text{ for } \quad i\neq j; i,j \in \{ 1,3,5,7,9,11,13,15 \} \\
0  \quad &\text{ otherwise },
\end{cases}
\end{align*}
which is inspired by the correlation structure from the simulations in \textcite{Janitza2016} and \textcite{Hornung2017}. Further, an option to include additional variables with zero effect is implemented as well. As such, another 15 covariates are added to the covariate space with $\beta_{16}=...=\beta_{30}=0$ from which 10 are again drawn from the normal distribution with zero mean and unit variance, i.e. $X^c_{i,0} \sim \mathcal{N}(0,1)$ and 5 are dummies drawn from the binomial distribution, i.e. $X^d_{i,0} \sim \mathcal{B}(0.5)$. As the performance of the \textit{Ordered Forest} estimator in high-dimensional settings is of particular interest, due to yet not fully understood theoretical properties in such settings, we include an option for additionally enlarging the covariate space with 1000 zero effect covariates $X_{i,0} \sim \mathcal{N}(0,1)$, effectively creating a setting with $p>>N$. In the high-dimensional case the ordered logit is excluded from the simulations for obvious reasons. Overall, considering all the possible combinations for specifying the DGP, we end up with 72 different DGPs.\footnote{For the low-dimensional setting we have $n=4$ options for the DGP settings, out of which we can choose from none to all of them, whereby the ordering does not matter, we end up with 16 possible combinations as given by the formula $\sum_{r=0}^{n} \binom{n}{r}$, each for 3 possible numbers of outcome classes resulting in 48 different DGPs. For the high-dimensional setting we have $n=3$ options as the additional noise variables are always considered. This for all 3 distinct numbers of outcome classes yields 24 different DGPs.} For all of them we simulate a training dataset of size $N=200$ and a testing dataset of size $N=10'000$ for evaluating the prediction performance of the considered methods. \ul{We simulate the large testing set for three main reasons. First, the large testing set enables us to reduce the prediction noise and thus provides a more reliable measure for average out-of-sample performance of the estimators. Second, the large testing set also helps to reduce the simulation noise and thus to obtain more precise estimates for the performance measures. Third, we choose the large testing set to ensure further comparability with the simulation studies performed by} \textcite{Janitza2016} and \textcite{Hornung2017}. Note that such a large testing set is also common choice in many other simulation studies \parencites(see e.g.)(){Jacob2020}[or][]{Knaus2021}. Further, we focus more closely on the simulation designs corresponding to the least and the most complex DGPs for which we simulate also a training set of size $N=800$. The former DGP (labelled as simple DGP henceforth) corresponds exactly to an ordered logit model as in \eqref{eq:ologit} with equal class widths, uncorrelated covariates with linear effects and without any additional zero effect variables. The latter DGP (labelled as complex DGP henceforth) features random class widths, correlated covariates with nonlinear effects and additional zero effect variables. For each replication, we estimate the model on the training set and evaluate the predictions on the testing set, for all tested methods.

\subsection{Evaluation Measures}\label{eval_measures}
In order to properly evaluate the prediction performance we use two measures of accuracy, namely the mean squared error (MSE) and the ranked probability score (RPS). The former evaluates the error of the estimated conditional choice probabilities as a squared difference from the true values of the conditional choice probabilities. Given our simulation design, we know these true values, which are given as in equation \eqref{eq:ologitprobs}. Hence, we can define the Monte Carlo average MSE as:
\begin{align*}
AMSE= \frac{1}{R}\sum^{R}_{j=1}\frac{1}{N}\sum^{N}_{i=1}\frac{1}{M}\sum^{M}_{m=1}\bigg(P[Y_{i,j}=m \mid X_{i,j}=x]-\hat{P}[Y_{i,j}=m \mid X_{i,j}=x]\bigg)^2 ,
\end{align*}
where $j$ refers to the $j$-th simulation replication, while $R$ being the total number of replications. The second measure, the RPS as developed by \textcite{Epstein1969} is arguably the preffered measure for the evaluation of probability forecasts for ordered outcomes as it takes the ordering information into account \parencites(see)(){Gneiting2007}[and][]{Constantinou2012}. The Monte Carlo average RPS can be defined as follows:
\begin{align*}
ARPS = \frac{1}{R}\sum^{R}_{j=1}\frac{1}{N}\sum^N_{i=1}\frac{1}{M-1}\sum^M_{m=1}\bigg(P[Y_{i,j}\leq m \mid X_{i,j}=x]-\hat{P}[Y_{i,j}\leq m \mid X_{i,j}=x]\bigg)^2 ,
\end{align*}
where on the contrary to the MSE, the difference between the cumulative choice probabilities is measured. The RPS can be seen as a generalization of the Brier Score  \parencite{Brier1950} for multiple, ordered outcomes. As such, it measures the discrepancy between the predicted cumulative distribution function and the true one. Nevertheless, although the ordering information is taken into account, the relative distance between the classes is not reflected \ul{as pointed out by} \textcite{Janitza2016}. 

\subsection{Simulation Results}\label{sec:boxplots}
\begin{figure}[h!]
\caption{Simulation Results: Simple DGP \& High Dimension}
\label{SimpleHigh}
\includegraphics[width=0.99\textwidth]{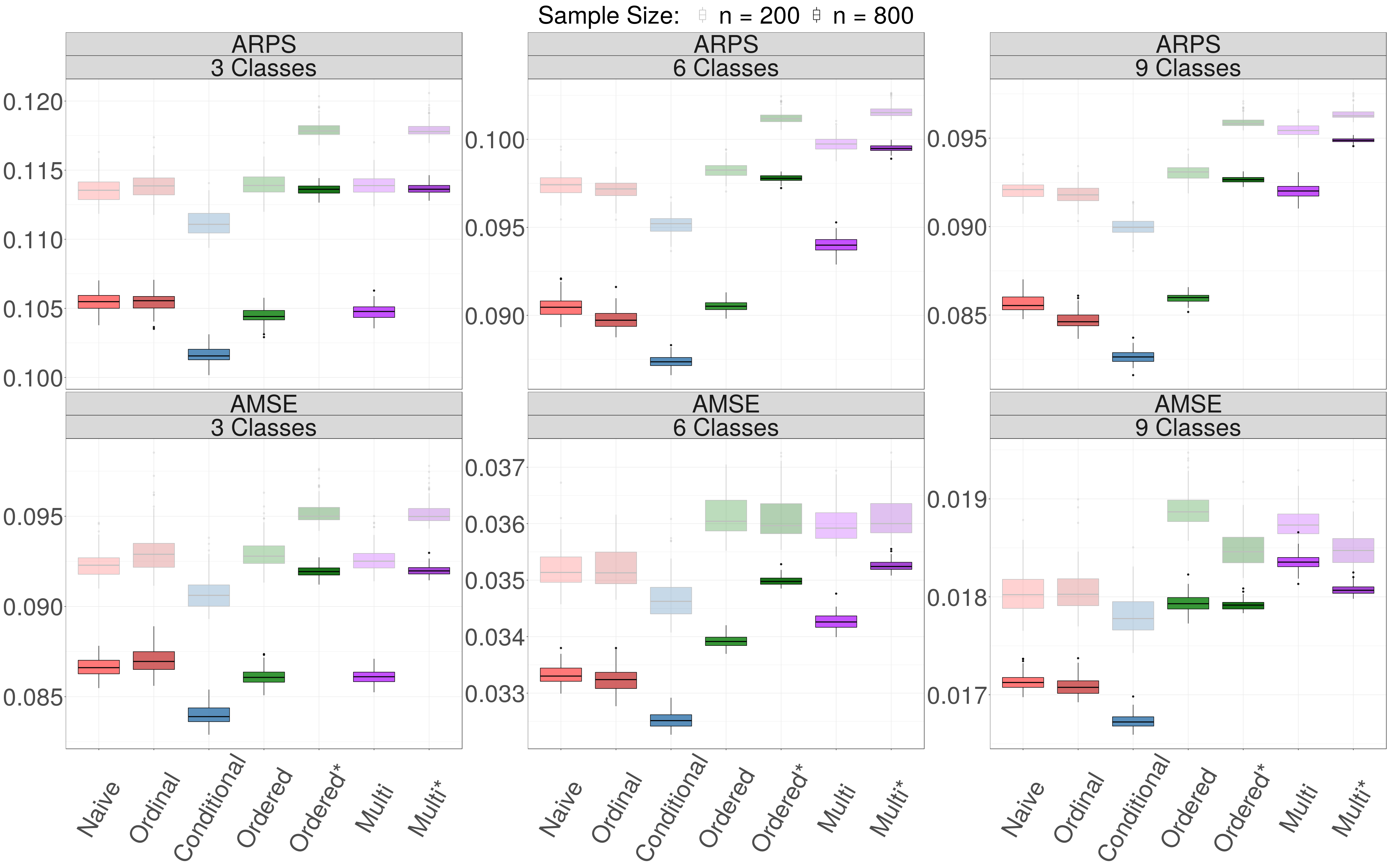}
\vspace{0.5cm}
\caption*{\footnotesize{\textit{Note: }Figure summarizes the prediction accuracy results based on 100 simulation replications. The upper panel contains the ARPS and the lower panel contains the AMSE. The boxplots show the median and the interquartile range of the respective measure. The transparent boxplots denote the results for the small sample size, while the bold boxplots denote the results for the big sample size. From left to right the results for 3, 6, and 9 outcome classes are displayed.}}
\end{figure}

\begin{figure}[h!]
\caption{Simulation Results: Complex DGP \& High Dimension}
\label{ComplexHigh}
\includegraphics[width=0.99\textwidth]{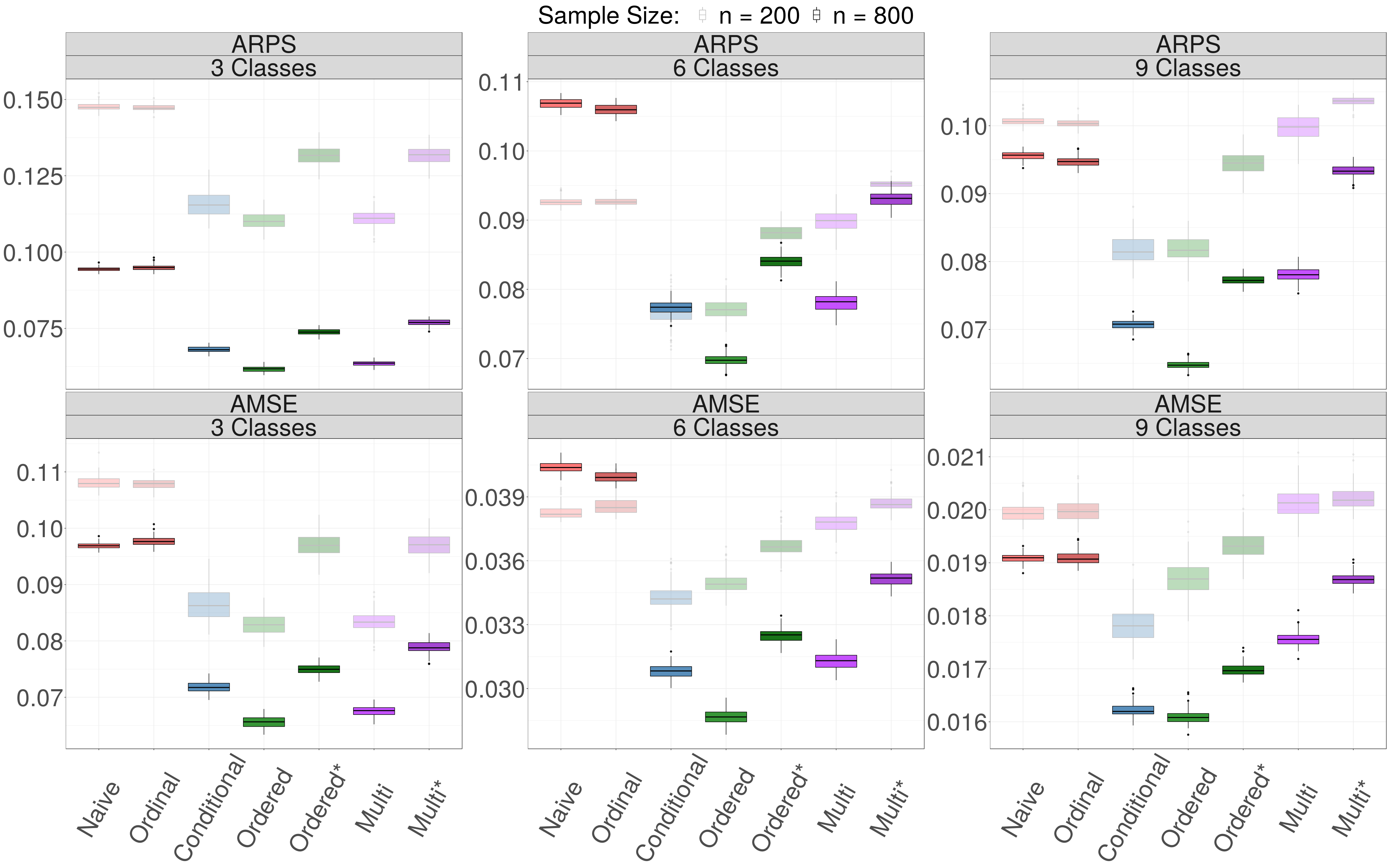}
\vspace{0.5cm}
\caption*{\footnotesize{\textit{Note: }Figure summarizes the prediction accuracy results based on 100 simulation replications. The upper panel contains the ARPS and the lower panel contains the AMSE. The boxplots show the median and the interquartile range of the respective measure. The transparent boxplots denote the results for the small sample size, while the bold boxplots denote the results for the big sample size. From left to right the results for 3, 6, and 9 outcome classes are displayed.}}
\end{figure}

For the sake of brevity, here we focus mainly on the simulation results obtained for the simple and for the complex DGP, while the results for all 72 DGPs are provided in Appendix \ref{AppendixSR}. Figures \ref{SimpleLow} and \ref{ComplexLow} summarize the results for the low-dimensional setting for the simple and the complex DGP, respectively. Similarly, Figures \ref{SimpleHigh} and \ref{ComplexHigh} present the results for the simple and the complex DGP for the high-dimensional setting. The upper panels of the figures show the ARPS, the preferred accuracy measure, whereas the lower panels show the AMSE as a complementary measure. Within the figures the transparent boxplots in the background show the results for the smaller sample size along with the bold boxplots in the foreground showing the results for the bigger sample size. From left to right the figures present the results for 3, 6 and 9 outcome classes, respectively. The figures compare the prediction accuracy of the ordered logit, naive ordinal forest, ordinal forest, conditional forest, \textit{Ordered Forest} and the multinomial forest, where the asterisk $(^*)$ denotes the honest version of the last two forests considered. Further tables with more detailed results and statistical tests for mean differences in the prediction errors are listed in Appendix \ref{Appendix:simtables}.

\begin{figure}[h!]
\caption{Simulation Results: Simple DGP \& Low Dimension}
\label{SimpleLow}
\includegraphics[width=0.99\textwidth]{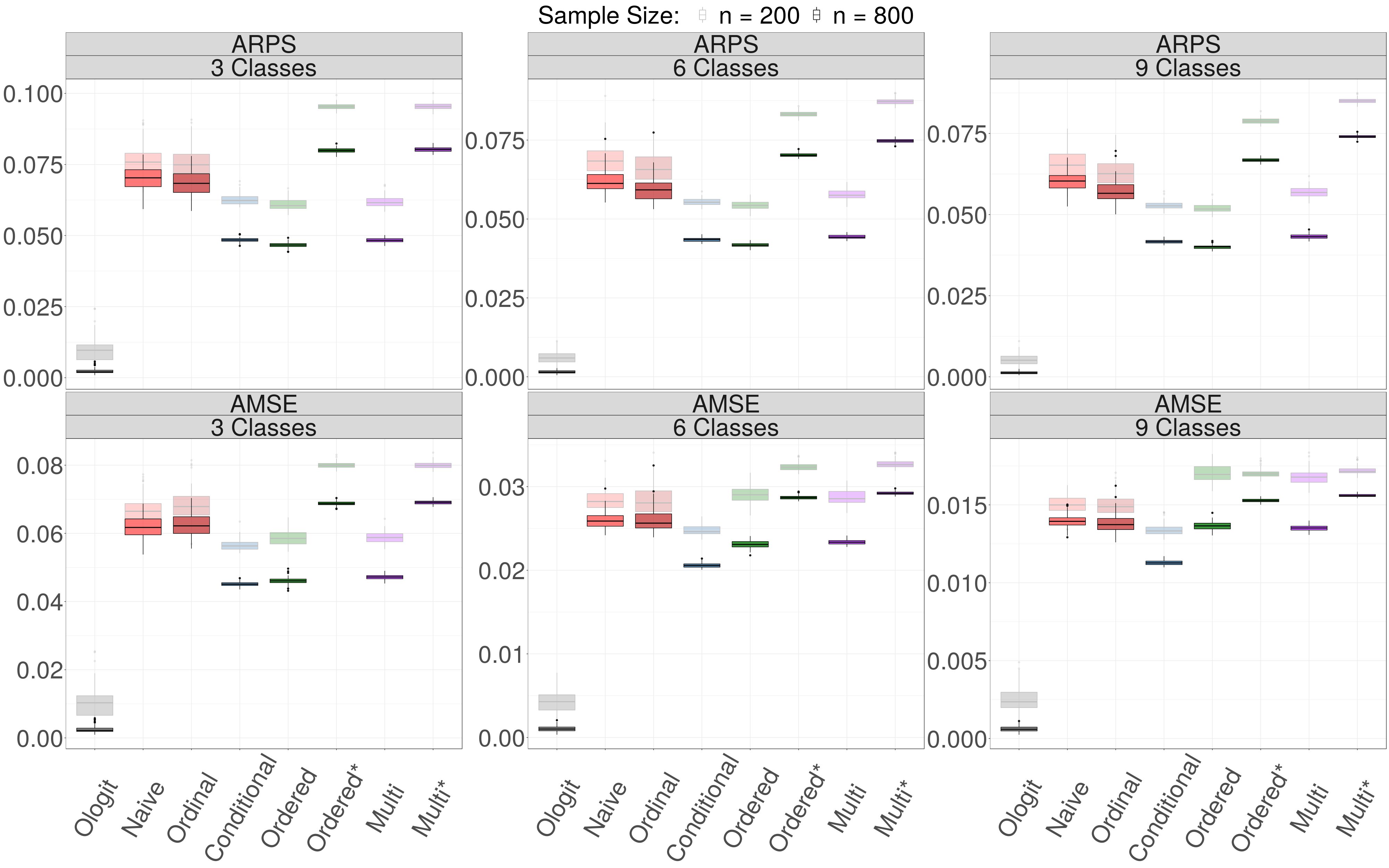}
\vspace{0.5cm}
\caption*{\footnotesize{\textit{Note: }Figure summarizes the prediction accuracy results based on 100 simulation replications. The upper panel contains the ARPS and the lower panel contains the AMSE. The boxplots show the median and the interquartile range of the respective measure. The transparent boxplots denote the results for the small sample size, while the bold boxplots denote the results for the big sample size. From left to right the results for 3, 6, and 9 outcome classes are displayed.}}
\end{figure}

In the low-dimensional setting with the simple DGP it is expected that the ordered logistic regression should perform best in terms of both the AMSE as well as the ARPS. Indeed, we do observe this results in Figure \ref{SimpleLow} as the ordered logit model performs unanimously best out of the considered models, reaching almost zero prediction error. Among the flexible forest-based estimators, the proposed \textit{Ordered Forest} belongs to those better performing methods in terms of both accuracy measures. The honest versions of the forests lag behind what points at the efficiency loss due to the additional sample splitting. Overall, the ranking of the estimators stays stable with regards to the number of outcome categories. Additional pattern common to all estimators is the lower prediction error and increased precision with growing sample size.

\begin{figure}[h!]
\caption{Simulation Results: Complex DGP \& Low Dimension}
\label{ComplexLow}
\includegraphics[width=0.99\textwidth]{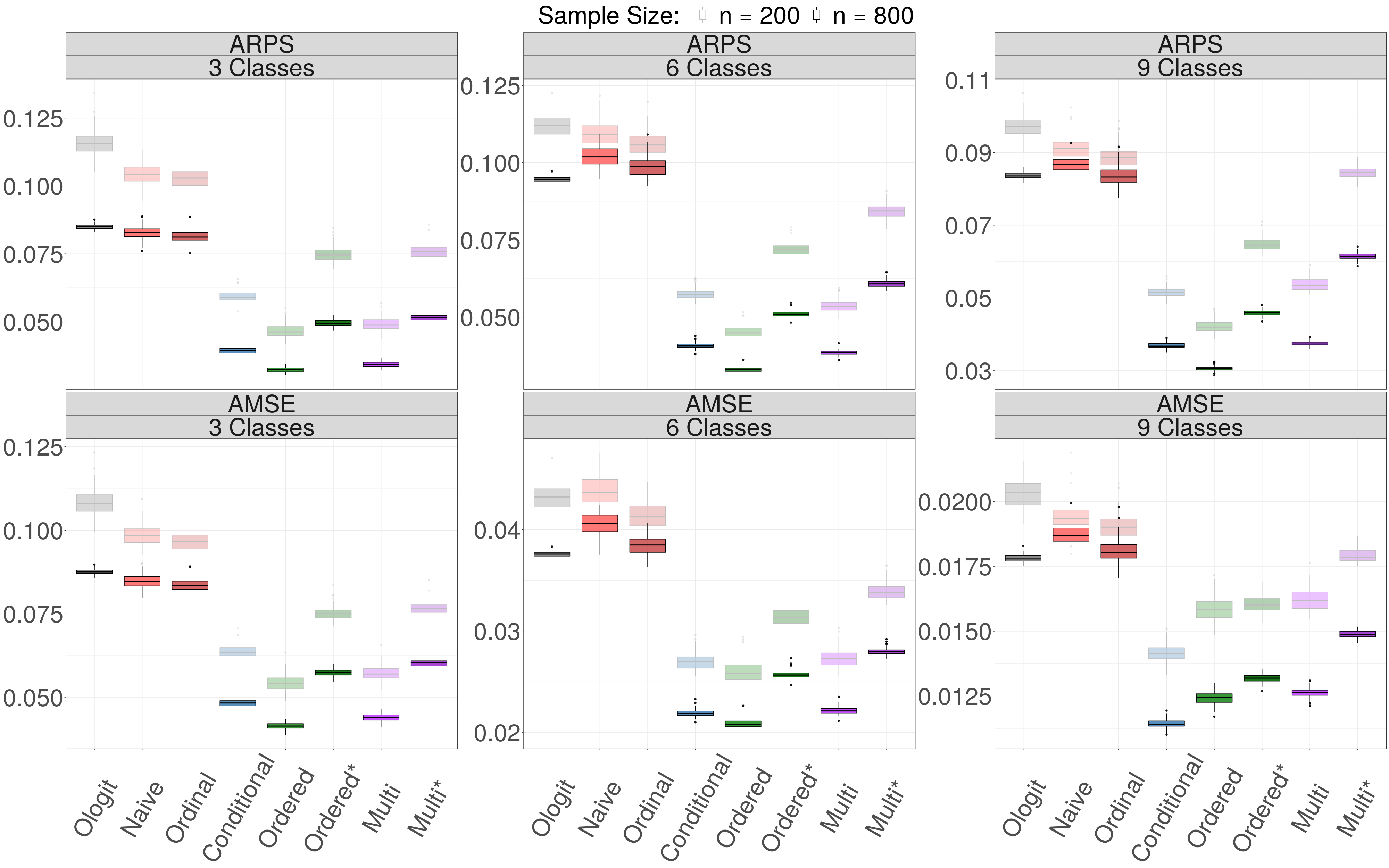}
\vspace{0.5cm}
\caption*{\footnotesize{\textit{Note: }Figure summarizes the prediction accuracy results based on 100 simulation replications. The upper panel contains the ARPS and the lower panel contains the AMSE. The boxplots show the median and the interquartile range of the respective measure. The transparent boxplots denote the results for the small sample size, while the bold boxplots denote the results for the big sample size. From left to right the results for 3, 6, and 9 outcome classes are displayed.}}
\end{figure}

In the case of the complex DGP, the performance of the flexible forest-based estimators is expected to be better in comparison to the parametric ordered logit. This can be seen in Figure \ref{ComplexLow} as the ordered logit lags behind the majority of the flexible methods in both accuracy measures. The somewhat higher prediction errors of the naive and the ordinal forest compared to the other forest-based methods might be due to their different primary target which are the ordered classes instead of the ordered probabilities as is the case for the other methods. In this respect the conditional forest exhibits considerably good prediction performance. The \textit{Ordered Forest} outperforms the competing forest-based estimators in terms of the ARPS throughout all outcome class scenarios and also in terms of the AMSE in two scenarios, being outperformed only by the conditional forest in case of 9 outcome classes. Interestingly, the multinomial forest performs very well across all scenarios. However, it is consistently worse than the \textit{Ordered Forest} with bigger discrepancy between the two the more outcome classes are considered. This points to the value of the ordering information and the ability of the \textit{Ordered Forest} to utilize it in the estimation. With regards to the sample size, we observe the same pattern as in Figure \ref{SimpleLow}.

Considering the high-dimensional setting for the case of the simple DGP, we see in Figure \ref{SimpleHigh} that the \textit{Ordered Forest} slightly lags behind the other methods, except the scenarios with 3 outcome classes. In comparison, the conditional forest performs best in terms of the ARPS as well as in terms of the AMSE. Also the naive and the ordinal forest exhibit better performance compared to the previous simulation designs. However, it should be noted that the overall differences in the magnitude of the prediction errors are much lower within this simulation design as compared to the previous designs. Further, taking a closer look at the ARPS results of the multinomial forest we clearly see that in the simple ordered design the ignorance of the ordering information really harms the predictive performance of the estimator the more outcome classes are considered. Additionally, it is interesting to see that the performance gain due to a bigger sample size seems to be much less for the honest version of the forests in the high-dimensional setting as opposed to the low-dimensional setting.

Lastly, the case of the complex DGP in the high-dimensional setting as in Figure \ref{ComplexHigh} shows some interesting patterns. In general, all of the methods exhibit good predictive performance as the loss in the prediction accuracy due to the high-dimensional covariate space is small. Additionally, although dealing with the most complex design, no substantial loss in the prediction accuracy can be observed in comparison to the less complex designs. This fact demonstrates the ability of the random forest algorithm as such to effectively cope with highly nonlinear functional forms even in high dimensions. Further, it seems that the role of the sample size is of particular importance in this complex design. On the contrary to the previous designs, where the prediction accuracy increases almost by a constant amount for all estimators and thus does not change their relative ranking, here it does not hold anymore. First, some estimators seem to learn faster than others, i.e. to have a faster rate of convergence. As such in the small sample size the \textit{Ordered Forest} has in some settings higher values of the ARPS as well as the AMSE than the conditional forest, however manages to outperform the conditional forest in the bigger training sample. This becomes most apparent in the case of 9 outcome classes. Here, the median of the ARPS is almost the same for the two methods based on the small training sample, but significantly lower for the \textit{Ordered Forest} based on the larger training sample.\footnote{See Appendix \ref{Appendix:simtables} for the detailed results of the statistical tests conducted.} Second, for \ul{the ordinal forest} the prediction accuracy even worsens with the bigger training sample, which might hint on possible convergence issues. \ul{This might possibly come from the fact that the estimator comprises multiple distinct optimization and partly nonlinear transformation steps that are tied together, but lack formal asymptotic arguments to analyse the impacts and propagation of the estimation errors into the final point estimator.} Overall, the \textit{Ordered Forest} achieves the lowest ARPS as well as AMSE within this design, closely followed by the conditional and the multinomial forest. \ul{However, the generally good performance of the conditional forest might be due to a different type of the stopping criterion, which enables growing very deep trees that are possibly deeper than the classical} \textcite{Breiman2001} \ul{trees with pre-specified minimum leaf size and as such might achieve lower bias which is then reflected in the lower values of ARPS as well as of AMSE.}

In addition to the four main simulation designs discussed above, we also inspect all 72 different DGPs to analyze the performance and the sensitivity of the \ul{\textit{Ordered Forest}} to the particular features of the simulated DGPs (for details see Appendix \ref{AppendixSR}). In case of both the low-dimensional setting, as well as the high-dimensional setting, the \textit{Ordered Forest} performs particularly well if there are nonlinear effects accompanied by near-multicollinearity of regressors as such as well as together with additional noise variables or randomly spaced thresholds. Furthermore, the honest version of the \textit{Ordered Forest} achieves consistently lower prediction accuracy in both settings. It seems that in small samples the increase in variance due to honesty dominates the reduction in the bias of the estimator. In order to further investigate the impact of the honesty feature in bigger samples as well as the convergence of the \textit{Ordered Forest}, we quadruple the size of the training set once again and repeat the main simulation for the \textit{Ordered Forest} and its honest version with $N=3'200$ observations (see Appendix \ref{Appendix:simtables} for the full results). Firstly, for both versions we observe that with growing sample size the prediction errors get lower and the precision increases. However, the rate of convergence seems to be slower than the parametric rate of $\sqrt{N}$. Secondly, we observe the same pattern as in the smaller sample sizes, namely slightly lower prediction accuracy for the honest version of the \textit{Ordered Forest} which stays roughly constant across all simulation designs. Hence, even in the biggest sample the additional variance dominates the bias reduction. However, it should be noted that for a prediction exercise honesty is an optional choice, while if inference is of interest, honesty becomes binding.

\subsection{Empirical Results}\label{sec:boxplotsemp}
Additionally to the above synthetic simulations, we explore the performance of the \textit{Ordered Forest} estimator based on real datasets\footnote{The here proposed algorithm has been already applied and is in use for predicting match outcomes in football, see \textcite{Okasa2018} and \href{https://sew.unisg.ch/de/empirische-wirtschaftsforschung/sports-economics-research-group/soccer-analytics}{\textsf{SEW Soccer Analytics}} for details.} \ul{previously used in} \textcite{Janitza2016} and \textcite{Hornung2017}. Table \ref{tab:data} summarizes the \ul{features of the} datasets and the descriptive statistics are provided in Appendix \ref{Appendix:desc}. We compare our estimator in terms of the prediction accuracy to all the estimators used in the above Monte Carlo simulation.

\begin{table}[ht]
\centering
\caption{Description of the Datasets}
\label{tab:data}
\resizebox{0.99\textwidth}{!}{
\begin{tabular}{lrllcrr}
\toprule
\multicolumn{7}{c}{\textbf{Datasets Summary}}\\
\midrule
Dataset & Sample Size & Outcome & \multicolumn{3}{c}{Class Range} & Covariates \\
\midrule
Wine Quality & 4893 & Quality Score & 1 (moderate) & - & 6 (high) & 11 \\
Mammography & 412 & Visits History & 1 (never) & - & 3 (over year) & 5 \\
Nhanes & 1914 & Health Status & 1 (excellent) & - & 5 (poor) & 26 \\
Vlbw & 218 & Physical Condition & 1 (threatening) & - & 9 (optimal) & 10 \\
Support Study & 798 & Disability Degree & 1 (none) & - & 5 (fatal) & 15 \\
\bottomrule
\end{tabular}
}
\end{table}

\begin{figure}[h!]
\caption{Cross-Validation: ARPS}
\label{cv_rps}
\includegraphics[width=0.99\textwidth]{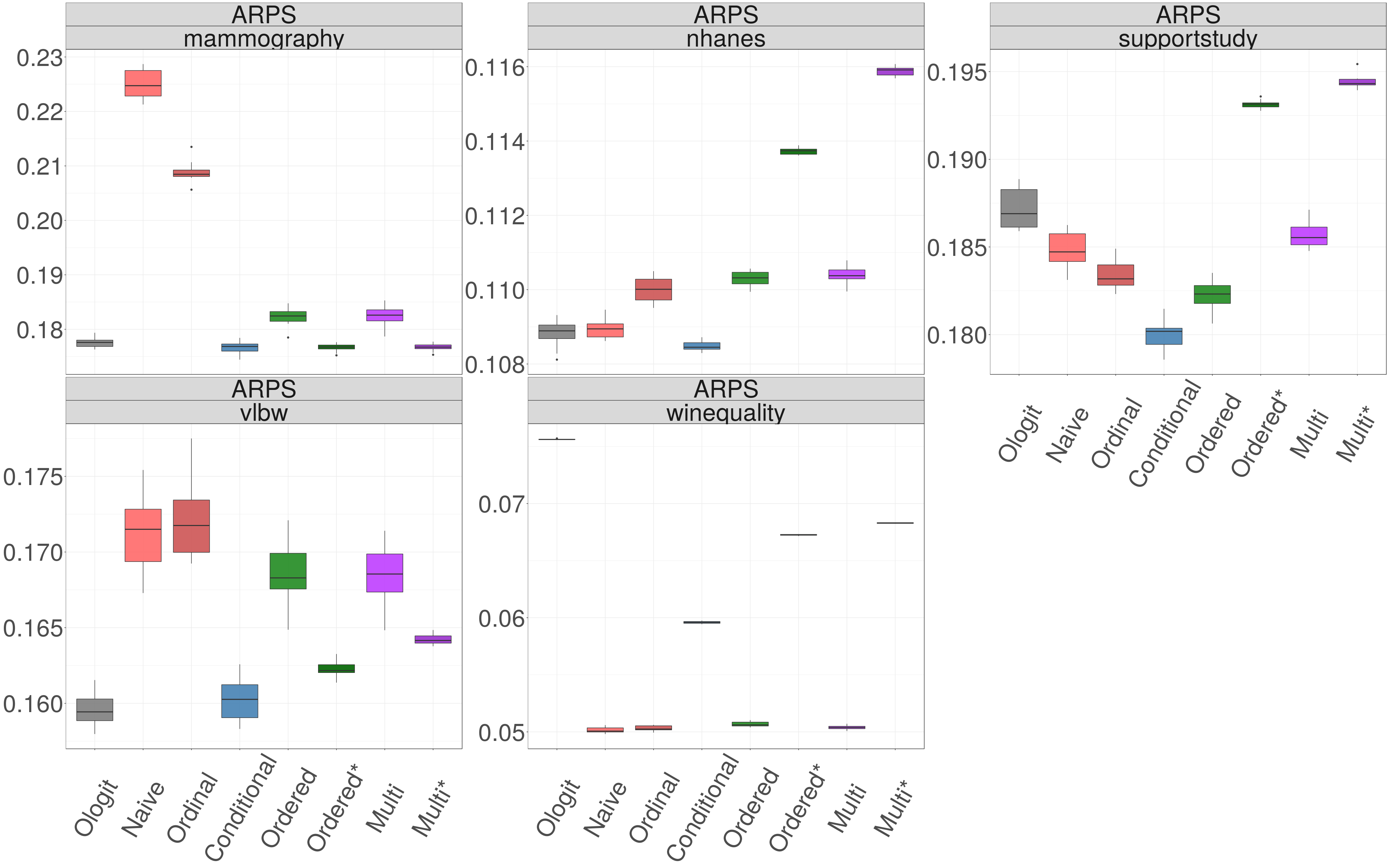}
\vspace{0.5cm}
\caption*{\footnotesize{\textit{Note: }Figure summarizes the prediction accuracy results in terms of the ARPS based on 10 repetitions of 10-fold cross-validation for respective datasets. The boxplots show the median and the interquartile range of the respective measure.}}
\end{figure}

\begin{figure}[h!]
\caption{Cross-Validation: AMSE}
\label{cv_mse}
\includegraphics[width=0.99\textwidth]{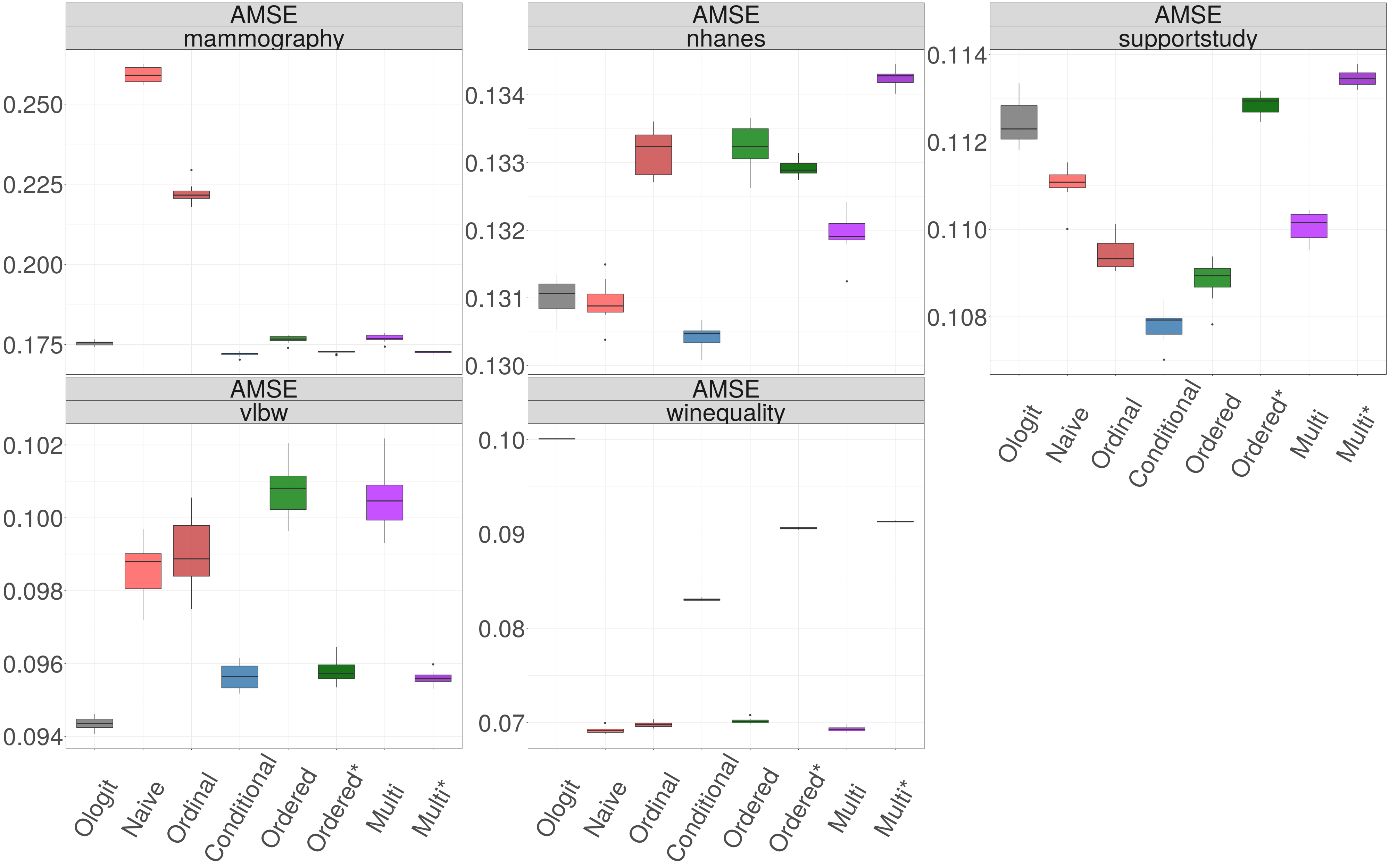}
\vspace{0.5cm}
\caption*{\footnotesize{\textit{Note: }Figure summarizes the prediction accuracy results in terms of the AMSE based on 10 repetitions of 10-fold cross-validation for respective datasets. The boxplots show the median and the interquartile range of the respective measure.}}
\end{figure}

Similarly to \textcite{Hornung2017} we evaluate the prediction accuracy based on a repeated cross-validation in order to reduce the dependency of the results on the particular training and test sample splits. As such we perform a 10-fold cross-validation on each dataset, i.e. we randomly split the dataset in 10 equally sized folds and use 9 folds for training the model and 1 fold for validation. This process is repeated such that each fold serves as a validation set exactly once. Lastly, we repeat this whole procedure 10 times and report average accuracy measures. The results of the cross-validation exercise for the ARPS as well as the AMSE are summarized in Figures \ref{cv_rps} and \ref{cv_mse}, respectively. Similarly as for the simulation results Appendix \ref{Appendix:empsim} contains more detailed statistics.

The main difference in evaluating the prediction accuracy in comparison to the simulation study is the fact that we do not observe the \ul{underlying ordered class} probabilities, but only the realized  \ul{ordered classes}. This affects the computation of the accuracy measures and it can be expected that the prediction errors are somewhat higher in comparison to the simulation data, which is also the case here. Overall, the results imply a substantial heterogeneity in the prediction accuracy across the considered datasets. On the one hand, the parametric ordered logit does well in small samples (\textit{vlbw}) whereas the forest-based methods are somewhat lagging behind. This is not surprising as a lower precision in small samples is the price to pay for the additional flexibility. On the other hand, in the largest sample (\textit{winequality}) the ordered logit is clearly the worst performing method and all forest-based methods perform substantially better. With respect to the \textit{Ordered Forest} estimator we observe relatively high prediction accuracy for three datasets (\textit{mammography}, \textit{supportstudy}, \textit{winequality}) and relatively low prediction accuracy for two datasets (\textit{nhanes}, \textit{vlbw}) in comparison to the competing methods. The good performance in the \textit{winequality} and the \textit{supportstudy} dataset is expected due to the large samples available. In case of the \textit{mammography} dataset, even when smaller in sample size, the \textit{Ordered Forest} maintains the good prediction performance, with its honest version doing even better. The worse performance for the \textit{vlbw} dataset might be due to the small sample size. However, the honest version of the \textit{Ordered Forest} performs rather well. The relatively poor performance in the case of the \textit{nhanes} dataset comes rather at surprise as the sample size is \ul{rather} large. Nevertheless, here the differences among all estimators are very small in magnitude, in fact the smallest among the considered datasets. \ul{Overall, the empirical results provide an evidence for a good predictive performance of the new \textit{Ordered Forest} estimator based on various real datasets.}

\section{Empirical Application}\label{sec:emp}
\ul{For an analysis of} the relationship between the covariates and the predicted \ul{ordered} choice probabilities we estimate the marginal effects for the \textit{Ordered Forest} and compare these to the marginal effects estimated by the ordered logit. We estimate both common measures for marginal effects, i.e. the mean marginal effects as well as the marginal effects at covariate means. The main difference between the ordered logit and the \textit{Ordered Forest} is the fact that the \textit{Ordered Forest} does not use any parametric link function in the estimation of the marginal effects and as such does not impose any functional form on these estimates. As a result, the \textit{Ordered Forest} does neither fix the sign of the marginal effects estimates nor revert it exactly once within the class range as is the case for the ordered logit \parencites(the so-called 'single crossing' feature, see i.e.)(){Boes2006}[or][]{Greene2010} but rather estimates these in a data-driven manner. Nevertheless, the \textit{Ordered Forest}, same as the ordered logit, still ensures that the marginal effects across the class range sum up to zero (being more likely to be in some particular classes must imply being less likely to be in some other classes). As such the \textit{Ordered Forest} not only enables a more flexible estimation of the \ul{ordered} choice probabilities but also of the marginal effects.

\ul{In order to showcase the \textit{Ordered Forest} estimation of marginal effects, we revisit the question of self-assessed health status and its relationship with socio-economic characteristics as for example analyzed previously by} \textcite{Case2002} and \textcite{Murasko2008}. \ul{In our empirical application we analyze the dataset from the 2009 National Health Interview Survey (NHIS) used in} \textcite{Angrist2014} which includes an ordered categorical outcome indicating a self-assessed health status. The specific survey question of interest reads as: \textit{'Would you say your health in general is excellent,
very good, good, fair, or poor?'} and is coded on an ordered scale ranging from 1 (poor) to 5 (excellent). We examine how the ordered choice probabilities of the self-assessed health status differ for individuals with and without a coverage by private health insurance \parencite[see][for a review of insurance effects on health]{Levy2008} as well as how these probabilities vary with further socio-demographic characteristics, namely age, race and family size as well as economic characteristics, namely education, employment status and family income. The considered dataset is well-suited for demonstrating the evaluation of marginal effects for several reasons. First, the dataset features an ordered categorical outcome with 5 distinct ordered categories, which are unevenly distributed and thus challenging for estimating the associated marginal effects. Second, the dataset includes both continuous as well as categorical covariates which enables an exhaustive demonstration of the evaluation of marginal effects for various variable types. Third, the dataset contains more than $18'000$ observations which allows for a precise estimation of the marginal effects. The descriptive statistics for the considered dataset are presented in Appendix \ref{Appendix:nhis}.\footnote{The dataset is freely accessible from the \textsf{R}-package \textsf{stevedata} \parencite{stevedata} \ul{or in the data appendix of} \textcite{Angrist2014} \ul{available} \href{http://masteringmetrics.com/wp-content/uploads/2015/01/Data.zip}{\textsf{online}}.} \ul{We follow the data preparation of} \textcite{Angrist2014} and \ul{discard all observations with missing values and retain only individuals from single family households and those of age between 26 and 59 years as those do not yet qualify for the public health insurance program Medicare.}

First of all, in order to describe the differences in the health status based on the health insurance we inspect the ordered class probabilities for the self-reported health status for individuals with and without a private health insurance contract. The descriptive results are reported in Table \ref{tab:nhismain} below, including statistical evidence for the differences between the two groups. The descriptive evidence suggests that individuals with health insurance have a higher probability to be in excellent or very good health condition and at the same time have a lower probability to be in good or fair health condition. This evidence is both statistically precise and economically relevant. Furthermore, individuals with health insurance seem to have also a lower probability to be in poor health condition. However the evidence for that is less pronounced, both in statistical as well as in economic terms.

Next, in order to investigate the differences in the health status based on the health insurance we estimate the ordered choice probabilities for the self-reported health status conditional on having a private health insurance contract and further socio-economic characteristics using the \textit{Ordered Forest} and the ordered logit and evaluate the corresponding marginal effects. Table \ref{tab:effects} contains the estimated \ul{mean marginal} effects for each outcome class for all covariates together with the associated standard errors, t-values, p-values as well as conventional significance levels for both the \textit{Ordered Forest} as well as the ordered logit.\footnote{The results for the marginal effects at mean are available in Appendix \ref{AppendixME}.}

\begin{table}[ht]
\centering
\caption{Differences in Health Status based on Health Insurance: NHIS Dataset}
\label{tab:nhismain}
\begin{tabular}{lrrrrr}
  \toprule
  \multicolumn{6}{c}{\textbf{NHIS Dataset}}\\
\midrule
& \multicolumn{5}{c}{Health Insurance} \\
\midrule
Health Status & Yes & No & Diff & tValue & pValue \\
\midrule
  Poor &  1.07 &  1.51 & -0.44 & -1.84 &  6.61 \\ 
  Fair &  4.81 & 10.19 & -5.38 & -9.28 &  0.00 \\ 
  Good &  23.26 &  35.14 & -11.88 & -12.66 &   0.00 \\ 
  Very good & 36.31 & 27.54 &  8.77 &  9.70 &  0.00 \\ 
  Excellent & 34.55 & 25.62 &  8.93 & 10.08 &  0.00 \\
  \midrule
  N & 15816 &  2974 &  &  &  \\ 
   \bottomrule
\end{tabular}
\end{table}

In general, we see similar patterns in terms of the effect sizes and effect direction for both the \textit{Ordered Forest} and the ordered logit. However, we do observe more variability in terms of the effect direction in case of the \textit{Ordered Forest} as we would also expect given the flexibility arguments discussed above. In terms of uncertainty of the effects the weight-based inference seems to be \ul{slightly} more conservative than the delta method used in the ordered logit. Nevertheless, \ul{the \textit{Ordered Forest}} also detects very precise effects which are not discovered by the ordered logit.

\begin{table}[!ht]
\centering
\caption{Mean Marginal Effects: NHIS Dataset}
\label{tab:effects}
\resizebox{0.99\textwidth}{!}{
\begin{threeparttable}
\begin{tabular}{l@{\phantom{......}}c@{\phantom{......}}@{ }r@{ }@{ }r@{ }@{ }r@{ }@{ }r@{ }@{ }l@{ }HHH@{ }r@{ }@{ }r@{ }@{ }r@{ }@{ }r@{ }@{ }l@{ }}
\toprule
\multicolumn{2}{c}{\normalsize{\textbf{Dataset}}} & \multicolumn{5}{c}{\normalsize{\textbf{Ordered Forest}}} & \multicolumn{3}{c}{} & \multicolumn{5}{c}{\normalsize{\textbf{Ordered Logit}}}\\
\midrule
 Variable & Class & Effect & Std.Error & t-Value & p-Value &  &  & Variable & Category & Effect & Std.Error & t-Value & p-Value &   \\ 
  \hline
  Health Insurance & 1 & 0.23 & 0.08 & 2.89 & 0.38 & *** &  & HealthInsurance & 1 & -0.11 & 0.05 & -2.19 & 2.85 & **  \\ 
   & 2 & -0.95 & 0.49 & -1.93 & 5.35 & *   &  &  & 2 & -0.49 & 0.22 & -2.22 & 2.68 & **  \\ 
   & 3 & -4.51 & 1.99 & -2.27 & 2.33 & **  &  &  & 3 & -1.32 & 0.59 & -2.25 & 2.44 & **  \\ 
   & 4 & 4.44 & 1.80 & 2.47 & 1.35 & **  &  &  & 4 & 0.02 & 0.03 & 0.65 & 51.88 &     \\ 
   & 5 & 0.78 & 2.47 & 0.32 & 75.21 &     &  &  & 5 & 1.90 & 0.83 & 2.29 & 2.22 & **  \\ 
   \hline
   Female & 1 & -0.19 & 0.12 & -1.59 & 11.16 &     &  & Female & 1 & 0.02 & 0.03 & 0.68 & 49.85 &     \\ 
   & 2 & 0.05 & 0.31 & 0.17 & 86.56 &     &  &  & 2 & 0.10 & 0.14 & 0.68 & 49.81 &     \\ 
   & 3 & 0.52 & 0.70 & 0.74 & 45.99 &     &  &  & 3 & 0.26 & 0.39 & 0.68 & 49.80 &     \\ 
   & 4 & 0.44 & 0.86 & 0.52 & 60.63 &     &  &  & 4 & 0.00 & 0.01 & 0.59 & 55.20 &     \\ 
   & 5 & -0.82 & 1.16 & -0.70 & 48.08 &     &  &  & 5 & -0.39 & 0.57 & -0.68 & 49.80 &     \\ 
   \hline
   Non White & 1 & 0.38 & 0.15 & 2.57 & 1.02 & **  &  & NonWhite & 1 & 0.36 & 0.05 & 7.02 & 0.00 & *** \\ 
   & 2 & 0.57 & 0.42 & 1.36 & 17.53 &     &  &  & 2 & 1.60 & 0.20 & 7.89 & 0.00 & *** \\ 
   & 3 & 5.97 & 1.12 & 5.32 & 0.00 & *** &  &  & 3 & 4.10 & 0.48 & 8.57 & 0.00 & *** \\ 
   & 4 & -4.23 & 1.09 & -3.87 & 0.01 & *** &  &  & 4 & -0.26 & 0.08 & -3.12 & 0.18 & *** \\ 
   & 5 & -2.69 & 1.57 & -1.72 & 8.57 & *   &  &  & 5 & -5.81 & 0.65 & -8.87 & 0.00 & *** \\ 
   \hline
Age & 1 & 0.04 & 0.01 & 4.22 & 0.00 & *** &  & Age & 1 & 0.04 & 0.00 & 12.60 & 0.00 & *** \\ 
   & 2 & 0.15 & 0.03 & 5.09 & 0.00 & *** &  &  & 2 & 0.20 & 0.01 & 19.77 & 0.00 & *** \\ 
   & 3 & 0.45 & 0.07 & 6.07 & 0.00 & *** &  &  & 3 & 0.54 & 0.02 & 23.87 & 0.00 & *** \\ 
   & 4 & -0.01 & 0.09 & -0.13 & 89.49 &     &  &  & 4 & 0.01 & 0.01 & 1.24 & 21.54 &     \\ 
   & 5 & -0.62 & 0.12 & -5.10 & 0.00 & *** &  &  & 5 & -0.78 & 0.03 & -24.15 & 0.00 & *** \\ 
   \hline
Education & 1 & 0.00 & 0.00 & 0.41 & 68.03 &     &  & Education & 1 & -0.11 & 0.01 & -11.61 & 0.00 & *** \\ 
   & 2 & -0.01 & 0.00 & -1.73 & 8.42 & *   &  &  & 2 & -0.51 & 0.03 & -16.80 & 0.00 & *** \\ 
   & 3 & -0.02 & 0.01 & -2.80 & 0.52 & *** &  &  & 3 & -1.39 & 0.07 & -18.94 & 0.00 & *** \\ 
   & 4 & 0.00 & 0.01 & 0.71 & 48.08 &     &  &  & 4 & -0.02 & 0.02 & -1.23 & 21.83 &     \\ 
   & 5 & 0.02 & 0.01 & 2.57 & 1.03 & **  &  &  & 5 & 2.04 & 0.11 & 18.85 & 0.00 & *** \\ 
   \hline
   Family Size & 1 & 0.00 & 0.00 & 0.32 & 74.77 &     &  & FamilySize & 1 & -0.01 & 0.01 & -0.81 & 42.01 &     \\ 
   & 2 & -0.00 & 0.01 & -0.21 & 83.33 &     &  &  & 2 & -0.04 & 0.05 & -0.81 & 41.95 &     \\ 
   & 3 & -0.06 & 0.02 & -3.49 & 0.05 & *** &  &  & 3 & -0.12 & 0.14 & -0.81 & 41.93 &     \\ 
   & 4 & -0.03 & 0.02 & -1.78 & 7.51 & *   &  &  & 4 & -0.00 & 0.00 & -0.67 & 50.41 &     \\ 
   & 5 & 0.10 & 0.02 & 4.97 & 0.00 & *** &  &  & 5 & 0.17 & 0.21 & 0.81 & 41.94 &     \\ 
   \hline
Employed & 1 & -3.99 & 0.50 & -7.94 & 0.00 & *** &  & Employed & 1 & -0.42 & 0.06 & -7.30 & 0.00 & *** \\ 
   & 2 & -3.81 & 0.73 & -5.19 & 0.00 & *** &  &  & 2 & -1.86 & 0.23 & -8.21 & 0.00 & *** \\ 
   & 3 & 2.58 & 1.15 & 2.25 & 2.44 & **  &  &  & 3 & -4.77 & 0.53 & -8.98 & 0.00 & *** \\ 
   & 4 & 4.37 & 1.24 & 3.51 & 0.04 & *** &  &  & 4 & 0.39 & 0.11 & 3.55 & 0.04 & *** \\ 
   & 5 & 0.84 & 1.82 & 0.46 & 64.34 &     &  &  & 5 & 6.66 & 0.71 & 9.42 & 0.00 & *** \\ 
   \hline
Income & 1 & -0.11 & 0.04 & -3.00 & 0.27 & *** &  & Income & 1 & -0.00 & 0.00 & -12.07 & 0.00 & *** \\ 
   & 2 & -0.46 & 0.14 & -3.27 & 0.11 & *** &  &  & 2 & -0.00 & 0.00 & -17.73 & 0.00 & *** \\ 
   & 3 & -0.06 & 0.51 & -0.12 & 90.68 &     &  &  & 3 & -0.00 & 0.00 & -20.61 & 0.00 & *** \\ 
   & 4 & 0.36 & 0.37 & 0.97 & 33.42 &     &  &  & 4 & -0.00 & 0.00 & -1.24 & 21.41 &     \\ 
   & 5 & 0.27 & 0.45 & 0.61 & 54.03 &     &  &  & 5 & 0.00 & 0.00 & 20.96 & 0.00 & *** \\ 
   \midrule
\multicolumn{13}{l}{\footnotesize{Significance levels correspond to: $*** .<0.01$, $** .<0.05$, $* .<0.1$.}}\\
   \bottomrule
\end{tabular}
\begin{tablenotes}\footnotesize
\item [] \hspace{-0.17cm} \textit{Notes:} Table shows the comparison of the mean marginal effects in \% points between the \textit{Ordered Forest} and the ordered logit. The effects are estimated for all classes, together with the corresponding standard errors, t-values and p-values. The standard errors for the \textit{Ordered Forest} are estimated using the weight-based inference and for the ordered logit are obtained via the delta method.
\end{tablenotes}
\end{threeparttable}
}
\end{table}

In particular, inspecting the variable of interest, namely the indicator for private health insurance, we immediately see the additional flexibility of the \textit{Ordered Forest}. While both methods estimate positive marginal effects of having a private health insurance on the probability of being in very good or excellent health condition and negative marginal effects for being in good or fair health condition, the \textit{Ordered Forest} estimates also a positive effect for being in poor health condition, whereas the ordered logit is forced to estimate a negative effect due to its above-mentioned single-crossing property. As such, the \textit{Ordered Forest} estimates a non-monotonic effect of having a private health insurance across the class probabilities. The results suggest that on one hand individuals with health insurance are less likely to be in good or fair health condition by $4.51$ or $0.95$ \% points, respectively. On the other hand, individuals with health insurance are more likely to be in very good or excellent health condition by $4.44$ or $0.78$ \% points, respectively, but they are also more likely to be  in poor health condition by $0.23$ \% points. As the decision to sign up for a private health insurance is not random, i.e. the data comes from a non-experimental setting, it is not possible to uncover the causal effect without strong assumptions. One might, however, argue that based on the partial correlation evidence, due to the regular medical care and prevention the health insurance increases the likelihood of being in rather good health condition, but also that individuals with rather poor health condition are more likely to sign up for a private health insurance to cover up for the expected medical care costs. As can be seen, the \textit{Ordered Forest} enables for such a non-monotonic effects analysis, while the ordered logit does not permit such mechanism to take place at all. Overall, in terms of effect sizes, for both estimators we observe smaller magnitudes in comparison to the unconditional differences presented in Table \ref{tab:nhismain}. However, the effect sizes estimated by the \textit{Ordered Forest} are slightly bigger than those of the ordered logit. With regards to the statistical uncertainty around the estimated marginal effects, both methods exhibit similar level of precision.

Inspecting the effects of the additional conditioning variables, we see that neither the \textit{Ordered Forest} nor the ordered logit find evidence for gender influencing the health class probabilities as the estimated effects are of small magnitude and lack statistical precision. In contrast, both methods estimate a higher probability of being in poor, fair or good health condition and conversely a lower probability of being in very good or excellent health condition for people of color, an effect that is sizeable and statistically precise. In this case, we note the slightly more conservative standard errors of the \textit{Ordered Forest}. Furthermore, both methods estimate a higher likelihood of being in rather bad health condition and a lower likelihood of being in rather good health condition for increasing age with similar effect sizes as well as with similar statistical precision. In terms of education, there seem to be a positive relationship with regard to the probability of being in an excellent health condition. However, this effect is less pronounced for the \textit{Ordered Forest} considering both the effect size and the precision in comparison to the ordered logit. The same positive relationship can be observed also for the family size and although the economic relevance of this effect is rather small, the \textit{Ordered Forest} estimates this effect with high statistical precision, whereas the ordered logit does not find statistical evidence in this respect. Considering the employment status, both methods estimate lower likelihood of being in rather bad health condition and a higher likelihood of being in good health condition with comparable effect sizes as well as statistical precision. Lastly, the \textit{Ordered Forest} and the ordered logit both estimate a positive relationship with regards to the income level. As such, individuals with higher income are less likely to be in rather bad health condition and more likely to be in rather good health condition. In case of the \textit{Ordered Forest}, the effect sizes are slightly larger, but with lower statistical precision, finding relevant evidence only for the negative effects on the fair and poor health status, whereas in case of the ordered logit the statistical precision is higher, however with effectively estimating a zero effect. This might be due to the somewhat higher collinearity between the education and income level ($0.45$), which suggests a better handling of near-multicollinearity among covariates of the \textit{Ordered Forest} as has been documented in the simulation study. Overall, however, the main advantage of the estimation of the marginal effects by the \textit{Ordered Forest} stems from a more flexible, data-driven approximation of possible nonlinearities in the functional form.

\section{Conclusion}\label{sec:conclusion}
In this paper, we develop and apply a new machine learning estimator of the econometric ordered choice models based on the random forest algorithm. The \textit{Ordered Forest} estimator is a flexible alternative to parametric ordered choice models such as the ordered logit or ordered probit which does not rely on any distributional assumptions and provides essentially the same output as the parametric models, including the estimation of the marginal effects as well as the associated inference. The proposed estimator utilizes the flexibility of random forests and can thus naturally deal with nonlinearities in the data and with a large-dimensional covariate space, while taking the ordering information of the categorical outcome variable into account. Hence, the estimator flexibly estimates the conditional \ul{ordered} choice probabilities without restrictive assumptions about the distribution of the error term, or other assumptions such as the single index and constant threshold assumptions as is the case for the parametric ordered choice models \parencite[see][for a discussion of these assumptions]{Boes2006}. Further, the estimator allows also the estimation of the marginal effects, i.e. how the estimated conditional \ul{ordered} choice probabilities vary with changes in covariates. The weighted representation of these effects enables the weight-based inference as suggested by \textcite{Lechner2019}. The fact that the estimator comprises of linear combinations of random forest predictions ensures \ul{that} the theoretical guarantees of \textcite{Wager2018} \ul{are satisfied}. Additionally, a free software implementation of the \textit{Ordered Forest} estimator in both \textsf{R} \parencite{rstats} and \textsf{Python} \parencite{python} is available in the package \textsf{orf} \ul{available on the official} \href{https://CRAN.R-project.org/package=orf}{\textsf{CRAN}} \parencite{orf2019} and \href{https://pypi.org/project/orf/}{\textsf{PyPI}} \parencite{orf2022} \ul{repositories} to enable the usage of the method by applied researchers.

The performance of the \textit{Ordered Forest} estimator is studied and compared to other competing estimators in an extensive Monte Carlo simulation as well as using real datasets. The simulation results suggest good performance of the estimator in finite samples, including also high-dimensional settings. The advantages of the machine learning estimation compared to a parametric method become apparent when dealing with near-multicollinearity and highly nonlinear functional forms. In such cases all of the considered forest-based estimators perform better than the ordered logit in terms of the prediction accuracy. Among the forest-based estimators the \textit{Ordered Forest} proposed in this paper performs well throughout all simulated DGPs and outperforms the competing methods in the most complex simulation designs. The empirical evidence using real datasets supports the findings from the Monte Carlo simulation. Additionally, the estimation of the marginal effects as well as the inference procedure seems to work well in the \ul{presented} empirical example.

Despite the attractive properties of the \textit{Ordered Forest} estimator, many interesting questions are left open. Particularly, a further extension of the Monte Carlo simulation to study the sensitivity of the \textit{Ordered Forest} in respect to tuning parameters of the underlying random forest as well as in respect to different simulation designs would be of interest. Similarly, the performance of the estimator with and without honesty for larger sample sizes should be further investigated. Also, the optimal choice of the size of the window for evaluating the marginal effects would be worth to explore. Additionally, \ul{besides the theoretical guarantees for the point estimator, a rigorous asymptotic analysis of the weight-based inference procedure for the estimation of standard errors would be beneficial to describe the exact theoretical properties.} Lastly, it would be of great interest to see more real data applications of the \textit{Ordered Forest} estimator \ul{such as for example in} \textcite{Kim2021}, especially for large samples.

\pagebreak
\printbibliography

\pagebreak

\appendix
\section{Other Machine Learning Estimators}\label{Appendix:methods}
\subsection{Multinomial Forest}\label{AppendixMRF}
Considering the \textit{Ordered Forest} estimator a possible modification for models with categorical outcome variable \textit{without} an inherent ordering appears to be straightforward. Instead of estimating cumulative probabilities and afterwards isolating the respective class probabilities, we can estimate the class probabilities $P_{m,i}=P[Y_{i}=m \mid X_i=x]$ directly. As such the binary outcomes are now constructed to indicate the particular outcome classes separately. Then the random forest predictions for each class yield the conditional choice probabilities which need to be afterwards normalized to sum up to 1. Formally, consider (un)ordered categorical outcome variable $Y_i \in \{1,...,M \}$ with classes $m$ and sample size $N(i=1,...,N)$. Then, the estimation procedure can be described as follows:
\begin{enumerate}
\item Create $M$ binary indicator variables such as
\begin{align}
Y_{m,i}=\mathbf{1}(Y_i = m) \text{\phantom{kkt} for \phantom{kkt}} m=1,...,M . \label{indicator}
\end{align}
where $m$ is known and given by the definition of the dependent variable.
\item Estimate regression random forest for each of the $M$ indicators as
\begin{align}
P[Y_{m,i}=1 \mid X_i=x] = \sum^N_{i=1}w_{m,i}(x)Y_{m,i} \text{\phantom{kkt} for \phantom{kkt}} m=1,...,M,
\end{align}
\vspace{-0.2cm}
where the forest weights are defined as $w_{m,i}(x)=\frac{1}{B}\sum^B_{b=1}w_{m,b,i}(x)$ with trees weights given by\\

\vspace{-0.4cm}
$w_{m,b,i}(x)=\frac{\mathbf{1}(\{X_i \in L_{b,m}(x) \})}{\mid \{ i:X_i \in L_{b,m}(x) \} \mid}$ with leaves $L_{b,m}(x)$ for a total of $B$ trees.
\item Obtain forest predictions for each of the $M$ indicators as
\begin{align}
\hat{Y}_{m,i}=\hat{P}[Y_{m,i}=1 \mid X_i=x] = \sum^N_{i=1}\hat{w}_{m,i}(x)Y_{m,i} \text{\phantom{kkt} for \phantom{kkt}} m=1,...,M,
\end{align}
\vspace{-0.2cm}
where $\hat{Y}_{m,i}$ are estimated probabilities.
\item Compute probabilities for each class as
\begin{align}
\hat{P}_{m,i}&=\hat{Y}_{m,i}  \text{\phantom{kkkkkkkkkt} for \phantom{kkt}} m=1,...,M \label{equ2} \\[5pt] 
\hat{P}_{m,i}&=\frac{\hat{P}_{m,i}}{\sum^{M}_{m=1}\hat{P}_{m,i}} \text{\phantom{kkkt} for \phantom{kkt}} m=1,...,M , \label{equ5}
\end{align}
\end{enumerate}
where the equation \eqref{equ2} defines the probabilities of all $M$ classes and subsequent equation \eqref{equ5} ensures that the probabilities sum up to 1 as this might not be the case otherwise. Similarly to the \textit{Ordered Forest} estimator, also the multinomial forest is a linear combination of the respective forest predictions and as such also inherits the theoretical properties stemming from random forest estimation as described in Section \ref{sec:RF} of the main text.

\subsection{Conditional Forest}\label{Appendix:hornik}
The conditional forest as discussed in Section \ref{sec:lit} of the main text is grown with the so-called conditional inference trees. The main idea is to provide an unbiased way of recursive splitting of the trees using a test statistic based on permutation tests \parencite{Strasser1999}. To describe the estimation procedure, consider an ordered categorical outcome $Y_i \in (1,...,M)$ with ordered classes $m$ and sample size $N(i=1,...,N)$. Further, define binary case weights $w_i \in \{0,1\}$ which determine if the observation is part of the current leaf. Then, the algorithm developed by \textcite{Hothorn2006} can be described as follows:
\begin{enumerate}
\item Test the global null hypothesis of independence between any of the $P$ covariates and the outcome, for the particular case weights, given a bootstrap sample $Z_b$. Afterwards, select the $p$-th covariate $X_{i,p}$ with the strongest association with the outcome $Y_i$, or stop if the null hypothesis cannot be rejected. The association is measured by a linear statistic $T$ given as:
\begin{align}\label{teststat}
T_p(Z_b,w)=\sum^N_{i=1}w_i g_p(X_{i,p})h(Y_i) ,
\end{align}
where $g_p(\cdot)$ and $h(\cdot)$ are specific transformation functions.\label{itm1}
\item Split the covariate sample space $\mathcal{X}_p$ into two disjoint sets $\mathcal{I}$ and $\mathcal{J}$ with adapted case weights $w_i\mathbf{1}(X_{i,p} \in \mathcal{I})$ and $w_i\mathbf{1}(X_{i,p} \in \mathcal{J})$ determining the observations falling into the subset $\mathcal{I}$ and $\mathcal{J}$, respectively. Then, the split is chosen by evaluating a two-sample statistic as a special case of \ref{teststat}:
\begin{align}\label{teststat2}
T^{\mathcal{I}}_p(Z_b,w)=\sum^N_{i=1}w_i\mathbf{1}(X_{i,p} \in \mathcal{I})h(Y_i)
\end{align}
for all possible subsets $\mathcal{I}$ of the covariate sample space $\mathcal{X}_p$.
\label{itm2}
\item Repeat steps \ref{itm1} and \ref{itm2} recursively with modified case weights.
\end{enumerate}
Hence, the above algorithm distinguishes between variable selection (step \ref{itm1}) and splitting rule (step \ref{itm2}), while both relying on the variations of the test statistic $T_p(Z_b,w)$. In practice, however, the distribution of this statistic under the null hypothesis is unknown and depends on the joint distribution of $Y_i$ and $X_{i,p}$. For this reason, the permutation tests are applied to abstract from the dependency by fixing the covariates and conditioning on all possible permutations of the outcomes. Then, the conditional mean and covariance of the test statistic can be derived and the asymptotic distribution can be approximated by Monte Carlo procedures, while \textcite{Strasser1999} proved its normality. Finally, variables and splits are chosen according to the lowest $p$-value of the test statistic $T_p(Z_b,w)$ and $T^{\mathcal{I}}_p(Z_b,w)$, respectively.

Besides the permutation tests, the choice of the tranformation functions $g_p(\cdot)$ and $h(\cdot)$ is important and depends on the type of the variables. For continuous outcome and covariates, identity transformation is suggested. For the case of an ordinal regression which is of interest here, the transformation function is given through the score function $s(m)$. If the underlying latent $Y_i^*$ is unobserved, it is suggested that $s(m)=m$ and thus $h(Y_i)=Y_i$. Hence, in the tree building the ordered outcome is treated as a continuous one \parencite{Janitza2016}. Then, however, the leaf predictions are the choice probabilities computed as proportions of the outcome classes falling within the leaf, instead of fitting a within leaf constant. The final conditional forest predictions for the choice probabilities are the averaged conditional tree probability predictions. Such obtained choice probabilities are analyzed in the Monte Carlo study in Section \ref{sec:MC} of the main text.

\subsection{Ordinal Forest}\label{Appendix:hornung}
In the following, the algorithm for the ordinal forest as developed by \textcite{Hornung2017} is described. To begin with, consider an ordered categorical outcome $Y_i \in (1,...,M)$ with ordered classes $m$ and sample size $N(i=1,...,N)$. Then, for a set of optimization forests $b=1,...,B_{sets}$:
\begin{enumerate}
\item Draw $M-1$ uniformly distributed variables $D_{b,m} \sim U(0,1)$ and sort them according to their values. Further, set $D_{b,1}=0$ and $D_{b,M+1}=1$.
\item Define a score set $S_{b,m}=\{S_{b,1},...,S_{b,M}\}$ with scores constructed as $S_{b,m}=\Phi^{-1}\big(\frac{D_{b,m}+D_{b,m+1}}{2}\big)$ for $m=1,...,M$, where $\Phi(\cdot)$ is the cdf of the standard normal.
\item Create a new continuous outcome $Z_{b,i}=(Z_{b,1},...,Z_{b,N})$ by replacing each class value $m$ of the original ordered categorical $Y_i$ by the $m$-th value of the score set $S_{b,m}$ for all $m=1,...,M$.
\item Use $Z_{b,i}$ as dependent variable and estimate a regression forest $RF_{S_{b,m}}$ with $B_{prior}$ trees.
\item Obtain the out-of-bag (OOB) predictions for the continuous $Z_{b,i}$ and transform them into predictions for $Y_i$ as follows: $\hat{Y}_{b,i}=m$ if $\hat{Z}_{b,i} \in \big]\Phi^{-1}(D_{b,m}, \Phi^{-1}(D_{b,m+1})\big]$ for all $i=1,...,N$.
\item Compute a performance measure for the given forest $\hat{RF}_{S_{b,m}}$ based on some performance function of type $f(Y_i,\hat{Y}_{b,i})$.
\end{enumerate}
After estimating $B_{sets}$ of optimization forests, take $S_{best}$ of these which achieved the best performance according to the performance function. Then, construct the final set of uniformly distributed variables $D_1,...,D_{M+1}$ as an average of those from $S_{best}$ for $m=1,...,M+1$. Finally, form the optimized score set $S_m=\{S_{1},...,S_{M}\}$ with scores constructed as $S_{m}=\Phi^{-1}\big(\frac{D_{m}+D_{m+1}}{2}\big)$ for $m=1,...,M$. The continuous outcome $Z_i=(Z_{1},...,Z_{N})$ is then similarly as in the optimization procedure constructed by replacing each $m$ value of the original outcome $Y_i$ by the $m$-th value of the optimized score set $S_m$ for all $m=1,...,M$. Finally, estimate the regression forest $RF_{final}$ using $Z_i$ as the dependent variable. On one hand, the class prediction of such an ordinal forest is one of the $M$ ordered classes which has been predicted the most by the respective trees of the forest. On the other hand, the probability prediction is obtained as a relative frequency of trees predicting the particular class. Such predicted choice probabilities are analyzed in the conducted Monte Carlo study in Section \ref{sec:MC} of the main text. Further, the so-called naive forest corresponds to the ordinal forest with omitting the above described optimization procedure.

\pagebreak

\section{Simulation Study}\label{Appendix:simulation}
\subsection{Main Simulation Results}\label{Appendix:simtables}
\begin{figure}[!ht]
\caption{Ordered Forest Simulation Results: Simple DGP \& Low Dimension}
\label{BigSimpleLow}
\includegraphics[width=0.99\textwidth]{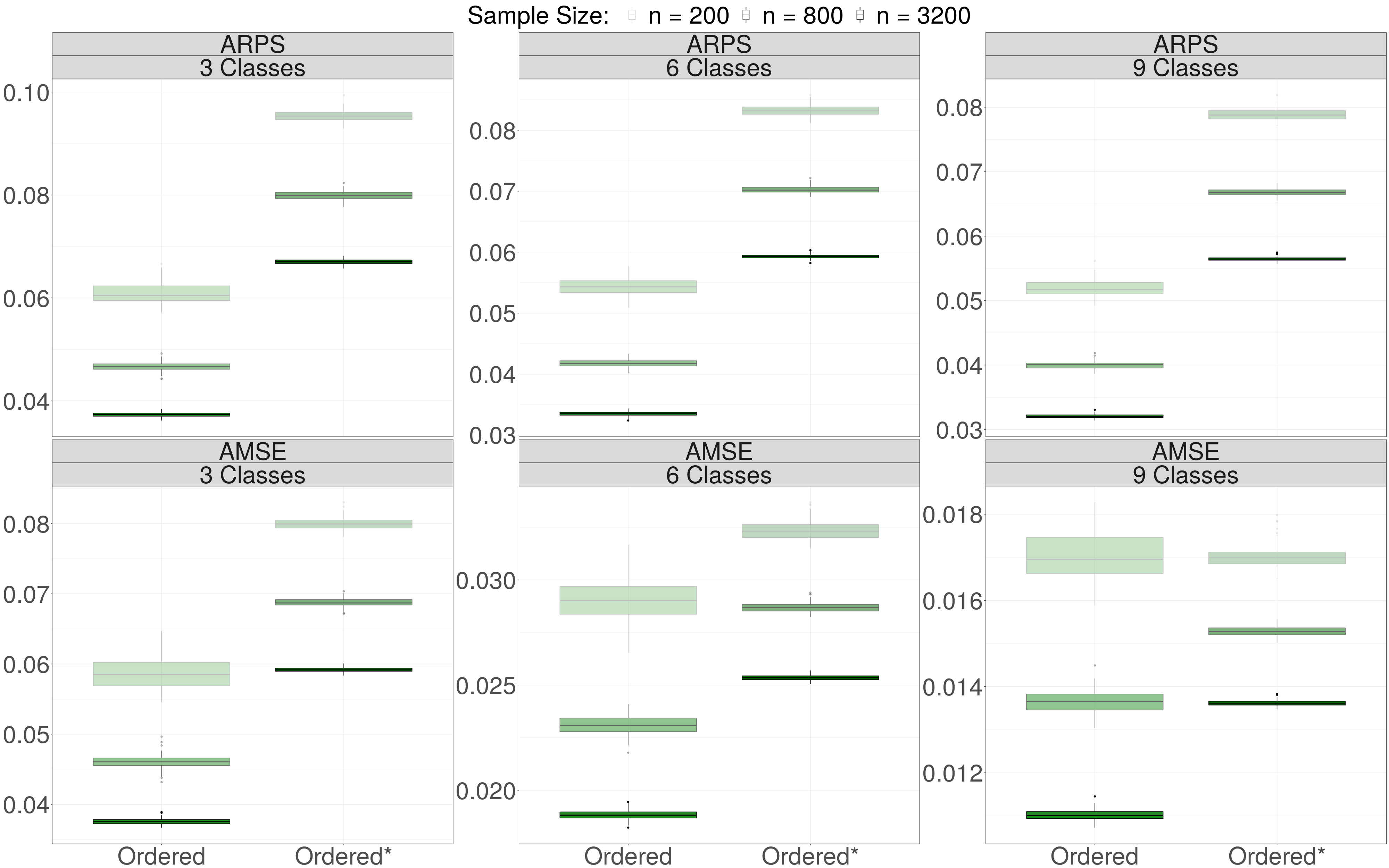}
\vspace{0.5cm}
\caption*{\footnotesize{\textit{Note: }Figure summarizes the prediction accuracy results based on 100 simulation replications. The upper panel contains the ARPS and the lower panel contains the AMSE. The boxplots show the median and the interquartile range of the respective measure. The transparent boxplots denote the results for the small sample size, the semi-transparent ones denote the medium sample size, while the bold boxplots denote the results for the big sample size. From left to right the results for 3, 6, and 9 outcome classes are displayed.}}
\end{figure}

\begin{figure}[!ht]
\caption{Ordered Forest Simulation Results: Complex DGP \& Low Dimension}
\label{BigComplexLow}
\includegraphics[width=0.99\textwidth]{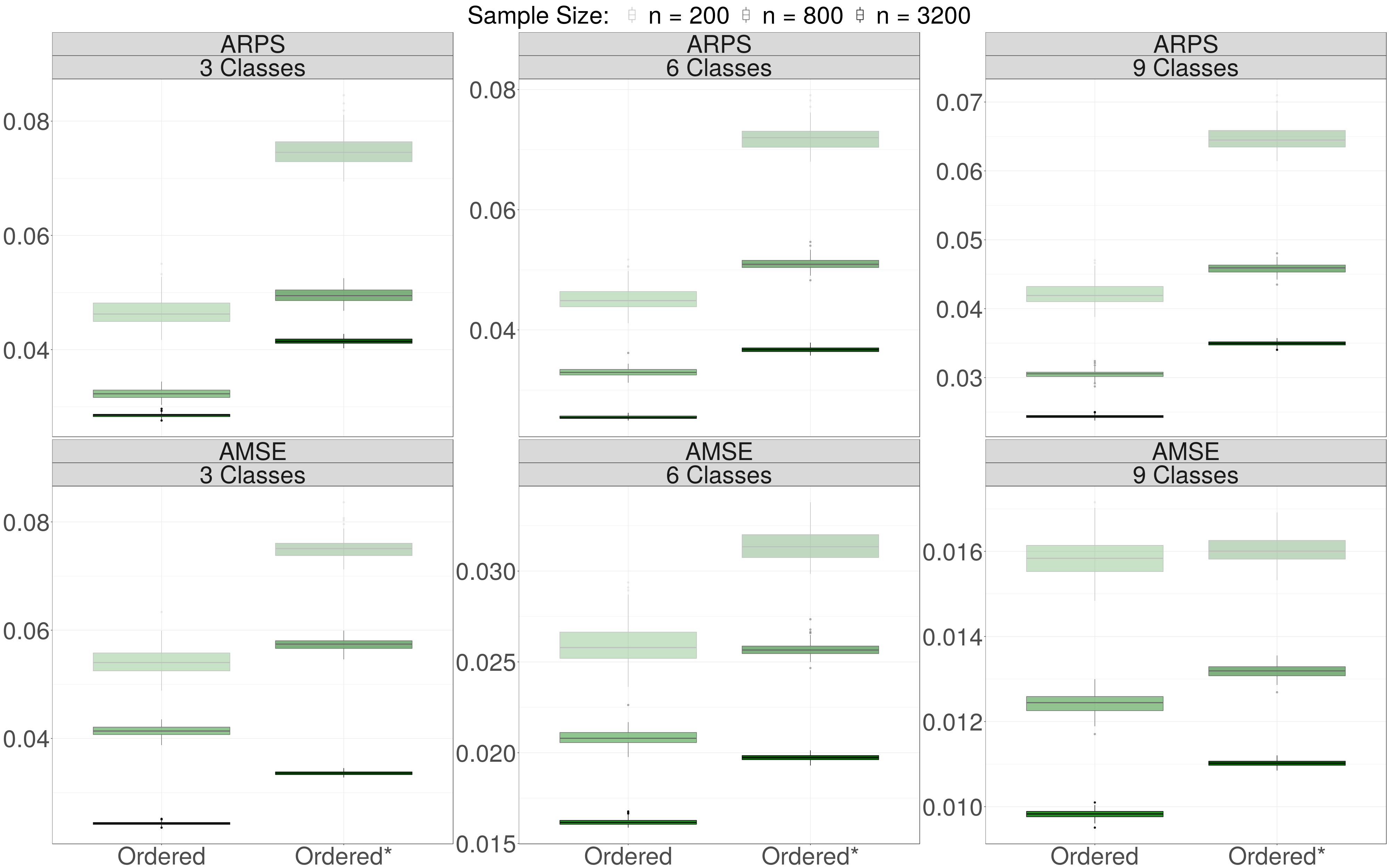}
\vspace{0.5cm}
\caption*{\footnotesize{\textit{Note: }Figure summarizes the prediction accuracy results based on 100 simulation replications. The upper panel contains the ARPS and the lower panel contains the AMSE. The boxplots show the median and the interquartile range of the respective measure. The transparent boxplots denote the results for the small sample size, the semi-transparent ones denote the medium sample size, while the bold boxplots denote the results for the big sample size. From left to right the results for 3, 6, and 9 outcome classes are displayed.}}
\end{figure}

\begin{figure}[!ht]
\caption{Ordered Forest Simulation Results: Simple DGP \& High Dimension}
\label{BigSimpleHigh}
\includegraphics[width=0.99\textwidth]{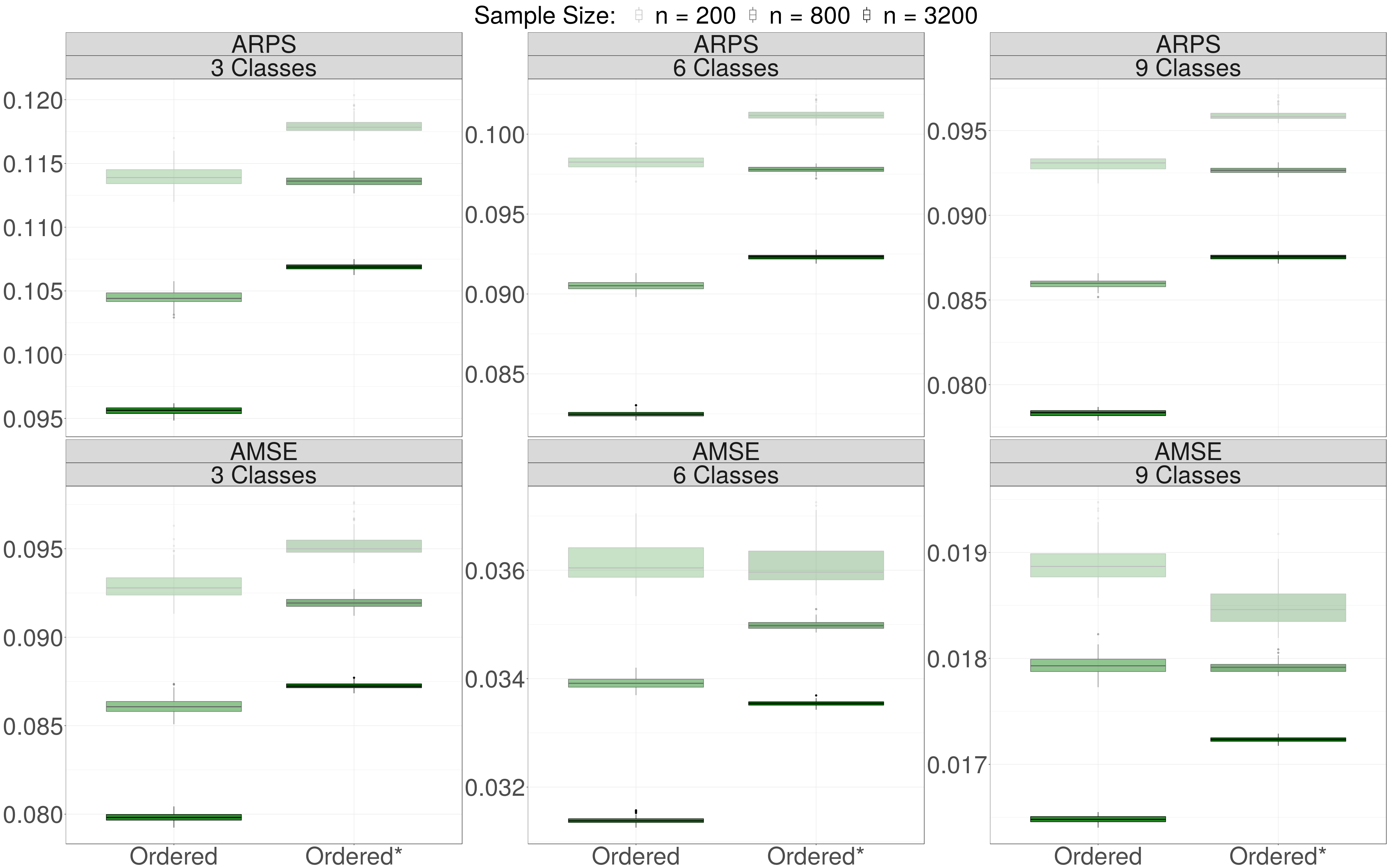}
\vspace{0.5cm}
\caption*{\footnotesize{\textit{Note: }Figure summarizes the prediction accuracy results based on 100 simulation replications. The upper panel contains the ARPS and the lower panel contains the AMSE. The boxplots show the median and the interquartile range of the respective measure. The transparent boxplots denote the results for the small sample size, the semi-transparent ones denote the medium sample size, while the bold boxplots denote the results for the big sample size. From left to right the results for 3, 6, and 9 outcome classes are displayed.}}
\end{figure}

\begin{figure}[!ht]
\caption{Ordered Forest Simulation Results: Complex DGP \& High Dimension}
\label{BigComplexHigh}
\includegraphics[width=0.99\textwidth]{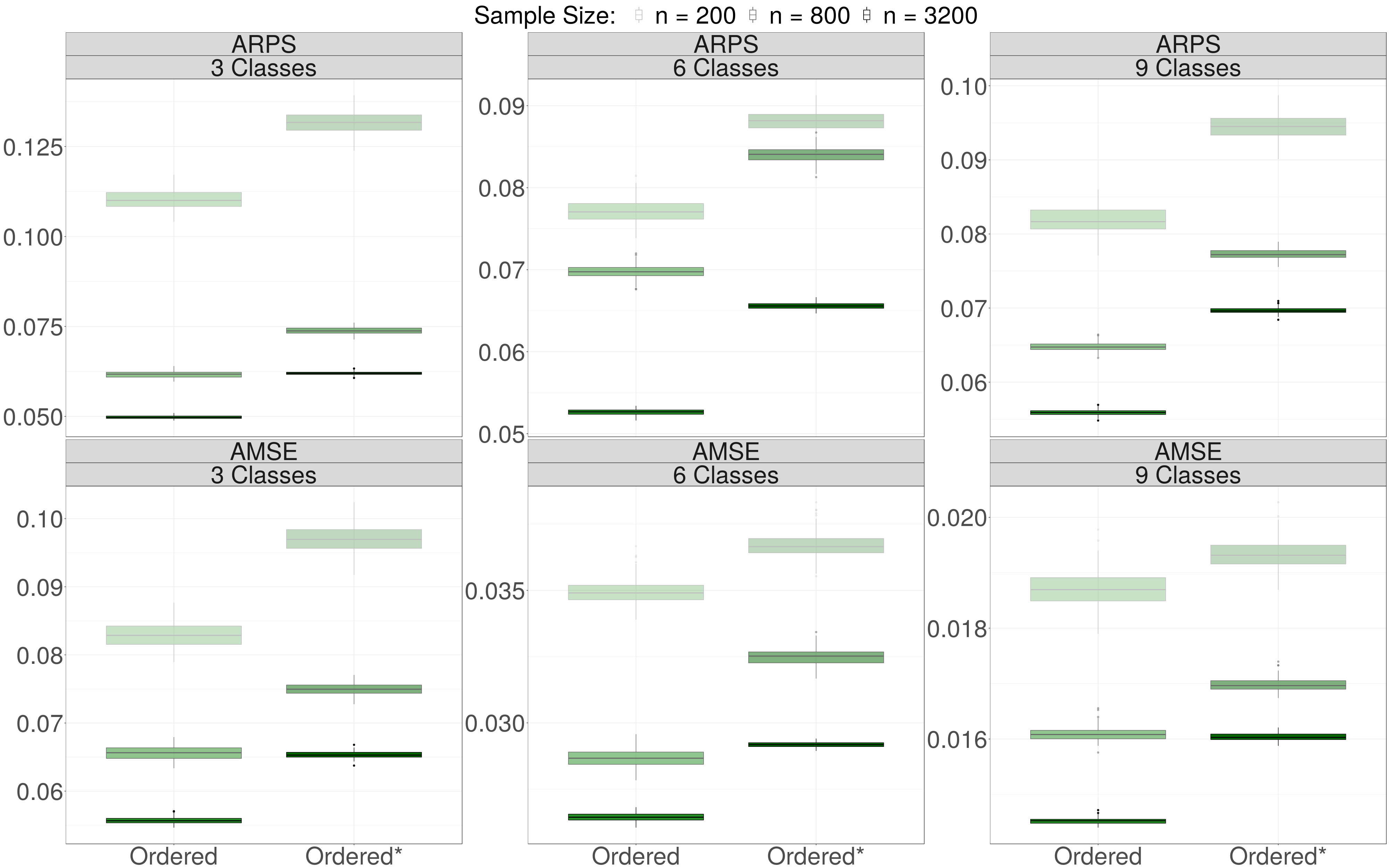}
\vspace{0.5cm}
\caption*{\footnotesize{\textit{Note: }Figure summarizes the prediction accuracy results based on 100 simulation replications. The upper panel contains the ARPS and the lower panel contains the AMSE. The boxplots show the median and the interquartile range of the respective measure. The transparent boxplots denote the results for the small sample size, the semi-transparent ones denote the medium sample size, while the bold boxplots denote the results for the big sample size. From left to right the results for 3, 6, and 9 outcome classes are displayed.}}
\end{figure}

In the following Tables \ref{tab:rps_200}, \ref{tab:mse_200}, \ref{tab:rps_800}, \ref{tab:mse_800} and \ref{tab:mse_3200} are summarized the simulation results presented in Section \ref{sec:boxplots} of the main text. Each table specifies the particular simulation design as follows: the column \textit{Class} indicates the number of outcome classes, \textit{Dim.} specifies the dimension, \textit{DGP} characterizes the data generating process as defined in the main text and \textit{Statistic} contains summary statistics of the simulation results. In particular, the mean of the respective accuracy measure and its standard deviation. Furthermore, rows \textit{t-test} and \textit{wilcox-test} contain the \textit{p}-values of the parametric t-test as well as the nonparametric Wilcoxon test for the equality of means between the results of the \textit{Ordered Forest} and all the other methods. The alternative hypothesis is that the mean of the \textit{Ordered Forest} is less than the mean of the other method to test if the \textit{Ordered Forest} achieves significantly lower prediction error than the other considered methods. Furthermore, Figures \ref{BigSimpleLow}, \ref{BigComplexLow}, \ref{BigSimpleHigh} and \ref{BigComplexHigh} complement the results presented in Section \ref{sec:boxplots} of the main text for the simulations with the increased sample size.

\pagebreak

\subsubsection{ARPS: Sample Size = 200}

\begin{table}[!h]
\footnotesize
\centering
\caption{Simulation results: Accuracy Measure = ARPS \& Sample Size = 200}\label{tab:rps_200}
\begin{threeparttable}
\begin{tabular}{clllrrrrrrrr}
\toprule
\midrule
\multicolumn{4}{c}{\normalsize{\textbf{Simulation Design}}} & \multicolumn{8}{c}{\normalsize{\textbf{Comparison of Methods}}}\\
\midrule
\midrule
 Class & Dim. & DGP & Statistic & Ologit & Naive & Ordinal & Cond. & Ordered & Ordered* & Multi & Multi* \\ 
  \midrule
 \midrule
3 & Low & Simple & mean & 0.0097 & 0.0765 & 0.0755 & 0.0625 & 0.0609 & 0.0954 & 0.0619 & 0.0954 \\ 
   &  &  & st.dev. & 0.0042 & 0.0056 & 0.0055 & 0.0018 & 0.0020 & 0.0011 & 0.0019 & 0.0012 \\ 
   &  &  & t-test & 1.0000 & 0.0000 & 0.0000 & 0.0000 &  & 0.0000 & 0.0002 & 0.0000 \\ 
   &  &  & wilcox-test & 1.0000 & 0.0000 & 0.0000 & 0.0000 &  & 0.0000 & 0.0000 & 0.0000 \\ 
   \midrule
3 & Low & Complex & mean & 0.1156 & 0.1044 & 0.1028 & 0.0593 & 0.0466 & 0.0748 & 0.0491 & 0.0760 \\ 
   &  &  & st.dev. & 0.0047 & 0.0039 & 0.0038 & 0.0023 & 0.0026 & 0.0028 & 0.0024 & 0.0027 \\ 
   &  &  & t-test & 0.0000 & 0.0000 & 0.0000 & 0.0000 &  & 0.0000 & 0.0000 & 0.0000 \\ 
   &  &  & wilcox-test & 0.0000 & 0.0000 & 0.0000 & 0.0000 &  & 0.0000 & 0.0000 & 0.0000 \\ 
   \midrule
3 & High & Simple & mean &  & 0.1135 & 0.1139 & 0.1112 & 0.1140 & 0.1180 & 0.1139 & 0.1179 \\ 
   &  &  & st.dev. &  & 0.0009 & 0.0010 & 0.0009 & 0.0008 & 0.0006 & 0.0008 & 0.0006 \\ 
   &  &  & t-test &  & 1.0000 & 0.7676 & 1.0000 &  & 0.0000 & 0.7268 & 0.0000 \\ 
   &  &  & wilcox-test &  & 0.9999 & 0.8438 & 1.0000 &  & 0.0000 & 0.7191 & 0.0000 \\ 
   \midrule
3 & High & Complex & mean &  & 0.1476 & 0.1474 & 0.1156 & 0.1102 & 0.1316 & 0.1110 & 0.1317 \\ 
   &  &  & st.dev. &  & 0.0013 & 0.0010 & 0.0041 & 0.0029 & 0.0031 & 0.0029 & 0.0031 \\ 
   &  &  & t-test &  & 0.0000 & 0.0000 & 0.0000 &  & 0.0000 & 0.0287 & 0.0000 \\ 
   &  &  & wilcox-test &  & 0.0000 & 0.0000 & 0.0000 &  & 0.0000 & 0.0272 & 0.0000 \\ 
   \midrule
 \midrule
6 & Low & Simple & mean & 0.0062 & 0.0687 & 0.0665 & 0.0554 & 0.0544 & 0.0833 & 0.0577 & 0.0872 \\ 
   &  &  & st.dev. & 0.0020 & 0.0048 & 0.0050 & 0.0012 & 0.0014 & 0.0009 & 0.0016 & 0.0010 \\ 
   &  &  & t-test & 1.0000 & 0.0000 & 0.0000 & 0.0000 &  & 0.0000 & 0.0000 & 0.0000 \\ 
   &  &  & wilcox-test & 1.0000 & 0.0000 & 0.0000 & 0.0000 &  & 0.0000 & 0.0000 & 0.0000 \\ 
   \midrule
6 & Low & Complex & mean & 0.1122 & 0.1093 & 0.1058 & 0.0574 & 0.0452 & 0.0719 & 0.0536 & 0.0842 \\ 
   &  &  & st.dev. & 0.0040 & 0.0045 & 0.0044 & 0.0017 & 0.0020 & 0.0022 & 0.0021 & 0.0024 \\ 
   &  &  & t-test & 0.0000 & 0.0000 & 0.0000 & 0.0000 &  & 0.0000 & 0.0000 & 0.0000 \\ 
   &  &  & wilcox-test & 0.0000 & 0.0000 & 0.0000 & 0.0000 &  & 0.0000 & 0.0000 & 0.0000 \\ 
   \midrule
6 & High & Simple & mean &  & 0.0974 & 0.0972 & 0.0951 & 0.0983 & 0.1012 & 0.0998 & 0.1016 \\ 
   &  &  & st.dev. &  & 0.0006 & 0.0006 & 0.0006 & 0.0005 & 0.0004 & 0.0005 & 0.0004 \\ 
   &  &  & t-test &  & 1.0000 & 1.0000 & 1.0000 &  & 0.0000 & 0.0000 & 0.0000 \\ 
   &  &  & wilcox-test &  & 1.0000 & 1.0000 & 1.0000 &  & 0.0000 & 0.0000 & 0.0000 \\ 
   \midrule
6 & High & Complex & mean &  & 0.0927 & 0.0927 & 0.0766 & 0.0772 & 0.0882 & 0.0898 & 0.0952 \\ 
   &  &  & st.dev. &  & 0.0006 & 0.0005 & 0.0020 & 0.0016 & 0.0014 & 0.0018 & 0.0006 \\ 
   &  &  & t-test &  & 0.0000 & 0.0000 & 0.9878 &  & 0.0000 & 0.0000 & 0.0000 \\ 
   &  &  & wilcox-test &  & 0.0000 & 0.0000 & 0.9887 &  & 0.0000 & 0.0000 & 0.0000 \\ 
   \midrule
 \midrule
9 & Low & Simple & mean & 0.0054 & 0.0653 & 0.0629 & 0.0528 & 0.0519 & 0.0789 & 0.0569 & 0.0850 \\ 
   &  &  & st.dev. & 0.0018 & 0.0042 & 0.0042 & 0.0012 & 0.0014 & 0.0009 & 0.0017 & 0.0009 \\ 
   &  &  & t-test & 1.0000 & 0.0000 & 0.0000 & 0.0000 &  & 0.0000 & 0.0000 & 0.0000 \\ 
   &  &  & wilcox-test & 1.0000 & 0.0000 & 0.0000 & 0.0000 &  & 0.0000 & 0.0000 & 0.0000 \\ 
   \midrule
9 & Low & Complex & mean & 0.0973 & 0.0912 & 0.0887 & 0.0515 & 0.0421 & 0.0647 & 0.0537 & 0.0845 \\ 
   &  &  & st.dev. & 0.0031 & 0.0033 & 0.0032 & 0.0015 & 0.0016 & 0.0019 & 0.0018 & 0.0017 \\ 
   &  &  & t-test & 0.0000 & 0.0000 & 0.0000 & 0.0000 &  & 0.0000 & 0.0000 & 0.0000 \\ 
   &  &  & wilcox-test & 0.0000 & 0.0000 & 0.0000 & 0.0000 &  & 0.0000 & 0.0000 & 0.0000 \\ 
   \midrule
9 & High & Simple & mean &  & 0.0921 & 0.0918 & 0.0900 & 0.0931 & 0.0959 & 0.0955 & 0.0964 \\ 
   &  &  & st.dev. &  & 0.0006 & 0.0006 & 0.0006 & 0.0005 & 0.0003 & 0.0004 & 0.0003 \\ 
   &  &  & t-test &  & 1.0000 & 1.0000 & 1.0000 &  & 0.0000 & 0.0000 & 0.0000 \\ 
   &  &  & wilcox-test &  & 1.0000 & 1.0000 & 1.0000 &  & 0.0000 & 0.0000 & 0.0000 \\ 
   \midrule
9 & High & Complex & mean &  & 0.1007 & 0.1004 & 0.0817 & 0.0819 & 0.0945 & 0.0997 & 0.1036 \\ 
   &  &  & st.dev. &  & 0.0007 & 0.0007 & 0.0020 & 0.0017 & 0.0015 & 0.0019 & 0.0006 \\ 
   &  &  & t-test &  & 0.0000 & 0.0000 & 0.7875 &  & 0.0000 & 0.0000 & 0.0000 \\ 
   &  &  & wilcox-test &  & 0.0000 & 0.0000 & 0.8473 &  & 0.0000 & 0.0000 & 0.0000 \\ 
   \midrule
 \bottomrule
\end{tabular}
\begin{tablenotes}\footnotesize
\item [] \hspace{-0.17cm} \textit{Notes:} Table reports the average measures of the RPS based on 100 simulation replications for the sample size of 200 observations. The first column denotes the number of outcome classes. Columns 2 and 3 specify the dimension and the DGP, respectively. The fourth column \textit{Statistic} shows the mean and the standard deviation of the accuracy measure for all methods. Additionally, \textit{t-test} and \textit{wilcox-test} contain the p-values of the parametric t-test as well as the nonparametric Wilcoxon test for the equality of means between the results of the \textit{Ordered Forest} and all the other methods.
\end{tablenotes}
\end{threeparttable}
\end{table}

\pagebreak

\subsubsection{AMSE: Sample Size = 200}

\begin{table}[h!]
\footnotesize
\centering
\caption{Simulation results: Accuracy Measure = AMSE \& Sample Size = 200}
\label{tab:mse_200}
\begin{threeparttable}
\begin{tabular}{clllrrrrrrrr}
\toprule
\midrule
\multicolumn{4}{c}{\normalsize{\textbf{Simulation Design}}} & \multicolumn{8}{c}{\normalsize{\textbf{Comparison of Methods}}}\\
\midrule
\midrule
 Class & Dim. & DGP & Statistic & Ologit & Naive & Ordinal & Cond. & Ordered & Ordered* & Multi & Multi* \\ 
  \midrule
 \midrule
3 & Low & Simple & mean & 0.0103 & 0.0669 & 0.0682 & 0.0565 & 0.0587 & 0.0800 & 0.0587 & 0.0800 \\ 
   &  &  & st.dev. & 0.0044 & 0.0041 & 0.0044 & 0.0015 & 0.0022 & 0.0009 & 0.0016 & 0.0010 \\ 
   &  &  & t-test & 1.0000 & 0.0000 & 0.0000 & 1.0000 &  & 0.0000 & 0.3900 & 0.0000 \\ 
   &  &  & wilcox-test & 1.0000 & 0.0000 & 0.0000 & 1.0000 &  & 0.0000 & 0.2614 & 0.0000 \\ 
   \midrule
3 & Low & Complex & mean & 0.1081 & 0.0985 & 0.0965 & 0.0637 & 0.0543 & 0.0752 & 0.0572 & 0.0768 \\ 
   &  &  & st.dev. & 0.0039 & 0.0034 & 0.0029 & 0.0020 & 0.0026 & 0.0021 & 0.0022 & 0.0019 \\ 
   &  &  & t-test & 0.0000 & 0.0000 & 0.0000 & 0.0000 &  & 0.0000 & 0.0000 & 0.0000 \\ 
   &  &  & wilcox-test & 0.0000 & 0.0000 & 0.0000 & 0.0000 &  & 0.0000 & 0.0000 & 0.0000 \\ 
   \midrule
3 & High & Simple & mean &  & 0.0923 & 0.0931 & 0.0908 & 0.0930 & 0.0952 & 0.0926 & 0.0952 \\ 
   &  &  & st.dev. &  & 0.0008 & 0.0013 & 0.0009 & 0.0009 & 0.0007 & 0.0007 & 0.0007 \\ 
   &  &  & t-test &  & 1.0000 & 0.2408 & 1.0000 &  & 0.0000 & 0.9980 & 0.0000 \\ 
   &  &  & wilcox-test &  & 1.0000 & 0.5433 & 1.0000 &  & 0.0000 & 0.9977 & 0.0000 \\ 
   \midrule
3 & High & Complex & mean &  & 0.1081 & 0.1079 & 0.0863 & 0.0828 & 0.0970 & 0.0834 & 0.0971 \\ 
   &  &  & st.dev. &  & 0.0012 & 0.0009 & 0.0028 & 0.0019 & 0.0021 & 0.0020 & 0.0021 \\ 
   &  &  & t-test &  & 0.0000 & 0.0000 & 0.0000 &  & 0.0000 & 0.0264 & 0.0000 \\ 
   &  &  & wilcox-test &  & 0.0000 & 0.0000 & 0.0000 &  & 0.0000 & 0.0364 & 0.0000 \\ 
   \midrule
 \midrule
6 & Low & Simple & mean & 0.0043 & 0.0284 & 0.0283 & 0.0248 & 0.0291 & 0.0324 & 0.0287 & 0.0327 \\ 
   &  &  & st.dev. & 0.0014 & 0.0012 & 0.0018 & 0.0007 & 0.0010 & 0.0005 & 0.0008 & 0.0005 \\ 
   &  &  & t-test & 1.0000 & 1.0000 & 0.9998 & 1.0000 &  & 0.0000 & 0.9958 & 0.0000 \\ 
   &  &  & wilcox-test & 1.0000 & 1.0000 & 1.0000 & 1.0000 &  & 0.0000 & 0.9953 & 0.0000 \\ 
   \midrule
6 & Low & Complex & mean & 0.0433 & 0.0438 & 0.0413 & 0.0270 & 0.0260 & 0.0314 & 0.0274 & 0.0339 \\ 
   &  &  & st.dev. & 0.0014 & 0.0017 & 0.0014 & 0.0008 & 0.0011 & 0.0009 & 0.0010 & 0.0008 \\ 
   &  &  & t-test & 0.0000 & 0.0000 & 0.0000 & 0.0000 &  & 0.0000 & 0.0000 & 0.0000 \\ 
   &  &  & wilcox-test & 0.0000 & 0.0000 & 0.0000 & 0.0000 &  & 0.0000 & 0.0000 & 0.0000 \\ 
   \midrule
6 & High & Simple & mean &  & 0.0352 & 0.0352 & 0.0347 & 0.0361 & 0.0361 & 0.0360 & 0.0361 \\ 
   &  &  & st.dev. &  & 0.0003 & 0.0004 & 0.0004 & 0.0004 & 0.0004 & 0.0003 & 0.0004 \\ 
   &  &  & t-test &  & 1.0000 & 1.0000 & 1.0000 &  & 0.8112 & 0.9994 & 0.6394 \\ 
   &  &  & wilcox-test &  & 1.0000 & 1.0000 & 1.0000 &  & 0.8788 & 0.9989 & 0.6579 \\ 
   \midrule
6 & High & Complex & mean &  & 0.0383 & 0.0386 & 0.0343 & 0.0350 & 0.0367 & 0.0378 & 0.0387 \\ 
   &  &  & st.dev. &  & 0.0003 & 0.0004 & 0.0006 & 0.0005 & 0.0005 & 0.0005 & 0.0004 \\ 
   &  &  & t-test &  & 0.0000 & 0.0000 & 1.0000 &  & 0.0000 & 0.0000 & 0.0000 \\ 
   &  &  & wilcox-test &  & 0.0000 & 0.0000 & 1.0000 &  & 0.0000 & 0.0000 & 0.0000 \\ 
   \midrule
 \midrule
9 & Low & Simple & mean & 0.0025 & 0.0150 & 0.0149 & 0.0134 & 0.0170 & 0.0170 & 0.0168 & 0.0172 \\ 
   &  &  & st.dev. & 0.0008 & 0.0005 & 0.0007 & 0.0004 & 0.0006 & 0.0003 & 0.0005 & 0.0002 \\ 
   &  &  & t-test & 1.0000 & 1.0000 & 1.0000 & 1.0000 &  & 0.5492 & 0.9993 & 0.0040 \\ 
   &  &  & wilcox-test & 1.0000 & 1.0000 & 1.0000 & 1.0000 &  & 0.3269 & 0.9985 & 0.0003 \\ 
   \midrule
9 & Low & Complex & mean & 0.0203 & 0.0194 & 0.0190 & 0.0142 & 0.0159 & 0.0161 & 0.0162 & 0.0179 \\ 
   &  &  & st.dev. & 0.0006 & 0.0006 & 0.0005 & 0.0003 & 0.0005 & 0.0003 & 0.0004 & 0.0003 \\ 
   &  &  & t-test & 0.0000 & 0.0000 & 0.0000 & 1.0000 &  & 0.0006 & 0.0000 & 0.0000 \\ 
   &  &  & wilcox-test & 0.0000 & 0.0000 & 0.0000 & 1.0000 &  & 0.0004 & 0.0000 & 0.0000 \\ 
   \midrule
9 & High & Simple & mean &  & 0.0180 & 0.0181 & 0.0178 & 0.0189 & 0.0185 & 0.0188 & 0.0185 \\ 
   &  &  & st.dev. &  & 0.0002 & 0.0002 & 0.0002 & 0.0002 & 0.0002 & 0.0002 & 0.0002 \\ 
   &  &  & t-test &  & 1.0000 & 1.0000 & 1.0000 &  & 1.0000 & 1.0000 & 1.0000 \\ 
   &  &  & wilcox-test &  & 1.0000 & 1.0000 & 1.0000 &  & 1.0000 & 1.0000 & 1.0000 \\ 
   \midrule
9 & High & Complex & mean &  & 0.0200 & 0.0200 & 0.0178 & 0.0187 & 0.0193 & 0.0201 & 0.0202 \\ 
   &  &  & st.dev. &  & 0.0002 & 0.0002 & 0.0003 & 0.0003 & 0.0003 & 0.0003 & 0.0002 \\ 
   &  &  & t-test &  & 0.0000 & 0.0000 & 1.0000 &  & 0.0000 & 0.0000 & 0.0000 \\ 
   &  &  & wilcox-test &  & 0.0000 & 0.0000 & 1.0000 &  & 0.0000 & 0.0000 & 0.0000 \\ 
   \midrule
 \bottomrule
\end{tabular}
\begin{tablenotes}\footnotesize
\item [] \hspace{-0.17cm} \textit{Notes:} Table reports the average measures of the MSE based on 100 simulation replications for the sample size of 200 observations. The first column denotes the number of outcome classes. Columns 2 and 3 specify the dimension and the DGP, respectively. The fourth column \textit{Statistic} shows the mean and the standard deviation of the accuracy measure for all methods. Additionally, \textit{t-test} and \textit{wilcox-test} contain the p-values of the parametric t-test as well as the nonparametric Wilcoxon test for the equality of means between the results of the \textit{Ordered Forest} and all the other methods.
\end{tablenotes}
\end{threeparttable}
\end{table}

\pagebreak

\subsubsection{ARPS: Sample Size = 800}

\begin{table}[h!]
\footnotesize
\centering
\caption{Simulation results: Accuracy Measure = ARPS \& Sample Size = 800}
\label{tab:rps_800}
\begin{threeparttable}
\begin{tabular}{clllrrrrrrrr}
\toprule
\midrule
\multicolumn{4}{c}{\normalsize{\textbf{Simulation Design}}} & \multicolumn{8}{c}{\normalsize{\textbf{Comparison of Methods}}}\\
\midrule
\midrule
 Class & Dim. & DGP & Statistic & Ologit & Naive & Ordinal & Cond. & Ordered & Ordered* & Multi & Multi* \\ 
  \midrule
 \midrule
3 & Low & Simple & mean & 0.0023 & 0.0701 & 0.0685 & 0.0484 & 0.0466 & 0.0799 & 0.0483 & 0.0803 \\ 
   &  &  & st.dev. & 0.0009 & 0.0043 & 0.0045 & 0.0007 & 0.0009 & 0.0008 & 0.0008 & 0.0008 \\ 
   &  &  & t-test & 1.0000 & 0.0000 & 0.0000 & 0.0000 &  & 0.0000 & 0.0000 & 0.0000 \\ 
   &  &  & wilcox-test & 1.0000 & 0.0000 & 0.0000 & 0.0000 &  & 0.0000 & 0.0000 & 0.0000 \\ 
   \midrule
3 & Low & Complex & mean & 0.0849 & 0.0828 & 0.0813 & 0.0394 & 0.0323 & 0.0495 & 0.0344 & 0.0516 \\ 
   &  &  & st.dev. & 0.0009 & 0.0024 & 0.0026 & 0.0012 & 0.0009 & 0.0013 & 0.0010 & 0.0012 \\ 
   &  &  & t-test & 0.0000 & 0.0000 & 0.0000 & 0.0000 &  & 0.0000 & 0.0000 & 0.0000 \\ 
   &  &  & wilcox-test & 0.0000 & 0.0000 & 0.0000 & 0.0000 &  & 0.0000 & 0.0000 & 0.0000 \\ 
   \midrule
3 & High & Simple & mean &  & 0.1055 & 0.1055 & 0.1017 & 0.1044 & 0.1136 & 0.1047 & 0.1136 \\ 
   &  &  & st.dev. &  & 0.0007 & 0.0007 & 0.0006 & 0.0005 & 0.0004 & 0.0005 & 0.0003 \\ 
   &  &  & t-test &  & 0.0000 & 0.0000 & 1.0000 &  & 0.0000 & 0.0000 & 0.0000 \\ 
   &  &  & wilcox-test &  & 0.0000 & 0.0000 & 1.0000 &  & 0.0000 & 0.0001 & 0.0000 \\ 
   \midrule
3 & High & Complex & mean &  & 0.0944 & 0.0949 & 0.0681 & 0.0616 & 0.0738 & 0.0635 & 0.0770 \\ 
   &  &  & st.dev. &  & 0.0007 & 0.0010 & 0.0010 & 0.0010 & 0.0010 & 0.0009 & 0.0011 \\ 
   &  &  & t-test &  & 0.0000 & 0.0000 & 0.0000 &  & 0.0000 & 0.0000 & 0.0000 \\ 
   &  &  & wilcox-test &  & 0.0000 & 0.0000 & 0.0000 &  & 0.0000 & 0.0000 & 0.0000 \\ 
   \midrule
 \midrule
6 & Low & Simple & mean & 0.0015 & 0.0619 & 0.0595 & 0.0435 & 0.0417 & 0.0702 & 0.0443 & 0.0748 \\ 
   &  &  & st.dev. & 0.0005 & 0.0037 & 0.0039 & 0.0006 & 0.0007 & 0.0007 & 0.0006 & 0.0006 \\ 
   &  &  & t-test & 1.0000 & 0.0000 & 0.0000 & 0.0000 &  & 0.0000 & 0.0000 & 0.0000 \\ 
   &  &  & wilcox-test & 1.0000 & 0.0000 & 0.0000 & 0.0000 &  & 0.0000 & 0.0000 & 0.0000 \\ 
   \midrule
6 & Low & Complex & mean & 0.0947 & 0.1020 & 0.0986 & 0.0408 & 0.0330 & 0.0510 & 0.0384 & 0.0608 \\ 
   &  &  & st.dev. & 0.0009 & 0.0031 & 0.0031 & 0.0009 & 0.0007 & 0.0010 & 0.0008 & 0.0012 \\ 
   &  &  & t-test & 0.0000 & 0.0000 & 0.0000 & 0.0000 &  & 0.0000 & 0.0000 & 0.0000 \\ 
   &  &  & wilcox-test & 0.0000 & 0.0000 & 0.0000 & 0.0000 &  & 0.0000 & 0.0000 & 0.0000 \\ 
   \midrule
6 & High & Simple & mean &  & 0.0905 & 0.0898 & 0.0874 & 0.0905 & 0.0978 & 0.0940 & 0.0995 \\ 
   &  &  & st.dev. &  & 0.0006 & 0.0005 & 0.0004 & 0.0003 & 0.0002 & 0.0004 & 0.0002 \\ 
   &  &  & t-test &  & 0.6597 & 1.0000 & 1.0000 &  & 0.0000 & 0.0000 & 0.0000 \\ 
   &  &  & wilcox-test &  & 0.8939 & 1.0000 & 1.0000 &  & 0.0000 & 0.0000 & 0.0000 \\ 
   \midrule
6 & High & Complex & mean &  & 0.1069 & 0.1060 & 0.0774 & 0.0698 & 0.0840 & 0.0781 & 0.0931 \\ 
   &  &  & st.dev. &  & 0.0007 & 0.0007 & 0.0010 & 0.0009 & 0.0010 & 0.0013 & 0.0011 \\ 
   &  &  & t-test &  & 0.0000 & 0.0000 & 0.0000 &  & 0.0000 & 0.0000 & 0.0000 \\ 
   &  &  & wilcox-test &  & 0.0000 & 0.0000 & 0.0000 &  & 0.0000 & 0.0000 & 0.0000 \\ 
   \midrule
 \midrule
9 & Low & Simple & mean & 0.0013 & 0.0603 & 0.0570 & 0.0417 & 0.0400 & 0.0668 & 0.0432 & 0.0741 \\ 
   &  &  & st.dev. & 0.0004 & 0.0032 & 0.0035 & 0.0006 & 0.0006 & 0.0006 & 0.0007 & 0.0006 \\ 
   &  &  & t-test & 1.0000 & 0.0000 & 0.0000 & 0.0000 &  & 0.0000 & 0.0000 & 0.0000 \\ 
   &  &  & wilcox-test & 1.0000 & 0.0000 & 0.0000 & 0.0000 &  & 0.0000 & 0.0000 & 0.0000 \\ 
   \midrule
9 & Low & Complex & mean & 0.0837 & 0.0867 & 0.0836 & 0.0368 & 0.0305 & 0.0459 & 0.0375 & 0.0614 \\ 
   &  &  & st.dev. & 0.0009 & 0.0023 & 0.0027 & 0.0008 & 0.0006 & 0.0008 & 0.0006 & 0.0009 \\ 
   &  &  & t-test & 0.0000 & 0.0000 & 0.0000 & 0.0000 &  & 0.0000 & 0.0000 & 0.0000 \\ 
   &  &  & wilcox-test & 0.0000 & 0.0000 & 0.0000 & 0.0000 &  & 0.0000 & 0.0000 & 0.0000 \\ 
   \midrule
9 & High & Simple & mean &  & 0.0857 & 0.0847 & 0.0826 & 0.0860 & 0.0927 & 0.0920 & 0.0949 \\ 
   &  &  & st.dev. &  & 0.0005 & 0.0005 & 0.0004 & 0.0003 & 0.0002 & 0.0004 & 0.0001 \\ 
   &  &  & t-test &  & 1.0000 & 1.0000 & 1.0000 &  & 0.0000 & 0.0000 & 0.0000 \\ 
   &  &  & wilcox-test &  & 1.0000 & 1.0000 & 1.0000 &  & 0.0000 & 0.0000 & 0.0000 \\ 
   \midrule
9 & High & Complex & mean &  & 0.0956 & 0.0947 & 0.0708 & 0.0648 & 0.0773 & 0.0781 & 0.0933 \\ 
   &  &  & st.dev. &  & 0.0006 & 0.0007 & 0.0007 & 0.0006 & 0.0007 & 0.0011 & 0.0009 \\ 
   &  &  & t-test &  & 0.0000 & 0.0000 & 0.0000 &  & 0.0000 & 0.0000 & 0.0000 \\ 
   &  &  & wilcox-test &  & 0.0000 & 0.0000 & 0.0000 &  & 0.0000 & 0.0000 & 0.0000 \\ 
   \midrule
 \bottomrule
\end{tabular}
\begin{tablenotes}\footnotesize
\item [] \hspace{-0.17cm} \textit{Notes:} Table reports the average measures of the RPS based on 100 simulation replications for the sample size of 800 observations. The first column denotes the number of outcome classes. Columns 2 and 3 specify the dimension and the DGP, respectively. The fourth column \textit{Statistic} shows the mean and the standard deviation of the accuracy measure for all methods. Additionally, \textit{t-test} and \textit{wilcox-test} contain the p-values of the parametric t-test as well as the nonparametric Wilcoxon test for the equality of means between the results of the \textit{Ordered Forest} and all the other methods.
\end{tablenotes}
\end{threeparttable}
\end{table}

\pagebreak

\subsubsection{AMSE: Sample Size = 800}

\begin{table}[h!]
\footnotesize
\centering
\caption{Simulation results: Accuracy Measure = AMSE \& Sample Size = 800}
\label{tab:mse_800}
\begin{threeparttable}
\begin{tabular}{clllrrrrrrrr}
\toprule
\midrule
\multicolumn{4}{c}{\normalsize{\textbf{Simulation Design}}} & \multicolumn{8}{c}{\normalsize{\textbf{Comparison of Methods}}}\\
\midrule
\midrule
 Class & Dim. & DGP & Statistic & Ologit & Naive & Ordinal & Cond. & Ordered & Ordered* & Multi & Multi* \\ 
  \midrule
 \midrule
3 & Low & Simple & mean & 0.0025 & 0.0618 & 0.0624 & 0.0451 & 0.0461 & 0.0688 & 0.0472 & 0.0691 \\ 
   &  &  & st.dev. & 0.0009 & 0.0032 & 0.0036 & 0.0006 & 0.0010 & 0.0006 & 0.0007 & 0.0006 \\ 
   &  &  & t-test & 1.0000 & 0.0000 & 0.0000 & 1.0000 &  & 0.0000 & 0.0000 & 0.0000 \\ 
   &  &  & wilcox-test & 1.0000 & 0.0000 & 0.0000 & 1.0000 &  & 0.0000 & 0.0000 & 0.0000 \\ 
   \midrule
3 & Low & Complex & mean & 0.0875 & 0.0848 & 0.0834 & 0.0482 & 0.0414 & 0.0574 & 0.0439 & 0.0602 \\ 
   &  &  & st.dev. & 0.0008 & 0.0020 & 0.0020 & 0.0011 & 0.0010 & 0.0011 & 0.0011 & 0.0010 \\ 
   &  &  & t-test & 0.0000 & 0.0000 & 0.0000 & 0.0000 &  & 0.0000 & 0.0000 & 0.0000 \\ 
   &  &  & wilcox-test & 0.0000 & 0.0000 & 0.0000 & 0.0000 &  & 0.0000 & 0.0000 & 0.0000 \\ 
   \midrule
3 & High & Simple & mean &  & 0.0866 & 0.0870 & 0.0840 & 0.0861 & 0.0920 & 0.0861 & 0.0920 \\ 
   &  &  & st.dev. &  & 0.0005 & 0.0007 & 0.0005 & 0.0005 & 0.0003 & 0.0004 & 0.0003 \\ 
   &  &  & t-test &  & 0.0000 & 0.0000 & 1.0000 &  & 0.0000 & 0.5234 & 0.0000 \\ 
   &  &  & wilcox-test &  & 0.0000 & 0.0000 & 1.0000 &  & 0.0000 & 0.4713 & 0.0000 \\ 
   \midrule
3 & High & Complex & mean &  & 0.0969 & 0.0977 & 0.0717 & 0.0656 & 0.0749 & 0.0675 & 0.0789 \\ 
   &  &  & st.dev. &  & 0.0006 & 0.0008 & 0.0009 & 0.0010 & 0.0009 & 0.0009 & 0.0011 \\ 
   &  &  & t-test &  & 0.0000 & 0.0000 & 0.0000 &  & 0.0000 & 0.0000 & 0.0000 \\ 
   &  &  & wilcox-test &  & 0.0000 & 0.0000 & 0.0000 &  & 0.0000 & 0.0000 & 0.0000 \\ 
   \midrule
 \midrule
6 & Low & Simple & mean & 0.0010 & 0.0260 & 0.0260 & 0.0206 & 0.0231 & 0.0287 & 0.0234 & 0.0292 \\ 
   &  &  & st.dev. & 0.0003 & 0.0010 & 0.0014 & 0.0003 & 0.0005 & 0.0002 & 0.0003 & 0.0002 \\ 
   &  &  & t-test & 1.0000 & 0.0000 & 0.0000 & 1.0000 &  & 0.0000 & 0.0000 & 0.0000 \\ 
   &  &  & wilcox-test & 1.0000 & 0.0000 & 0.0000 & 1.0000 &  & 0.0000 & 0.0000 & 0.0000 \\ 
   \midrule
6 & Low & Complex & mean & 0.0376 & 0.0406 & 0.0384 & 0.0219 & 0.0208 & 0.0257 & 0.0221 & 0.0280 \\ 
   &  &  & st.dev. & 0.0003 & 0.0010 & 0.0009 & 0.0004 & 0.0004 & 0.0004 & 0.0004 & 0.0003 \\ 
   &  &  & t-test & 0.0000 & 0.0000 & 0.0000 & 0.0000 &  & 0.0000 & 0.0000 & 0.0000 \\ 
   &  &  & wilcox-test & 0.0000 & 0.0000 & 0.0000 & 0.0000 &  & 0.0000 & 0.0000 & 0.0000 \\ 
   \midrule
6 & High & Simple & mean &  & 0.0333 & 0.0332 & 0.0325 & 0.0339 & 0.0350 & 0.0343 & 0.0353 \\ 
   &  &  & st.dev. &  & 0.0002 & 0.0002 & 0.0001 & 0.0001 & 0.0001 & 0.0001 & 0.0001 \\ 
   &  &  & t-test &  & 1.0000 & 1.0000 & 1.0000 &  & 0.0000 & 0.0000 & 0.0000 \\ 
   &  &  & wilcox-test &  & 1.0000 & 1.0000 & 1.0000 &  & 0.0000 & 0.0000 & 0.0000 \\ 
   \midrule
6 & High & Complex & mean &  & 0.0404 & 0.0399 & 0.0308 & 0.0287 & 0.0325 & 0.0313 & 0.0352 \\ 
   &  &  & st.dev. &  & 0.0002 & 0.0002 & 0.0003 & 0.0003 & 0.0004 & 0.0004 & 0.0003 \\ 
   &  &  & t-test &  & 0.0000 & 0.0000 & 0.0000 &  & 0.0000 & 0.0000 & 0.0000 \\ 
   &  &  & wilcox-test &  & 0.0000 & 0.0000 & 0.0000 &  & 0.0000 & 0.0000 & 0.0000 \\ 
   \midrule
 \midrule
9 & Low & Simple & mean & 0.0006 & 0.0140 & 0.0138 & 0.0113 & 0.0136 & 0.0153 & 0.0135 & 0.0156 \\ 
   &  &  & st.dev. & 0.0002 & 0.0004 & 0.0006 & 0.0002 & 0.0003 & 0.0001 & 0.0002 & 0.0001 \\ 
   &  &  & t-test & 1.0000 & 0.0000 & 0.0121 & 1.0000 &  & 0.0000 & 1.0000 & 0.0000 \\ 
   &  &  & wilcox-test & 1.0000 & 0.0000 & 0.0241 & 1.0000 &  & 0.0000 & 1.0000 & 0.0000 \\ 
   \midrule
9 & Low & Complex & mean & 0.0178 & 0.0187 & 0.0181 & 0.0114 & 0.0124 & 0.0132 & 0.0126 & 0.0149 \\ 
   &  &  & st.dev. & 0.0001 & 0.0004 & 0.0005 & 0.0002 & 0.0003 & 0.0002 & 0.0002 & 0.0001 \\ 
   &  &  & t-test & 0.0000 & 0.0000 & 0.0000 & 1.0000 &  & 0.0000 & 0.0000 & 0.0000 \\ 
   &  &  & wilcox-test & 0.0000 & 0.0000 & 0.0000 & 1.0000 &  & 0.0000 & 0.0000 & 0.0000 \\ 
   \midrule
9 & High & Simple & mean &  & 0.0171 & 0.0171 & 0.0167 & 0.0179 & 0.0179 & 0.0184 & 0.0181 \\ 
   &  &  & st.dev. &  & 0.0001 & 0.0001 & 0.0001 & 0.0001 & 0.0001 & 0.0001 & 0.0001 \\ 
   &  &  & t-test &  & 1.0000 & 1.0000 & 1.0000 &  & 0.9803 & 0.0000 & 0.0000 \\ 
   &  &  & wilcox-test &  & 1.0000 & 1.0000 & 1.0000 &  & 0.9670 & 0.0000 & 0.0000 \\ 
   \midrule
9 & High & Complex & mean &  & 0.0191 & 0.0191 & 0.0162 & 0.0161 & 0.0170 & 0.0176 & 0.0187 \\ 
   &  &  & st.dev. &  & 0.0001 & 0.0001 & 0.0001 & 0.0001 & 0.0001 & 0.0001 & 0.0001 \\ 
   &  &  & t-test &  & 0.0000 & 0.0000 & 0.0000 &  & 0.0000 & 0.0000 & 0.0000 \\ 
   &  &  & wilcox-test &  & 0.0000 & 0.0000 & 0.0000 &  & 0.0000 & 0.0000 & 0.0000 \\ 
   \midrule
 \bottomrule
\end{tabular}
\begin{tablenotes}\footnotesize
\item [] \hspace{-0.17cm} \textit{Notes:} Table reports the average measures of the MSE based on 100 simulation replications for the sample size of 800 observations. The first column denotes the number of outcome classes. Columns 2 and 3 specify the dimension and the DGP, respectively. The fourth column \textit{Statistic} shows the mean and the standard deviation of the accuracy measure for all methods. Additionally, \textit{t-test} and \textit{wilcox-test} contain the p-values of the parametric t-test as well as the nonparametric Wilcoxon test for the equality of means between the results of the \textit{Ordered Forest} and all the other methods.
\end{tablenotes}
\end{threeparttable}
\end{table}

\pagebreak

\subsubsection{ARPS \& AMSE: Sample Size = 3200}

\begin{table}[h!]
\footnotesize
\centering
\caption{Simulation results: Accuracy Measure = ARPS/AMSE \& Sample Size = 3200} 
\label{tab:mse_3200}
\begin{threeparttable}
\begin{tabular}{clllrrrr}
\toprule
\midrule
\multicolumn{4}{c}{\normalsize{\textbf{Simulation Design}}} & \multicolumn{2}{c}{\normalsize{\textbf{ARPS}}} & \multicolumn{2}{c}{\normalsize{\textbf{AMSE}}}\\
\midrule
\midrule
 Class & Dim. & DGP & Statistic  & Ordered & Ordered* & Ordered & Ordered* \\ 
  \midrule
 \midrule
3 & Low & Simple & mean & 0.0373 & 0.0670 & 0.0376 & 0.0591 \\ 
   &  &  & st.dev. & 0.0004 & 0.0005 & 0.0005 & 0.0004 \\ 
   &  &  & t-test &  & 0.0000 &  & 0.0000 \\ 
   &  &  & wilcox-test &  & 0.0000 &  & 0.0000 \\ 
   \midrule
3 & Low & Complex & mean & 0.0285 & 0.0415 & 0.0243 & 0.0336 \\ 
   &  &  & st.dev. & 0.0004 & 0.0005 & 0.0003 & 0.0004 \\ 
   &  &  & t-test &  & 0.0000 &  & 0.0000 \\ 
   &  &  & wilcox-test &  & 0.0000 &  & 0.0000 \\ 
   \midrule
3 & High & Simple & mean & 0.0956 & 0.1069 & 0.0798 & 0.0872 \\ 
   &  &  & st.dev. & 0.0003 & 0.0002 & 0.0002 & 0.0002 \\ 
   &  &  & t-test &  & 0.0000 &  & 0.0000 \\ 
   &  &  & wilcox-test &  & 0.0000 &  & 0.0000 \\ 
   \midrule
3 & High & Complex & mean & 0.0498 & 0.0620 & 0.0557 & 0.0653 \\ 
   &  &  & st.dev. & 0.0004 & 0.0005 & 0.0005 & 0.0005 \\ 
   &  &  & t-test &  & 0.0000 &  & 0.0000 \\ 
   &  &  & wilcox-test &  & 0.0000 &  & 0.0000 \\ 
   \midrule
 \midrule
6 & Low & Simple & mean & 0.0335 & 0.0593 & 0.0188 & 0.0253 \\ 
   &  &  & st.dev. & 0.0004 & 0.0004 & 0.0002 & 0.0001 \\ 
   &  &  & t-test &  & 0.0000 &  & 0.0000 \\ 
   &  &  & wilcox-test &  & 0.0000 &  & 0.0000 \\ 
   \midrule
6 & Low & Complex & mean & 0.0255 & 0.0367 & 0.0162 & 0.0197 \\ 
   &  &  & st.dev. & 0.0003 & 0.0004 & 0.0002 & 0.0002 \\ 
   &  &  & t-test &  & 0.0000 &  & 0.0000 \\ 
   &  &  & wilcox-test &  & 0.0000 &  & 0.0000 \\ 
   \midrule
6 & High & Simple & mean & 0.0825 & 0.0923 & 0.0314 & 0.0335 \\ 
   &  &  & st.dev. & 0.0002 & 0.0002 & 0.0001 & 0.0000 \\ 
   &  &  & t-test &  & 0.0000 &  & 0.0000 \\ 
   &  &  & wilcox-test &  & 0.0000 &  & 0.0000 \\ 
   \midrule
6 & High & Complex & mean & 0.0526 & 0.0656 & 0.0264 & 0.0292 \\ 
   &  &  & st.dev. & 0.0004 & 0.0004 & 0.0002 & 0.0001 \\ 
   &  &  & t-test &  & 0.0000 &  & 0.0000 \\ 
   &  &  & wilcox-test &  & 0.0000 &  & 0.0000 \\ 
   \midrule
 \midrule
9 & Low & Simple & mean & 0.0321 & 0.0565 & 0.0110 & 0.0136 \\ 
   &  &  & st.dev. & 0.0003 & 0.0003 & 0.0001 & 0.0001 \\ 
   &  &  & t-test &  & 0.0000 &  & 0.0000 \\ 
   &  &  & wilcox-test &  & 0.0000 &  & 0.0000 \\ 
   \midrule
9 & Low & Complex & mean & 0.0244 & 0.0350 & 0.0098 & 0.0110 \\ 
   &  &  & st.dev. & 0.0002 & 0.0003 & 0.0001 & 0.0001 \\ 
   &  &  & t-test &  & 0.0000 &  & 0.0000 \\ 
   &  &  & wilcox-test &  & 0.0000 &  & 0.0000 \\ 
   \midrule
9 & High & Simple & mean & 0.0783 & 0.0875 & 0.0165 & 0.0172 \\ 
   &  &  & st.dev. & 0.0002 & 0.0002 & 0.0000 & 0.0000 \\ 
   &  &  & t-test &  & 0.0000 &  & 0.0000 \\ 
   &  &  & wilcox-test &  & 0.0000 &  & 0.0000 \\ 
   \midrule
9 & High & Complex & mean & 0.0559 & 0.0697 & 0.0145 & 0.0160 \\ 
   &  &  & st.dev. & 0.0004 & 0.0004 & 0.0001 & 0.0001 \\ 
   &  &  & t-test &  & 0.0000 &  & 0.0000 \\ 
   &  &  & wilcox-test &  & 0.0000 &  & 0.0000 \\ 
   \midrule
 \bottomrule
\end{tabular}
\begin{tablenotes}\footnotesize
\item [] \hspace{-0.17cm} \textit{Notes:} Table reports the average measures of the RPS and MSE based on 100 simulation replications for the sample size of 3200 observations. The first column denotes the number of outcome classes. Columns 2 and 3 specify the dimension and the DGP, respectively. The fourth column \textit{Statistic} shows the mean and the standard deviation of the accuracy measure for all methods. Additionally, \textit{t-test} and \textit{wilcox-test} contain the p-values of the parametric t-test as well as the nonparametric Wilcoxon test for the equality of means between the results of the \textit{Ordered Forest} and the honest version of the \textit{Ordered Forest}.
\end{tablenotes}
\end{threeparttable}
\end{table}

\pagebreak

\pagebreak

\pagebreak

\pagebreak

\pagebreak

\subsection{Complete Simulation Results}\label{AppendixSR}
Tables \ref{tab:allsim1} to \ref{tab:allsim12} below summarize the simulation results for all 72 different DGPs, complementing the main results presented in Section \ref{sec:boxplots} of the main text. Each table specifies the particular simulation design as follows: the first column \textit{DGP} provides the identifier for the data generating process. Columns 2 to 5 specify the particular characteristics of the respective DGP, namely if the DGP features additional noise variables (\textit{noise}), 15 in the low-dimensional case and 1000 in the high-dimensional case, nonlinear effects (\textit{nonlin}), multicollinearity among covariates (\textit{multi}), and randomly spaced thresholds (\textit{random}). The sixth column \textit{Statistic} contains summary statistics of the simulation results. In particular, the mean of the respective accuracy measure (\textit{mean}) and its standard deviation (\textit{st.dev.}). Furthermore, rows \textit{t-test} and \textit{wilcox-test} contain the \textit{p}-values of the parametric t-test as well as the nonparametric Wilcoxon test for the equality of means between the results of the \textit{Ordered Forest} and all the other methods. The alternative hypothesis is that the mean of the \textit{Ordered Forest} is less than the mean of the other method to test if the \textit{Ordered Forest} achieves significantly lower prediction error than the other considered methods.

\pagebreak
\subsubsection{ARPS: Low Dimension with 3 Classes}
\begin{table}[h!]
\footnotesize
\centering
\caption{Simulation Results: Accuracy Measure = ARPS \& Low Dimension with 3 Classes} \label{tab:allsim1}
\resizebox{0.94\textwidth}{!}{
\begin{threeparttable}
\begin{tabular}{c|cccc|lrrrrrrrr}
\toprule
\midrule
\multicolumn{6}{c}{\normalsize{\textbf{Simulation Design}}} & \multicolumn{8}{c}{\normalsize{\textbf{Comparison of Methods}}}\\
\midrule
\midrule
 DGP & noise & nonlin & multi & random & Statistic & Ologit & Naive & Ordinal & Cond. & Ordered & Ordered* & Multi & Multi* \\ 
  \midrule
 \midrule
1 & \xmark & \xmark & \xmark & \xmark & mean & 0.0097 & 0.0765 & 0.0755 & 0.0625 & 0.0609 & 0.0954 & 0.0619 & 0.0954 \\ 
   &  &  &  &  & st.dev. & 0.0042 & 0.0056 & 0.0055 & 0.0018 & 0.0020 & 0.0011 & 0.0019 & 0.0012 \\ 
   &  &  &  &  & t-test & 1.0000 & 0.0000 & 0.0000 & 0.0000 &  & 0.0000 & 0.0002 & 0.0000 \\ 
   &  &  &  &  & wilcox-test & 1.0000 & 0.0000 & 0.0000 & 0.0000 &  & 0.0000 & 0.0000 & 0.0000 \\ 
   \midrule
2 & \cmark & \xmark & \xmark & \xmark & mean & 0.0216 & 0.0840 & 0.0832 & 0.0738 & 0.0754 & 0.1041 & 0.0763 & 0.1041 \\ 
   &  &  &  &  & st.dev. & 0.0054 & 0.0046 & 0.0048 & 0.0015 & 0.0016 & 0.0013 & 0.0016 & 0.0013 \\ 
   &  &  &  &  & t-test & 1.0000 & 0.0000 & 0.0000 & 1.0000 &  & 0.0000 & 0.0001 & 0.0000 \\ 
   &  &  &  &  & wilcox-test & 1.0000 & 0.0000 & 0.0000 & 1.0000 &  & 0.0000 & 0.0001 & 0.0000 \\ 
   \midrule
3 & \xmark & \cmark & \xmark & \xmark & mean & 0.0904 & 0.0715 & 0.0726 & 0.0688 & 0.0681 & 0.0824 & 0.0672 & 0.0824 \\ 
   &  &  &  &  & st.dev. & 0.0045 & 0.0031 & 0.0033 & 0.0021 & 0.0022 & 0.0013 & 0.0020 & 0.0013 \\ 
   &  &  &  &  & t-test & 0.0000 & 0.0000 & 0.0000 & 0.0132 &  & 0.0000 & 0.9988 & 0.0000 \\ 
   &  &  &  &  & wilcox-test & 0.0000 & 0.0000 & 0.0000 & 0.0070 &  & 0.0000 & 0.9976 & 0.0000 \\ 
   \midrule
4 & \xmark & \xmark & \cmark & \xmark & mean & 0.0097 & 0.1236 & 0.1194 & 0.0316 & 0.0297 & 0.0449 & 0.0297 & 0.0493 \\ 
   &  &  &  &  & st.dev. & 0.0031 & 0.0079 & 0.0079 & 0.0015 & 0.0015 & 0.0013 & 0.0014 & 0.0013 \\ 
   &  &  &  &  & t-test & 1.0000 & 0.0000 & 0.0000 & 0.0000 &  & 0.0000 & 0.4099 & 0.0000 \\ 
   &  &  &  &  & wilcox-test & 1.0000 & 0.0000 & 0.0000 & 0.0000 &  & 0.0000 & 0.3721 & 0.0000 \\ 
   \midrule
5 & \xmark & \xmark & \xmark & \cmark & mean & 0.0104 & 0.0730 & 0.0698 & 0.0611 & 0.0594 & 0.0942 & 0.0607 & 0.0948 \\ 
   &  &  &  &  & st.dev. & 0.0035 & 0.0072 & 0.0066 & 0.0017 & 0.0020 & 0.0015 & 0.0021 & 0.0016 \\ 
   &  &  &  &  & t-test & 1.0000 & 0.0000 & 0.0000 & 0.0000 &  & 0.0000 & 0.0000 & 0.0000 \\ 
   &  &  &  &  & wilcox-test & 1.0000 & 0.0000 & 0.0000 & 0.0000 &  & 0.0000 & 0.0000 & 0.0000 \\ 
   \midrule
6 & \cmark & \cmark & \xmark & \xmark & mean & 0.1052 & 0.0772 & 0.0781 & 0.0763 & 0.0768 & 0.0863 & 0.0759 & 0.0862 \\ 
   &  &  &  &  & st.dev. & 0.0066 & 0.0025 & 0.0030 & 0.0021 & 0.0020 & 0.0011 & 0.0019 & 0.0011 \\ 
   &  &  &  &  & t-test & 0.0000 & 0.1612 & 0.0004 & 0.9717 &  & 0.0000 & 0.9998 & 0.0000 \\ 
   &  &  &  &  & wilcox-test & 0.0000 & 0.1979 & 0.0004 & 0.9589 &  & 0.0000 & 0.9996 & 0.0000 \\ 
   \midrule
7 & \cmark & \xmark & \cmark & \xmark & mean & 0.0221 & 0.1349 & 0.1321 & 0.0344 & 0.0335 & 0.0502 & 0.0353 & 0.0569 \\ 
   &  &  &  &  & st.dev. & 0.0064 & 0.0060 & 0.0057 & 0.0013 & 0.0011 & 0.0021 & 0.0013 & 0.0022 \\ 
   &  &  &  &  & t-test & 1.0000 & 0.0000 & 0.0000 & 0.0000 &  & 0.0000 & 0.0000 & 0.0000 \\ 
   &  &  &  &  & wilcox-test & 1.0000 & 0.0000 & 0.0000 & 0.0000 &  & 0.0000 & 0.0000 & 0.0000 \\ 
   \midrule
8 & \cmark & \xmark & \xmark & \cmark & mean & 0.0196 & 0.0750 & 0.0753 & 0.0669 & 0.0694 & 0.0938 & 0.0699 & 0.0940 \\ 
   &  &  &  &  & st.dev. & 0.0056 & 0.0036 & 0.0040 & 0.0017 & 0.0019 & 0.0010 & 0.0018 & 0.0010 \\ 
   &  &  &  &  & t-test & 1.0000 & 0.0000 & 0.0000 & 1.0000 &  & 0.0000 & 0.0412 & 0.0000 \\ 
   &  &  &  &  & wilcox-test & 1.0000 & 0.0000 & 0.0000 & 1.0000 &  & 0.0000 & 0.0339 & 0.0000 \\ 
   \midrule
9 & \xmark & \cmark & \cmark & \xmark & mean & 0.1116 & 0.1204 & 0.1170 & 0.0486 & 0.0401 & 0.0706 & 0.0422 & 0.0722 \\ 
   &  &  &  &  & st.dev. & 0.0030 & 0.0077 & 0.0075 & 0.0022 & 0.0021 & 0.0025 & 0.0021 & 0.0024 \\ 
   &  &  &  &  & t-test & 0.0000 & 0.0000 & 0.0000 & 0.0000 &  & 0.0000 & 0.0000 & 0.0000 \\ 
   &  &  &  &  & wilcox-test & 0.0000 & 0.0000 & 0.0000 & 0.0000 &  & 0.0000 & 0.0000 & 0.0000 \\ 
   \midrule
10 & \xmark & \cmark & \xmark & \cmark & mean & 0.0905 & 0.0703 & 0.0693 & 0.0673 & 0.0668 & 0.0808 & 0.0672 & 0.0809 \\ 
   &  &  &  &  & st.dev. & 0.0047 & 0.0042 & 0.0042 & 0.0023 & 0.0023 & 0.0013 & 0.0023 & 0.0013 \\ 
   &  &  &  &  & t-test & 0.0000 & 0.0000 & 0.0000 & 0.0650 &  & 0.0000 & 0.0923 & 0.0000 \\ 
   &  &  &  &  & wilcox-test & 0.0000 & 0.0000 & 0.0000 & 0.0862 &  & 0.0000 & 0.0921 & 0.0000 \\ 
   \midrule
11 & \xmark & \xmark & \cmark & \cmark & mean & 0.0111 & 0.1299 & 0.1284 & 0.0312 & 0.0295 & 0.0428 & 0.0298 & 0.0463 \\ 
   &  &  &  &  & st.dev. & 0.0042 & 0.0115 & 0.0121 & 0.0017 & 0.0017 & 0.0014 & 0.0016 & 0.0015 \\ 
   &  &  &  &  & t-test & 1.0000 & 0.0000 & 0.0000 & 0.0000 &  & 0.0000 & 0.0868 & 0.0000 \\ 
   &  &  &  &  & wilcox-test & 1.0000 & 0.0000 & 0.0000 & 0.0000 &  & 0.0000 & 0.0901 & 0.0000 \\ 
   \midrule
12 & \cmark & \cmark & \cmark & \xmark & mean & 0.1297 & 0.1232 & 0.1209 & 0.0639 & 0.0483 & 0.0809 & 0.0512 & 0.0819 \\ 
   &  &  &  &  & st.dev. & 0.0051 & 0.0058 & 0.0055 & 0.0024 & 0.0025 & 0.0028 & 0.0023 & 0.0027 \\ 
   &  &  &  &  & t-test & 0.0000 & 0.0000 & 0.0000 & 0.0000 &  & 0.0000 & 0.0000 & 0.0000 \\ 
   &  &  &  &  & wilcox-test & 0.0000 & 0.0000 & 0.0000 & 0.0000 &  & 0.0000 & 0.0000 & 0.0000 \\ 
   \midrule
13 & \cmark & \cmark & \xmark & \cmark & mean & 0.0915 & 0.0682 & 0.0697 & 0.0675 & 0.0689 & 0.0764 & 0.0677 & 0.0764 \\ 
   &  &  &  &  & st.dev. & 0.0063 & 0.0022 & 0.0024 & 0.0020 & 0.0020 & 0.0011 & 0.0019 & 0.0011 \\ 
   &  &  &  &  & t-test & 0.0000 & 0.9877 & 0.0036 & 1.0000 &  & 0.0000 & 1.0000 & 0.0000 \\ 
   &  &  &  &  & wilcox-test & 0.0000 & 0.9813 & 0.0032 & 1.0000 &  & 0.0000 & 1.0000 & 0.0000 \\ 
   \midrule
14 & \cmark & \xmark & \cmark & \cmark & mean & 0.0235 & 0.1219 & 0.1194 & 0.0319 & 0.0312 & 0.0468 & 0.0324 & 0.0524 \\ 
   &  &  &  &  & st.dev. & 0.0068 & 0.0052 & 0.0050 & 0.0015 & 0.0014 & 0.0020 & 0.0014 & 0.0021 \\ 
   &  &  &  &  & t-test & 1.0000 & 0.0000 & 0.0000 & 0.0012 &  & 0.0000 & 0.0000 & 0.0000 \\ 
   &  &  &  &  & wilcox-test & 1.0000 & 0.0000 & 0.0000 & 0.0008 &  & 0.0000 & 0.0000 & 0.0000 \\ 
   \midrule
15 & \xmark & \cmark & \cmark & \cmark & mean & 0.1118 & 0.1222 & 0.1204 & 0.0482 & 0.0396 & 0.0688 & 0.0411 & 0.0712 \\ 
   &  &  &  &  & st.dev. & 0.0042 & 0.0087 & 0.0092 & 0.0024 & 0.0025 & 0.0026 & 0.0024 & 0.0026 \\ 
   &  &  &  &  & t-test & 0.0000 & 0.0000 & 0.0000 & 0.0000 &  & 0.0000 & 0.0000 & 0.0000 \\ 
   &  &  &  &  & wilcox-test & 0.0000 & 0.0000 & 0.0000 & 0.0000 &  & 0.0000 & 0.0000 & 0.0000 \\ 
   \midrule
16 & \cmark & \cmark & \cmark & \cmark & mean & 0.1156 & 0.1044 & 0.1028 & 0.0593 & 0.0466 & 0.0748 & 0.0491 & 0.0760 \\ 
   &  &  &  &  & st.dev. & 0.0047 & 0.0039 & 0.0038 & 0.0023 & 0.0026 & 0.0028 & 0.0024 & 0.0027 \\ 
   &  &  &  &  & t-test & 0.0000 & 0.0000 & 0.0000 & 0.0000 &  & 0.0000 & 0.0000 & 0.0000 \\ 
   &  &  &  &  & wilcox-test & 0.0000 & 0.0000 & 0.0000 & 0.0000 &  & 0.0000 & 0.0000 & 0.0000 \\ 
   \midrule
 \bottomrule
\end{tabular}
\begin{tablenotes}\footnotesize
\item [] \hspace{-0.17cm} \textit{Notes:} Table reports the average measures of the RPS based on 100 simulation replications for the sample size of 200 observations with 3 outcome classes. Columns 1 to 5 specify the DGP identifier and its features, namely 15 additional noise variables (\textit{noise}), nonlinear effects (\textit{nonlin}), multicollinearity among covariates (\textit{multi}), and randomly spaced thresholds (\textit{random}). The sixth column \textit{Statistic} shows the mean and the standard deviation of the accuracy measure for all methods. Additionally, \textit{t-test} and \textit{wilcox-test} contain the p-values of the parametric t-test as well as the nonparametric Wilcoxon test for the equality of means between the results of the \textit{Ordered Forest} and all the other methods.
\end{tablenotes}
\end{threeparttable}
}
\end{table}

\pagebreak

\subsubsection{ARPS: Low Dimension with 6 Classes}
\begin{table}[h!]
\footnotesize
\centering
\caption{Simulation Results: Accuracy Measure = ARPS \& Low Dimension with 6 Classes}\label{tab:allsim2} 
\resizebox{0.94\textwidth}{!}{
\begin{threeparttable}
\begin{tabular}{c|cccc|lrrrrrrrr}
\toprule
\midrule
\multicolumn{6}{c}{\normalsize{\textbf{Simulation Design}}} & \multicolumn{8}{c}{\normalsize{\textbf{Comparison of Methods}}}\\
\midrule
\midrule
 DGP & noise & nonlin & multi & random & Statistic & Ologit & Naive & Ordinal & Cond. & Ordered & Ordered* & Multi & Multi* \\ 
  \midrule
 \midrule
17 & \xmark & \xmark & \xmark & \xmark & mean & 0.0062 & 0.0687 & 0.0665 & 0.0554 & 0.0544 & 0.0833 & 0.0577 & 0.0872 \\ 
   &  &  &  &  & st.dev. & 0.0020 & 0.0048 & 0.0050 & 0.0012 & 0.0014 & 0.0009 & 0.0016 & 0.0010 \\ 
   &  &  &  &  & t-test & 1.0000 & 0.0000 & 0.0000 & 0.0000 &  & 0.0000 & 0.0000 & 0.0000 \\ 
   &  &  &  &  & wilcox-test & 1.0000 & 0.0000 & 0.0000 & 0.0000 &  & 0.0000 & 0.0000 & 0.0000 \\ 
   \midrule
18 & \cmark & \xmark & \xmark & \xmark & mean & 0.0129 & 0.0726 & 0.0708 & 0.0645 & 0.0669 & 0.0901 & 0.0709 & 0.0932 \\ 
   &  &  &  &  & st.dev. & 0.0034 & 0.0026 & 0.0028 & 0.0013 & 0.0012 & 0.0007 & 0.0013 & 0.0007 \\ 
   &  &  &  &  & t-test & 1.0000 & 0.0000 & 0.0000 & 1.0000 &  & 0.0000 & 0.0000 & 0.0000 \\ 
   &  &  &  &  & wilcox-test & 1.0000 & 0.0000 & 0.0000 & 1.0000 &  & 0.0000 & 0.0000 & 0.0000 \\ 
   \midrule
19 & \xmark & \cmark & \xmark & \xmark & mean & 0.0749 & 0.0610 & 0.0608 & 0.0585 & 0.0593 & 0.0707 & 0.0597 & 0.0725 \\ 
   &  &  &  &  & st.dev. & 0.0022 & 0.0030 & 0.0027 & 0.0016 & 0.0018 & 0.0010 & 0.0020 & 0.0010 \\ 
   &  &  &  &  & t-test & 0.0000 & 0.0000 & 0.0000 & 0.9996 &  & 0.0000 & 0.0947 & 0.0000 \\ 
   &  &  &  &  & wilcox-test & 0.0000 & 0.0000 & 0.0000 & 0.9995 &  & 0.0000 & 0.0966 & 0.0000 \\ 
   \midrule
20 & \xmark & \xmark & \cmark & \xmark & mean & 0.0059 & 0.1111 & 0.1071 & 0.0285 & 0.0273 & 0.0407 & 0.0292 & 0.0539 \\ 
   &  &  &  &  & st.dev. & 0.0016 & 0.0050 & 0.0061 & 0.0010 & 0.0009 & 0.0011 & 0.0010 & 0.0015 \\ 
   &  &  &  &  & t-test & 1.0000 & 0.0000 & 0.0000 & 0.0000 &  & 0.0000 & 0.0000 & 0.0000 \\ 
   &  &  &  &  & wilcox-test & 1.0000 & 0.0000 & 0.0000 & 0.0000 &  & 0.0000 & 0.0000 & 0.0000 \\ 
   \midrule
21 & \xmark & \xmark & \xmark & \cmark & mean & 0.0062 & 0.0670 & 0.0648 & 0.0544 & 0.0537 & 0.0816 & 0.0569 & 0.0853 \\ 
   &  &  &  &  & st.dev. & 0.0022 & 0.0044 & 0.0044 & 0.0013 & 0.0014 & 0.0009 & 0.0015 & 0.0009 \\ 
   &  &  &  &  & t-test & 1.0000 & 0.0000 & 0.0000 & 0.0000 &  & 0.0000 & 0.0000 & 0.0000 \\ 
   &  &  &  &  & wilcox-test & 1.0000 & 0.0000 & 0.0000 & 0.0000 &  & 0.0000 & 0.0000 & 0.0000 \\ 
   \midrule
22 & \cmark & \cmark & \xmark & \xmark & mean & 0.0853 & 0.0650 & 0.0651 & 0.0644 & 0.0664 & 0.0735 & 0.0675 & 0.0748 \\ 
   &  &  &  &  & st.dev. & 0.0049 & 0.0022 & 0.0022 & 0.0016 & 0.0014 & 0.0008 & 0.0014 & 0.0006 \\ 
   &  &  &  &  & t-test & 0.0000 & 1.0000 & 1.0000 & 1.0000 &  & 0.0000 & 0.0000 & 0.0000 \\ 
   &  &  &  &  & wilcox-test & 0.0000 & 1.0000 & 1.0000 & 1.0000 &  & 0.0000 & 0.0000 & 0.0000 \\ 
   \midrule
23 & \cmark & \xmark & \cmark & \xmark & mean & 0.0106 & 0.1177 & 0.1145 & 0.0313 & 0.0307 & 0.0462 & 0.0377 & 0.0640 \\ 
   &  &  &  &  & st.dev. & 0.0028 & 0.0038 & 0.0049 & 0.0010 & 0.0008 & 0.0014 & 0.0011 & 0.0018 \\ 
   &  &  &  &  & t-test & 1.0000 & 0.0000 & 0.0000 & 0.0000 &  & 0.0000 & 0.0000 & 0.0000 \\ 
   &  &  &  &  & wilcox-test & 1.0000 & 0.0000 & 0.0000 & 0.0000 &  & 0.0000 & 0.0000 & 0.0000 \\ 
   \midrule
24 & \cmark & \xmark & \xmark & \cmark & mean & 0.0148 & 0.0745 & 0.0722 & 0.0655 & 0.0677 & 0.0919 & 0.0718 & 0.0946 \\ 
   &  &  &  &  & st.dev. & 0.0040 & 0.0032 & 0.0029 & 0.0012 & 0.0013 & 0.0009 & 0.0014 & 0.0008 \\ 
   &  &  &  &  & t-test & 1.0000 & 0.0000 & 0.0000 & 1.0000 &  & 0.0000 & 0.0000 & 0.0000 \\ 
   &  &  &  &  & wilcox-test & 1.0000 & 0.0000 & 0.0000 & 1.0000 &  & 0.0000 & 0.0000 & 0.0000 \\ 
   \midrule
25 & \xmark & \cmark & \cmark & \xmark & mean & 0.0952 & 0.0995 & 0.0961 & 0.0439 & 0.0372 & 0.0630 & 0.0418 & 0.0747 \\ 
   &  &  &  &  & st.dev. & 0.0020 & 0.0041 & 0.0043 & 0.0016 & 0.0016 & 0.0017 & 0.0016 & 0.0020 \\ 
   &  &  &  &  & t-test & 0.0000 & 0.0000 & 0.0000 & 0.0000 &  & 0.0000 & 0.0000 & 0.0000 \\ 
   &  &  &  &  & wilcox-test & 0.0000 & 0.0000 & 0.0000 & 0.0000 &  & 0.0000 & 0.0000 & 0.0000 \\ 
   \midrule
26 & \xmark & \cmark & \xmark & \cmark & mean & 0.0733 & 0.0590 & 0.0594 & 0.0573 & 0.0582 & 0.0691 & 0.0586 & 0.0707 \\ 
   &  &  &  &  & st.dev. & 0.0024 & 0.0021 & 0.0020 & 0.0015 & 0.0015 & 0.0010 & 0.0015 & 0.0009 \\ 
   &  &  &  &  & t-test & 0.0000 & 0.0017 & 0.0000 & 1.0000 &  & 0.0000 & 0.0660 & 0.0000 \\ 
   &  &  &  &  & wilcox-test & 0.0000 & 0.0041 & 0.0000 & 1.0000 &  & 0.0000 & 0.0809 & 0.0000 \\ 
   \midrule
27 & \xmark & \xmark & \cmark & \cmark & mean & 0.0053 & 0.1069 & 0.1046 & 0.0278 & 0.0266 & 0.0401 & 0.0286 & 0.0533 \\ 
   &  &  &  &  & st.dev. & 0.0014 & 0.0048 & 0.0056 & 0.0010 & 0.0009 & 0.0011 & 0.0009 & 0.0015 \\ 
   &  &  &  &  & t-test & 1.0000 & 0.0000 & 0.0000 & 0.0000 &  & 0.0000 & 0.0000 & 0.0000 \\ 
   &  &  &  &  & wilcox-test & 1.0000 & 0.0000 & 0.0000 & 0.0000 &  & 0.0000 & 0.0000 & 0.0000 \\ 
   \midrule
28 & \cmark & \cmark & \cmark & \xmark & mean & 0.1090 & 0.1022 & 0.1001 & 0.0564 & 0.0447 & 0.0709 & 0.0527 & 0.0843 \\ 
   &  &  &  &  & st.dev. & 0.0041 & 0.0031 & 0.0030 & 0.0015 & 0.0018 & 0.0020 & 0.0018 & 0.0024 \\ 
   &  &  &  &  & t-test & 0.0000 & 0.0000 & 0.0000 & 0.0000 &  & 0.0000 & 0.0000 & 0.0000 \\ 
   &  &  &  &  & wilcox-test & 0.0000 & 0.0000 & 0.0000 & 0.0000 &  & 0.0000 & 0.0000 & 0.0000 \\ 
   \midrule
29 & \cmark & \cmark & \xmark & \cmark & mean & 0.0881 & 0.0666 & 0.0662 & 0.0658 & 0.0676 & 0.0751 & 0.0697 & 0.0764 \\ 
   &  &  &  &  & st.dev. & 0.0051 & 0.0024 & 0.0022 & 0.0016 & 0.0015 & 0.0008 & 0.0015 & 0.0006 \\ 
   &  &  &  &  & t-test & 0.0000 & 0.9997 & 1.0000 & 1.0000 &  & 0.0000 & 0.0000 & 0.0000 \\ 
   &  &  &  &  & wilcox-test & 0.0000 & 1.0000 & 1.0000 & 1.0000 &  & 0.0000 & 0.0000 & 0.0000 \\ 
   \midrule
30 & \cmark & \xmark & \cmark & \cmark & mean & 0.0118 & 0.1214 & 0.1161 & 0.0317 & 0.0309 & 0.0469 & 0.0378 & 0.0642 \\ 
   &  &  &  &  & st.dev. & 0.0032 & 0.0046 & 0.0055 & 0.0009 & 0.0008 & 0.0014 & 0.0012 & 0.0019 \\ 
   &  &  &  &  & t-test & 1.0000 & 0.0000 & 0.0000 & 0.0000 &  & 0.0000 & 0.0000 & 0.0000 \\ 
   &  &  &  &  & wilcox-test & 1.0000 & 0.0000 & 0.0000 & 0.0000 &  & 0.0000 & 0.0000 & 0.0000 \\ 
   \midrule
31 & \xmark & \cmark & \cmark & \cmark & mean & 0.0931 & 0.0956 & 0.0925 & 0.0434 & 0.0368 & 0.0619 & 0.0414 & 0.0731 \\ 
   &  &  &  &  & st.dev. & 0.0019 & 0.0044 & 0.0045 & 0.0015 & 0.0014 & 0.0016 & 0.0014 & 0.0020 \\ 
   &  &  &  &  & t-test & 0.0000 & 0.0000 & 0.0000 & 0.0000 &  & 0.0000 & 0.0000 & 0.0000 \\ 
   &  &  &  &  & wilcox-test & 0.0000 & 0.0000 & 0.0000 & 0.0000 &  & 0.0000 & 0.0000 & 0.0000 \\ 
   \midrule
32 & \cmark & \cmark & \cmark & \cmark & mean & 0.1122 & 0.1093 & 0.1058 & 0.0574 & 0.0452 & 0.0719 & 0.0536 & 0.0842 \\ 
   &  &  &  &  & st.dev. & 0.0040 & 0.0045 & 0.0044 & 0.0017 & 0.0020 & 0.0022 & 0.0021 & 0.0024 \\ 
   &  &  &  &  & t-test & 0.0000 & 0.0000 & 0.0000 & 0.0000 &  & 0.0000 & 0.0000 & 0.0000 \\ 
   &  &  &  &  & wilcox-test & 0.0000 & 0.0000 & 0.0000 & 0.0000 &  & 0.0000 & 0.0000 & 0.0000 \\ 
   \midrule
 \bottomrule
\end{tabular}
\begin{tablenotes}\footnotesize
\item [] \hspace{-0.17cm} \textit{Notes:} Table reports the average measures of the RPS based on 100 simulation replications for the sample size of 200 observations with 6 outcome classes. Columns 1 to 5 specify the DGP identifier and its features, namely 15 additional noise variables (\textit{noise}), nonlinear effects (\textit{nonlin}), multicollinearity among covariates (\textit{multi}), and randomly spaced thresholds (\textit{random}). The sixth column \textit{Statistic} shows the mean and the standard deviation of the accuracy measure for all methods. Additionally, \textit{t-test} and \textit{wilcox-test} contain the p-values of the parametric t-test as well as the nonparametric Wilcoxon test for the equality of means between the results of the \textit{Ordered Forest} and all the other methods.
\end{tablenotes}
\end{threeparttable}
}
\end{table}
\pagebreak
\subsubsection{ARPS: Low Dimension with 9 Classes}
\begin{table}[h!]
\footnotesize
\centering
\caption{Simulation Results: Accuracy Measure = ARPS \& Low Dimension with 9 Classes}\label{tab:allsim3} 
\resizebox{0.94\textwidth}{!}{
\begin{threeparttable}
\begin{tabular}{c|cccc|lrrrrrrrr}
\toprule
\midrule
\multicolumn{6}{c}{\normalsize{\textbf{Simulation Design}}} & \multicolumn{8}{c}{\normalsize{\textbf{Comparison of Methods}}}\\
\midrule
\midrule
 DGP & noise & nonlin & multi & random & Statistic & Ologit & Naive & Ordinal & Cond. & Ordered & Ordered* & Multi & Multi* \\ 
  \midrule
 \midrule
33 & \xmark & \xmark & \xmark & \xmark & mean & 0.0054 & 0.0653 & 0.0629 & 0.0528 & 0.0519 & 0.0789 & 0.0569 & 0.0850 \\ 
   &  &  &  &  & st.dev. & 0.0018 & 0.0042 & 0.0042 & 0.0012 & 0.0014 & 0.0009 & 0.0017 & 0.0009 \\ 
   &  &  &  &  & t-test & 1.0000 & 0.0000 & 0.0000 & 0.0000 &  & 0.0000 & 0.0000 & 0.0000 \\ 
   &  &  &  &  & wilcox-test & 1.0000 & 0.0000 & 0.0000 & 0.0000 &  & 0.0000 & 0.0000 & 0.0000 \\ 
   \midrule
34 & \cmark & \xmark & \xmark & \xmark & mean & 0.0112 & 0.0693 & 0.0672 & 0.0609 & 0.0638 & 0.0855 & 0.0704 & 0.0901 \\ 
   &  &  &  &  & st.dev. & 0.0028 & 0.0027 & 0.0027 & 0.0012 & 0.0012 & 0.0007 & 0.0013 & 0.0006 \\ 
   &  &  &  &  & t-test & 1.0000 & 0.0000 & 0.0000 & 1.0000 &  & 0.0000 & 0.0000 & 0.0000 \\ 
   &  &  &  &  & wilcox-test & 1.0000 & 0.0000 & 0.0000 & 1.0000 &  & 0.0000 & 0.0000 & 0.0000 \\ 
   \midrule
35 & \xmark & \cmark & \xmark & \xmark & mean & 0.0706 & 0.0573 & 0.0572 & 0.0555 & 0.0567 & 0.0669 & 0.0590 & 0.0698 \\ 
   &  &  &  &  & st.dev. & 0.0023 & 0.0026 & 0.0027 & 0.0014 & 0.0015 & 0.0009 & 0.0016 & 0.0007 \\ 
   &  &  &  &  & t-test & 0.0000 & 0.0220 & 0.0445 & 1.0000 &  & 0.0000 & 0.0000 & 0.0000 \\ 
   &  &  &  &  & wilcox-test & 0.0000 & 0.0788 & 0.2389 & 1.0000 &  & 0.0000 & 0.0000 & 0.0000 \\ 
   \midrule
36 & \xmark & \xmark & \cmark & \xmark & mean & 0.0052 & 0.1057 & 0.1047 & 0.0277 & 0.0263 & 0.0396 & 0.0303 & 0.0601 \\ 
   &  &  &  &  & st.dev. & 0.0014 & 0.0046 & 0.0056 & 0.0009 & 0.0009 & 0.0010 & 0.0010 & 0.0014 \\ 
   &  &  &  &  & t-test & 1.0000 & 0.0000 & 0.0000 & 0.0000 &  & 0.0000 & 0.0000 & 0.0000 \\ 
   &  &  &  &  & wilcox-test & 1.0000 & 0.0000 & 0.0000 & 0.0000 &  & 0.0000 & 0.0000 & 0.0000 \\ 
   \midrule
37 & \xmark & \xmark & \xmark & \cmark & mean & 0.0054 & 0.0627 & 0.0608 & 0.0518 & 0.0511 & 0.0774 & 0.0558 & 0.0835 \\ 
   &  &  &  &  & st.dev. & 0.0019 & 0.0036 & 0.0035 & 0.0012 & 0.0014 & 0.0009 & 0.0016 & 0.0010 \\ 
   &  &  &  &  & t-test & 1.0000 & 0.0000 & 0.0000 & 0.0001 &  & 0.0000 & 0.0000 & 0.0000 \\ 
   &  &  &  &  & wilcox-test & 1.0000 & 0.0000 & 0.0000 & 0.0002 &  & 0.0000 & 0.0000 & 0.0000 \\ 
   \midrule
38 & \cmark & \cmark & \xmark & \xmark & mean & 0.0806 & 0.0607 & 0.0608 & 0.0606 & 0.0629 & 0.0695 & 0.0661 & 0.0715 \\ 
   &  &  &  &  & st.dev. & 0.0036 & 0.0016 & 0.0018 & 0.0013 & 0.0012 & 0.0008 & 0.0014 & 0.0007 \\ 
   &  &  &  &  & t-test & 0.0000 & 1.0000 & 1.0000 & 1.0000 &  & 0.0000 & 0.0000 & 0.0000 \\ 
   &  &  &  &  & wilcox-test & 0.0000 & 1.0000 & 1.0000 & 1.0000 &  & 0.0000 & 0.0000 & 0.0000 \\ 
   \midrule
39 & \cmark & \xmark & \cmark & \xmark & mean & 0.0086 & 0.1122 & 0.1102 & 0.0301 & 0.0295 & 0.0443 & 0.0408 & 0.0710 \\ 
   &  &  &  &  & st.dev. & 0.0017 & 0.0036 & 0.0041 & 0.0009 & 0.0008 & 0.0012 & 0.0011 & 0.0017 \\ 
   &  &  &  &  & t-test & 1.0000 & 0.0000 & 0.0000 & 0.0000 &  & 0.0000 & 0.0000 & 0.0000 \\ 
   &  &  &  &  & wilcox-test & 1.0000 & 0.0000 & 0.0000 & 0.0000 &  & 0.0000 & 0.0000 & 0.0000 \\ 
   \midrule
40 & \cmark & \xmark & \xmark & \cmark & mean & 0.0106 & 0.0663 & 0.0646 & 0.0586 & 0.0615 & 0.0820 & 0.0679 & 0.0866 \\ 
   &  &  &  &  & st.dev. & 0.0028 & 0.0026 & 0.0026 & 0.0011 & 0.0012 & 0.0008 & 0.0012 & 0.0007 \\ 
   &  &  &  &  & t-test & 1.0000 & 0.0000 & 0.0000 & 1.0000 &  & 0.0000 & 0.0000 & 0.0000 \\ 
   &  &  &  &  & wilcox-test & 1.0000 & 0.0000 & 0.0000 & 1.0000 &  & 0.0000 & 0.0000 & 0.0000 \\ 
   \midrule
41 & \xmark & \cmark & \cmark & \xmark & mean & 0.0897 & 0.0929 & 0.0897 & 0.0417 & 0.0356 & 0.0596 & 0.0424 & 0.0776 \\ 
   &  &  &  &  & st.dev. & 0.0017 & 0.0037 & 0.0038 & 0.0014 & 0.0013 & 0.0015 & 0.0014 & 0.0018 \\ 
   &  &  &  &  & t-test & 0.0000 & 0.0000 & 0.0000 & 0.0000 &  & 0.0000 & 0.0000 & 0.0000 \\ 
   &  &  &  &  & wilcox-test & 0.0000 & 0.0000 & 0.0000 & 0.0000 &  & 0.0000 & 0.0000 & 0.0000 \\ 
   \midrule
42 & \xmark & \cmark & \xmark & \cmark & mean & 0.0701 & 0.0565 & 0.0564 & 0.0545 & 0.0556 & 0.0657 & 0.0579 & 0.0685 \\ 
   &  &  &  &  & st.dev. & 0.0025 & 0.0024 & 0.0024 & 0.0015 & 0.0014 & 0.0008 & 0.0016 & 0.0007 \\ 
   &  &  &  &  & t-test & 0.0000 & 0.0006 & 0.0010 & 1.0000 &  & 0.0000 & 0.0000 & 0.0000 \\ 
   &  &  &  &  & wilcox-test & 0.0000 & 0.0028 & 0.0066 & 1.0000 &  & 0.0000 & 0.0000 & 0.0000 \\ 
   \midrule
43 & \xmark & \xmark & \cmark & \cmark & mean & 0.0051 & 0.1034 & 0.1025 & 0.0273 & 0.0258 & 0.0394 & 0.0298 & 0.0593 \\ 
   &  &  &  &  & st.dev. & 0.0015 & 0.0040 & 0.0045 & 0.0008 & 0.0007 & 0.0010 & 0.0009 & 0.0014 \\ 
   &  &  &  &  & t-test & 1.0000 & 0.0000 & 0.0000 & 0.0000 &  & 0.0000 & 0.0000 & 0.0000 \\ 
   &  &  &  &  & wilcox-test & 1.0000 & 0.0000 & 0.0000 & 0.0000 &  & 0.0000 & 0.0000 & 0.0000 \\ 
   \midrule
44 & \cmark & \cmark & \cmark & \xmark & mean & 0.1018 & 0.0956 & 0.0933 & 0.0534 & 0.0432 & 0.0673 & 0.0550 & 0.0873 \\ 
   &  &  &  &  & st.dev. & 0.0035 & 0.0031 & 0.0031 & 0.0013 & 0.0016 & 0.0017 & 0.0019 & 0.0021 \\ 
   &  &  &  &  & t-test & 0.0000 & 0.0000 & 0.0000 & 0.0000 &  & 0.0000 & 0.0000 & 0.0000 \\ 
   &  &  &  &  & wilcox-test & 0.0000 & 0.0000 & 0.0000 & 0.0000 &  & 0.0000 & 0.0000 & 0.0000 \\ 
   \midrule
45 & \cmark & \cmark & \xmark & \cmark & mean & 0.0763 & 0.0587 & 0.0588 & 0.0582 & 0.0605 & 0.0664 & 0.0638 & 0.0684 \\ 
   &  &  &  &  & st.dev. & 0.0040 & 0.0019 & 0.0018 & 0.0014 & 0.0012 & 0.0007 & 0.0011 & 0.0006 \\ 
   &  &  &  &  & t-test & 0.0000 & 1.0000 & 1.0000 & 1.0000 &  & 0.0000 & 0.0000 & 0.0000 \\ 
   &  &  &  &  & wilcox-test & 0.0000 & 1.0000 & 1.0000 & 1.0000 &  & 0.0000 & 0.0000 & 0.0000 \\ 
   \midrule
46 & \cmark & \xmark & \cmark & \cmark & mean & 0.0084 & 0.1079 & 0.1066 & 0.0292 & 0.0286 & 0.0432 & 0.0391 & 0.0699 \\ 
   &  &  &  &  & st.dev. & 0.0021 & 0.0034 & 0.0040 & 0.0008 & 0.0007 & 0.0012 & 0.0012 & 0.0017 \\ 
   &  &  &  &  & t-test & 1.0000 & 0.0000 & 0.0000 & 0.0000 &  & 0.0000 & 0.0000 & 0.0000 \\ 
   &  &  &  &  & wilcox-test & 1.0000 & 0.0000 & 0.0000 & 0.0000 &  & 0.0000 & 0.0000 & 0.0000 \\ 
   \midrule
47 & \xmark & \cmark & \cmark & \cmark & mean & 0.0881 & 0.0915 & 0.0887 & 0.0411 & 0.0352 & 0.0588 & 0.0414 & 0.0765 \\ 
   &  &  &  &  & st.dev. & 0.0017 & 0.0039 & 0.0041 & 0.0014 & 0.0012 & 0.0014 & 0.0014 & 0.0016 \\ 
   &  &  &  &  & t-test & 0.0000 & 0.0000 & 0.0000 & 0.0000 &  & 0.0000 & 0.0000 & 0.0000 \\ 
   &  &  &  &  & wilcox-test & 0.0000 & 0.0000 & 0.0000 & 0.0000 &  & 0.0000 & 0.0000 & 0.0000 \\ 
   \midrule
48 & \cmark & \cmark & \cmark & \cmark & mean & 0.0973 & 0.0912 & 0.0887 & 0.0515 & 0.0421 & 0.0647 & 0.0537 & 0.0845 \\ 
   &  &  &  &  & st.dev. & 0.0031 & 0.0033 & 0.0032 & 0.0015 & 0.0016 & 0.0019 & 0.0018 & 0.0017 \\ 
   &  &  &  &  & t-test & 0.0000 & 0.0000 & 0.0000 & 0.0000 &  & 0.0000 & 0.0000 & 0.0000 \\ 
   &  &  &  &  & wilcox-test & 0.0000 & 0.0000 & 0.0000 & 0.0000 &  & 0.0000 & 0.0000 & 0.0000 \\ 
   \midrule
 \bottomrule
\end{tabular}
\begin{tablenotes}\footnotesize
\item [] \hspace{-0.17cm} \textit{Notes:} Table reports the average measures of the RPS based on 100 simulation replications for the sample size of 200 observations with 9 outcome classes. Columns 1 to 5 specify the DGP identifier and its features, namely 15 additional noise variables (\textit{noise}), nonlinear effects (\textit{nonlin}), multicollinearity among covariates (\textit{multi}), and randomly spaced thresholds (\textit{random}). The sixth column \textit{Statistic} shows the mean and the standard deviation of the accuracy measure for all methods. Additionally, \textit{t-test} and \textit{wilcox-test} contain the p-values of the parametric t-test as well as the nonparametric Wilcoxon test for the equality of means between the results of the \textit{Ordered Forest} and all the other methods.
\end{tablenotes}
\end{threeparttable}
}
\end{table}
\pagebreak

\subsubsection{ARPS: High Dimension with 3 Classes}
\begin{table}[h!]
\footnotesize
\centering
\caption{Simulation Results: Accuracy Measure = ARPS \& High Dimension with 3 Classes}\label{tab:allsim4}
\resizebox{0.94\textwidth}{!}{
\begin{threeparttable}
\begin{tabular}{c|cccc|lrrrrrrrr}
\toprule
\midrule
\multicolumn{6}{c}{\normalsize{\textbf{Simulation Design}}} & \multicolumn{7}{c}{\normalsize{\textbf{Comparison of Methods}}}\\
\midrule
\midrule
 DGP & noise & nonlin & multi & random & Statistic & Naive & Ordinal & Cond. & Ordered & Ordered* & Multi & Multi* \\ 
  \midrule
 \midrule
49 & \cmark & \xmark & \xmark & \xmark & mean & 0.1135 & 0.1139 & 0.1112 & 0.1140 & 0.1180 & 0.1139 & 0.1179 \\ 
   &  &  &  &  & st.dev. & 0.0009 & 0.0010 & 0.0009 & 0.0008 & 0.0006 & 0.0008 & 0.0006 \\ 
   &  &  &  &  & t-test & 1.0000 & 0.7676 & 1.0000 &  & 0.0000 & 0.7268 & 0.0000 \\ 
   &  &  &  &  & wilcox-test & 0.9999 & 0.8438 & 1.0000 &  & 0.0000 & 0.7191 & 0.0000 \\ 
   \midrule
50 & \cmark & \cmark & \xmark & \xmark & mean & 0.0896 & 0.0899 & 0.0901 & 0.0903 & 0.0907 & 0.0901 & 0.0907 \\ 
   &  &  &  &  & st.dev. & 0.0008 & 0.0010 & 0.0008 & 0.0007 & 0.0007 & 0.0007 & 0.0006 \\ 
   &  &  &  &  & t-test & 1.0000 & 0.9997 & 0.9840 &  & 0.0002 & 0.9973 & 0.0004 \\ 
   &  &  &  &  & wilcox-test & 1.0000 & 1.0000 & 0.9929 &  & 0.0000 & 0.9989 & 0.0000 \\ 
   \midrule
51 & \cmark & \xmark & \cmark & \xmark & mean & 0.1534 & 0.1529 & 0.0827 & 0.0766 & 0.1082 & 0.0867 & 0.1134 \\ 
   &  &  &  &  & st.dev. & 0.0011 & 0.0012 & 0.0024 & 0.0025 & 0.0029 & 0.0024 & 0.0026 \\ 
   &  &  &  &  & t-test & 0.0000 & 0.0000 & 0.0000 &  & 0.0000 & 0.0000 & 0.0000 \\ 
   &  &  &  &  & wilcox-test & 0.0000 & 0.0000 & 0.0000 &  & 0.0000 & 0.0000 & 0.0000 \\ 
   \midrule
52 & \cmark & \xmark & \xmark & \cmark & mean & 0.1253 & 0.1252 & 0.1224 & 0.1248 & 0.1296 & 0.1250 & 0.1296 \\ 
   &  &  &  &  & st.dev. & 0.0013 & 0.0013 & 0.0010 & 0.0009 & 0.0007 & 0.0009 & 0.0007 \\ 
   &  &  &  &  & t-test & 0.0011 & 0.0115 & 1.0000 &  & 0.0000 & 0.1664 & 0.0000 \\ 
   &  &  &  &  & wilcox-test & 0.0013 & 0.0140 & 1.0000 &  & 0.0000 & 0.1515 & 0.0000 \\ 
   \midrule
53 & \cmark & \cmark & \cmark & \xmark & mean & 0.1299 & 0.1300 & 0.1048 & 0.1016 & 0.1200 & 0.1021 & 0.1202 \\ 
   &  &  &  &  & st.dev. & 0.0011 & 0.0012 & 0.0034 & 0.0027 & 0.0026 & 0.0027 & 0.0025 \\ 
   &  &  &  &  & t-test & 0.0000 & 0.0000 & 0.0000 &  & 0.0000 & 0.0674 & 0.0000 \\ 
   &  &  &  &  & wilcox-test & 0.0000 & 0.0000 & 0.0000 &  & 0.0000 & 0.0494 & 0.0000 \\ 
   \midrule
54 & \cmark & \cmark & \xmark & \cmark & mean & 0.0997 & 0.0996 & 0.0999 & 0.0998 & 0.1004 & 0.0997 & 0.1004 \\ 
   &  &  &  &  & st.dev. & 0.0012 & 0.0013 & 0.0012 & 0.0012 & 0.0011 & 0.0011 & 0.0012 \\ 
   &  &  &  &  & t-test & 0.5772 & 0.8438 & 0.3065 &  & 0.0000 & 0.6432 & 0.0000 \\ 
   &  &  &  &  & wilcox-test & 0.6792 & 0.9705 & 0.2427 &  & 0.0000 & 0.7183 & 0.0000 \\ 
   \midrule
55 & \cmark & \xmark & \cmark & \cmark & mean & 0.1678 & 0.1667 & 0.0862 & 0.0836 & 0.1167 & 0.0906 & 0.1195 \\ 
   &  &  &  &  & st.dev. & 0.0015 & 0.0013 & 0.0026 & 0.0030 & 0.0029 & 0.0029 & 0.0029 \\ 
   &  &  &  &  & t-test & 0.0000 & 0.0000 & 0.0000 &  & 0.0000 & 0.0000 & 0.0000 \\ 
   &  &  &  &  & wilcox-test & 0.0000 & 0.0000 & 0.0000 &  & 0.0000 & 0.0000 & 0.0000 \\ 
   \midrule
56 & \cmark & \cmark & \cmark & \cmark & mean & 0.1476 & 0.1474 & 0.1156 & 0.1102 & 0.1316 & 0.1110 & 0.1317 \\ 
   &  &  &  &  & st.dev. & 0.0013 & 0.0010 & 0.0041 & 0.0029 & 0.0031 & 0.0029 & 0.0031 \\ 
   &  &  &  &  & t-test & 0.0000 & 0.0000 & 0.0000 &  & 0.0000 & 0.0287 & 0.0000 \\ 
   &  &  &  &  & wilcox-test & 0.0000 & 0.0000 & 0.0000 &  & 0.0000 & 0.0272 & 0.0000 \\ 
   \midrule
 \bottomrule
\end{tabular}
\begin{tablenotes}\footnotesize
\item [] \hspace{-0.17cm} \textit{Notes:} Table reports the average measures of the RPS based on 100 simulation replications for the sample size of 200 observations with 3 outcome classes. Columns 1 to 5 specify the DGP identifier and its features, namely 1000 additional noise variables (\textit{noise}), nonlinear effects (\textit{nonlin}), multicollinearity among covariates (\textit{multi}), and randomly spaced thresholds (\textit{random}). The sixth column \textit{Statistic} shows the mean and the standard deviation of the accuracy measure for all methods. Additionally, \textit{t-test} and \textit{wilcox-test} contain the p-values of the parametric t-test as well as the nonparametric Wilcoxon test for the equality of means between the results of the \textit{Ordered Forest} and all the other methods.
\end{tablenotes}
\end{threeparttable}
}
\end{table}
\pagebreak
\subsubsection{ARPS: High Dimension with 6 Classes}
\begin{table}[h!]
\footnotesize
\centering
\caption{Simulation Results: Accuracy Measure = ARPS \& High Dimension with 6 Classes}\label{tab:allsim5} 
\resizebox{0.94\textwidth}{!}{
\begin{threeparttable}
\begin{tabular}{c|cccc|lrrrrrrrr}
\toprule
\midrule
\multicolumn{6}{c}{\normalsize{\textbf{Simulation Design}}} & \multicolumn{7}{c}{\normalsize{\textbf{Comparison of Methods}}}\\
\midrule
\midrule
 DGP & noise & nonlin & multi & random & Statistic & Naive & Ordinal & Cond. & Ordered & Ordered* & Multi & Multi* \\ 
  \midrule
 \midrule
57 & \cmark & \xmark & \xmark & \xmark & mean & 0.0974 & 0.0972 & 0.0951 & 0.0983 & 0.1012 & 0.0998 & 0.1016 \\ 
   &  &  &  &  & st.dev. & 0.0006 & 0.0006 & 0.0006 & 0.0005 & 0.0004 & 0.0005 & 0.0004 \\ 
   &  &  &  &  & t-test & 1.0000 & 1.0000 & 1.0000 &  & 0.0000 & 0.0000 & 0.0000 \\ 
   &  &  &  &  & wilcox-test & 1.0000 & 1.0000 & 1.0000 &  & 0.0000 & 0.0000 & 0.0000 \\ 
   \midrule
58 & \cmark & \cmark & \xmark & \xmark & mean & 0.0762 & 0.0762 & 0.0765 & 0.0773 & 0.0772 & 0.0776 & 0.0773 \\ 
   &  &  &  &  & st.dev. & 0.0006 & 0.0006 & 0.0006 & 0.0005 & 0.0005 & 0.0004 & 0.0004 \\ 
   &  &  &  &  & t-test & 1.0000 & 1.0000 & 1.0000 &  & 0.9803 & 0.0000 & 0.7833 \\ 
   &  &  &  &  & wilcox-test & 1.0000 & 1.0000 & 1.0000 &  & 0.9838 & 0.0000 & 0.7449 \\ 
   \midrule
59 & \cmark & \xmark & \cmark & \xmark & mean & 0.1336 & 0.1327 & 0.0747 & 0.0675 & 0.0968 & 0.0912 & 0.1152 \\ 
   &  &  &  &  & st.dev. & 0.0008 & 0.0010 & 0.0013 & 0.0016 & 0.0015 & 0.0016 & 0.0017 \\ 
   &  &  &  &  & t-test & 0.0000 & 0.0000 & 0.0000 &  & 0.0000 & 0.0000 & 0.0000 \\ 
   &  &  &  &  & wilcox-test & 0.0000 & 0.0000 & 0.0000 &  & 0.0000 & 0.0000 & 0.0000 \\ 
   \midrule
60 & \cmark & \xmark & \xmark & \cmark & mean & 0.0845 & 0.0845 & 0.0826 & 0.0857 & 0.0880 & 0.0872 & 0.0883 \\ 
   &  &  &  &  & st.dev. & 0.0005 & 0.0005 & 0.0006 & 0.0004 & 0.0003 & 0.0004 & 0.0003 \\ 
   &  &  &  &  & t-test & 1.0000 & 1.0000 & 1.0000 &  & 0.0000 & 0.0000 & 0.0000 \\ 
   &  &  &  &  & wilcox-test & 1.0000 & 1.0000 & 1.0000 &  & 0.0000 & 0.0000 & 0.0000 \\ 
   \midrule
61 & \cmark & \cmark & \cmark & \xmark & mean & 0.1091 & 0.1088 & 0.0891 & 0.0885 & 0.1026 & 0.1010 & 0.1105 \\ 
   &  &  &  &  & st.dev. & 0.0009 & 0.0008 & 0.0025 & 0.0021 & 0.0018 & 0.0023 & 0.0010 \\ 
   &  &  &  &  & t-test & 0.0000 & 0.0000 & 0.0547 &  & 0.0000 & 0.0000 & 0.0000 \\ 
   &  &  &  &  & wilcox-test & 0.0000 & 0.0000 & 0.0626 &  & 0.0000 & 0.0000 & 0.0000 \\ 
   \midrule
62 & \cmark & \cmark & \xmark & \cmark & mean & 0.0658 & 0.0659 & 0.0660 & 0.0669 & 0.0665 & 0.0672 & 0.0666 \\ 
   &  &  &  &  & st.dev. & 0.0006 & 0.0006 & 0.0006 & 0.0006 & 0.0006 & 0.0005 & 0.0005 \\ 
   &  &  &  &  & t-test & 1.0000 & 1.0000 & 1.0000 &  & 1.0000 & 0.0006 & 0.9998 \\ 
   &  &  &  &  & wilcox-test & 1.0000 & 1.0000 & 1.0000 &  & 1.0000 & 0.0000 & 1.0000 \\ 
   \midrule
63 & \cmark & \xmark & \cmark & \cmark & mean & 0.1167 & 0.1163 & 0.0682 & 0.0606 & 0.0872 & 0.0820 & 0.1052 \\ 
   &  &  &  &  & st.dev. & 0.0007 & 0.0008 & 0.0014 & 0.0016 & 0.0015 & 0.0018 & 0.0015 \\ 
   &  &  &  &  & t-test & 0.0000 & 0.0000 & 0.0000 &  & 0.0000 & 0.0000 & 0.0000 \\ 
   &  &  &  &  & wilcox-test & 0.0000 & 0.0000 & 0.0000 &  & 0.0000 & 0.0000 & 0.0000 \\ 
   \midrule
64 & \cmark & \cmark & \cmark & \cmark & mean & 0.0927 & 0.0927 & 0.0766 & 0.0772 & 0.0882 & 0.0898 & 0.0952 \\ 
   &  &  &  &  & st.dev. & 0.0006 & 0.0005 & 0.0020 & 0.0016 & 0.0014 & 0.0018 & 0.0006 \\ 
   &  &  &  &  & t-test & 0.0000 & 0.0000 & 0.9878 &  & 0.0000 & 0.0000 & 0.0000 \\ 
   &  &  &  &  & wilcox-test & 0.0000 & 0.0000 & 0.9887 &  & 0.0000 & 0.0000 & 0.0000 \\ 
   \midrule
 \bottomrule
\end{tabular}
\begin{tablenotes}\footnotesize
\item [] \hspace{-0.17cm} \textit{Notes:} Table reports the average measures of the RPS based on 100 simulation replications for the sample size of 200 observations with 6 outcome classes. Columns 1 to 5 specify the DGP identifier and its features, namely 1000 additional noise variables (\textit{noise}), nonlinear effects (\textit{nonlin}), multicollinearity among covariates (\textit{multi}), and randomly spaced thresholds (\textit{random}). The sixth column \textit{Statistic} shows the mean and the standard deviation of the accuracy measure for all methods. Additionally, \textit{t-test} and \textit{wilcox-test} contain the p-values of the parametric t-test as well as the nonparametric Wilcoxon test for the equality of means between the results of the \textit{Ordered Forest} and all the other methods.
\end{tablenotes}
\end{threeparttable}
}
\end{table}
\pagebreak
\subsubsection{ARPS: High Dimension with 9 Classes}
\begin{table}[h!]
\footnotesize
\centering
\caption{Simulation Results: Accuracy Measure = ARPS \& High Dimension with 9 Classes}\label{tab:allsim6} 
\resizebox{0.94\textwidth}{!}{
\begin{threeparttable}
\begin{tabular}{c|cccc|lrrrrrrrr}
\toprule
\midrule
\multicolumn{6}{c}{\normalsize{\textbf{Simulation Design}}} & \multicolumn{7}{c}{\normalsize{\textbf{Comparison of Methods}}}\\
\midrule
\midrule
 DGP & noise & nonlin & multi & random & Statistic & Naive & Ordinal & Cond. & Ordered & Ordered* & Multi & Multi* \\ 
  \midrule
 \midrule
65 & \cmark & \xmark & \xmark & \xmark & mean & 0.0921 & 0.0918 & 0.0900 & 0.0931 & 0.0959 & 0.0955 & 0.0964 \\ 
   &  &  &  &  & st.dev. & 0.0006 & 0.0006 & 0.0006 & 0.0005 & 0.0003 & 0.0004 & 0.0003 \\ 
   &  &  &  &  & t-test & 1.0000 & 1.0000 & 1.0000 &  & 0.0000 & 0.0000 & 0.0000 \\ 
   &  &  &  &  & wilcox-test & 1.0000 & 1.0000 & 1.0000 &  & 0.0000 & 0.0000 & 0.0000 \\ 
   \midrule
66 & \cmark & \cmark & \xmark & \xmark & mean & 0.0721 & 0.0720 & 0.0724 & 0.0732 & 0.0730 & 0.0739 & 0.0731 \\ 
   &  &  &  &  & st.dev. & 0.0006 & 0.0005 & 0.0006 & 0.0005 & 0.0004 & 0.0004 & 0.0004 \\ 
   &  &  &  &  & t-test & 1.0000 & 1.0000 & 1.0000 &  & 0.9959 & 0.0000 & 0.8717 \\ 
   &  &  &  &  & wilcox-test & 1.0000 & 1.0000 & 1.0000 &  & 0.9991 & 0.0000 & 0.9308 \\ 
   \midrule
67 & \cmark & \xmark & \cmark & \xmark & mean & 0.1268 & 0.1260 & 0.0713 & 0.0648 & 0.0926 & 0.0979 & 0.1175 \\ 
   &  &  &  &  & st.dev. & 0.0008 & 0.0009 & 0.0013 & 0.0013 & 0.0014 & 0.0017 & 0.0015 \\ 
   &  &  &  &  & t-test & 0.0000 & 0.0000 & 0.0000 &  & 0.0000 & 0.0000 & 0.0000 \\ 
   &  &  &  &  & wilcox-test & 0.0000 & 0.0000 & 0.0000 &  & 0.0000 & 0.0000 & 0.0000 \\ 
   \midrule
68 & \cmark & \xmark & \xmark & \cmark & mean & 0.0904 & 0.0902 & 0.0884 & 0.0915 & 0.0941 & 0.0937 & 0.0946 \\ 
   &  &  &  &  & st.dev. & 0.0006 & 0.0006 & 0.0005 & 0.0005 & 0.0003 & 0.0004 & 0.0003 \\ 
   &  &  &  &  & t-test & 1.0000 & 1.0000 & 1.0000 &  & 0.0000 & 0.0000 & 0.0000 \\ 
   &  &  &  &  & wilcox-test & 1.0000 & 1.0000 & 1.0000 &  & 0.0000 & 0.0000 & 0.0000 \\ 
   \midrule
69 & \cmark & \cmark & \cmark & \xmark & mean & 0.1031 & 0.1028 & 0.0838 & 0.0838 & 0.0967 & 0.1024 & 0.1061 \\ 
   &  &  &  &  & st.dev. & 0.0007 & 0.0007 & 0.0021 & 0.0017 & 0.0016 & 0.0016 & 0.0005 \\ 
   &  &  &  &  & t-test & 0.0000 & 0.0000 & 0.4695 &  & 0.0000 & 0.0000 & 0.0000 \\ 
   &  &  &  &  & wilcox-test & 0.0000 & 0.0000 & 0.5044 &  & 0.0000 & 0.0000 & 0.0000 \\ 
   \midrule
70 & \cmark & \cmark & \xmark & \cmark & mean & 0.0706 & 0.0707 & 0.0710 & 0.0718 & 0.0716 & 0.0724 & 0.0717 \\ 
   &  &  &  &  & st.dev. & 0.0007 & 0.0007 & 0.0006 & 0.0006 & 0.0005 & 0.0005 & 0.0006 \\ 
   &  &  &  &  & t-test & 1.0000 & 1.0000 & 1.0000 &  & 0.9903 & 0.0000 & 0.8186 \\ 
   &  &  &  &  & wilcox-test & 1.0000 & 1.0000 & 1.0000 &  & 0.9983 & 0.0000 & 0.8723 \\ 
   \midrule
71 & \cmark & \xmark & \cmark & \cmark & mean & 0.1246 & 0.1238 & 0.0704 & 0.0636 & 0.0911 & 0.0966 & 0.1153 \\ 
   &  &  &  &  & st.dev. & 0.0007 & 0.0008 & 0.0014 & 0.0013 & 0.0014 & 0.0016 & 0.0018 \\ 
   &  &  &  &  & t-test & 0.0000 & 0.0000 & 0.0000 &  & 0.0000 & 0.0000 & 0.0000 \\ 
   &  &  &  &  & wilcox-test & 0.0000 & 0.0000 & 0.0000 &  & 0.0000 & 0.0000 & 0.0000 \\ 
   \midrule
72 & \cmark & \cmark & \cmark & \cmark & mean & 0.1007 & 0.1004 & 0.0817 & 0.0819 & 0.0945 & 0.0997 & 0.1036 \\ 
   &  &  &  &  & st.dev. & 0.0007 & 0.0007 & 0.0020 & 0.0017 & 0.0015 & 0.0019 & 0.0006 \\ 
   &  &  &  &  & t-test & 0.0000 & 0.0000 & 0.7875 &  & 0.0000 & 0.0000 & 0.0000 \\ 
   &  &  &  &  & wilcox-test & 0.0000 & 0.0000 & 0.8473 &  & 0.0000 & 0.0000 & 0.0000 \\ 
   \midrule
 \bottomrule
\end{tabular}
\begin{tablenotes}\footnotesize
\item [] \hspace{-0.17cm} \textit{Notes:} Table reports the average measures of the RPS based on 100 simulation replications for the sample size of 200 observations with 9 outcome classes. Columns 1 to 5 specify the DGP identifier and its features, namely 1000 additional noise variables (\textit{noise}), nonlinear effects (\textit{nonlin}), multicollinearity among covariates (\textit{multi}), and randomly spaced thresholds (\textit{random}). The sixth column \textit{Statistic} shows the mean and the standard deviation of the accuracy measure for all methods. Additionally, \textit{t-test} and \textit{wilcox-test} contain the p-values of the parametric t-test as well as the nonparametric Wilcoxon test for the equality of means between the results of the \textit{Ordered Forest} and all the other methods.
\end{tablenotes}
\end{threeparttable}
}
\end{table}

\pagebreak

\subsubsection{AMSE: Low Dimension with 3 Classes}
\begin{table}[h!]
\footnotesize
\centering
\caption{Simulation Results: Accuracy Measure = AMSE \& Low Dimension with 3 Classes}\label{tab:allsim7} 
\resizebox{0.94\textwidth}{!}{
\begin{threeparttable}
\begin{tabular}{c|cccc|lrrrrrrrr}
\toprule
\midrule
\multicolumn{6}{c}{\normalsize{\textbf{Simulation Design}}} & \multicolumn{8}{c}{\normalsize{\textbf{Comparison of Methods}}}\\
\midrule
\midrule
 DGP & noise & nonlin & multi & random & Statistic & Ologit & Naive & Ordinal & Cond. & Ordered & Ordered* & Multi & Multi* \\ 
  \midrule
 \midrule
1 & \xmark & \xmark & \xmark & \xmark & mean & 0.0103 & 0.0669 & 0.0682 & 0.0565 & 0.0587 & 0.0800 & 0.0587 & 0.0800 \\ 
   &  &  &  &  & st.dev. & 0.0044 & 0.0041 & 0.0044 & 0.0015 & 0.0022 & 0.0009 & 0.0016 & 0.0010 \\ 
   &  &  &  &  & t-test & 1.0000 & 0.0000 & 0.0000 & 1.0000 &  & 0.0000 & 0.3900 & 0.0000 \\ 
   &  &  &  &  & wilcox-test & 1.0000 & 0.0000 & 0.0000 & 1.0000 &  & 0.0000 & 0.2614 & 0.0000 \\ 
   \midrule
2 & \cmark & \xmark & \xmark & \xmark & mean & 0.0227 & 0.0723 & 0.0727 & 0.0648 & 0.0682 & 0.0859 & 0.0684 & 0.0859 \\ 
   &  &  &  &  & st.dev. & 0.0056 & 0.0034 & 0.0038 & 0.0013 & 0.0015 & 0.0010 & 0.0014 & 0.0010 \\ 
   &  &  &  &  & t-test & 1.0000 & 0.0000 & 0.0000 & 1.0000 &  & 0.0000 & 0.1609 & 0.0000 \\ 
   &  &  &  &  & wilcox-test & 1.0000 & 0.0000 & 0.0000 & 1.0000 &  & 0.0000 & 0.1287 & 0.0000 \\ 
   \midrule
3 & \xmark & \cmark & \xmark & \xmark & mean & 0.0700 & 0.0576 & 0.0609 & 0.0552 & 0.0586 & 0.0644 & 0.0565 & 0.0644 \\ 
   &  &  &  &  & st.dev. & 0.0032 & 0.0024 & 0.0032 & 0.0016 & 0.0021 & 0.0011 & 0.0016 & 0.0011 \\ 
   &  &  &  &  & t-test & 0.0000 & 0.9980 & 0.0000 & 1.0000 &  & 0.0000 & 1.0000 & 0.0000 \\ 
   &  &  &  &  & wilcox-test & 0.0000 & 0.9954 & 0.0000 & 1.0000 &  & 0.0000 & 1.0000 & 0.0000 \\ 
   \midrule
4 & \xmark & \xmark & \cmark & \xmark & mean & 0.0124 & 0.1217 & 0.1166 & 0.0378 & 0.0370 & 0.0500 & 0.0367 & 0.0554 \\ 
   &  &  &  &  & st.dev. & 0.0040 & 0.0068 & 0.0068 & 0.0017 & 0.0018 & 0.0014 & 0.0017 & 0.0016 \\ 
   &  &  &  &  & t-test & 1.0000 & 0.0000 & 0.0000 & 0.0005 &  & 0.0000 & 0.8458 & 0.0000 \\ 
   &  &  &  &  & wilcox-test & 1.0000 & 0.0000 & 0.0000 & 0.0003 &  & 0.0000 & 0.8530 & 0.0000 \\ 
   \midrule
5 & \xmark & \xmark & \xmark & \cmark & mean & 0.0096 & 0.0594 & 0.0567 & 0.0495 & 0.0511 & 0.0726 & 0.0517 & 0.0732 \\ 
   &  &  &  &  & st.dev. & 0.0032 & 0.0057 & 0.0047 & 0.0015 & 0.0018 & 0.0011 & 0.0017 & 0.0011 \\ 
   &  &  &  &  & t-test & 1.0000 & 0.0000 & 0.0000 & 1.0000 &  & 0.0000 & 0.0044 & 0.0000 \\ 
   &  &  &  &  & wilcox-test & 1.0000 & 0.0000 & 0.0000 & 1.0000 &  & 0.0000 & 0.0024 & 0.0000 \\ 
   \midrule
6 & \cmark & \cmark & \xmark & \xmark & mean & 0.0809 & 0.0612 & 0.0636 & 0.0604 & 0.0638 & 0.0671 & 0.0617 & 0.0670 \\ 
   &  &  &  &  & st.dev. & 0.0048 & 0.0019 & 0.0030 & 0.0016 & 0.0017 & 0.0010 & 0.0015 & 0.0010 \\ 
   &  &  &  &  & t-test & 0.0000 & 1.0000 & 0.7436 & 1.0000 &  & 0.0000 & 1.0000 & 0.0000 \\ 
   &  &  &  &  & wilcox-test & 0.0000 & 1.0000 & 0.9265 & 1.0000 &  & 0.0000 & 1.0000 & 0.0000 \\ 
   \midrule
7 & \cmark & \xmark & \cmark & \xmark & mean & 0.0283 & 0.1297 & 0.1262 & 0.0411 & 0.0407 & 0.0548 & 0.0427 & 0.0634 \\ 
   &  &  &  &  & st.dev. & 0.0083 & 0.0052 & 0.0049 & 0.0015 & 0.0015 & 0.0020 & 0.0017 & 0.0022 \\ 
   &  &  &  &  & t-test & 1.0000 & 0.0000 & 0.0000 & 0.0734 &  & 0.0000 & 0.0000 & 0.0000 \\ 
   &  &  &  &  & wilcox-test & 1.0000 & 0.0000 & 0.0000 & 0.0494 &  & 0.0000 & 0.0000 & 0.0000 \\ 
   \midrule
8 & \cmark & \xmark & \xmark & \cmark & mean & 0.0230 & 0.0722 & 0.0746 & 0.0660 & 0.0705 & 0.0855 & 0.0701 & 0.0857 \\ 
   &  &  &  &  & st.dev. & 0.0065 & 0.0028 & 0.0038 & 0.0014 & 0.0018 & 0.0008 & 0.0015 & 0.0008 \\ 
   &  &  &  &  & t-test & 1.0000 & 0.0000 & 0.0000 & 1.0000 &  & 0.0000 & 0.9630 & 0.0000 \\ 
   &  &  &  &  & wilcox-test & 1.0000 & 0.0000 & 0.0000 & 1.0000 &  & 0.0000 & 0.9578 & 0.0000 \\ 
   \midrule
9 & \xmark & \cmark & \cmark & \xmark & mean & 0.0968 & 0.1066 & 0.1012 & 0.0493 & 0.0443 & 0.0660 & 0.0465 & 0.0680 \\ 
   &  &  &  &  & st.dev. & 0.0024 & 0.0070 & 0.0060 & 0.0018 & 0.0021 & 0.0018 & 0.0019 & 0.0016 \\ 
   &  &  &  &  & t-test & 0.0000 & 0.0000 & 0.0000 & 0.0000 &  & 0.0000 & 0.0000 & 0.0000 \\ 
   &  &  &  &  & wilcox-test & 0.0000 & 0.0000 & 0.0000 & 0.0000 &  & 0.0000 & 0.0000 & 0.0000 \\ 
   \midrule
10 & \xmark & \cmark & \xmark & \cmark & mean & 0.0667 & 0.0538 & 0.0533 & 0.0507 & 0.0530 & 0.0599 & 0.0529 & 0.0600 \\ 
   &  &  &  &  & st.dev. & 0.0034 & 0.0033 & 0.0031 & 0.0017 & 0.0019 & 0.0010 & 0.0018 & 0.0010 \\ 
   &  &  &  &  & t-test & 0.0000 & 0.0119 & 0.1801 & 1.0000 &  & 0.0000 & 0.6401 & 0.0000 \\ 
   &  &  &  &  & wilcox-test & 0.0000 & 0.0332 & 0.2314 & 1.0000 &  & 0.0000 & 0.7041 & 0.0000 \\ 
   \midrule
11 & \xmark & \xmark & \cmark & \cmark & mean & 0.0132 & 0.1201 & 0.1172 & 0.0328 & 0.0326 & 0.0427 & 0.0327 & 0.0472 \\ 
   &  &  &  &  & st.dev. & 0.0050 & 0.0111 & 0.0113 & 0.0017 & 0.0018 & 0.0013 & 0.0018 & 0.0016 \\ 
   &  &  &  &  & t-test & 1.0000 & 0.0000 & 0.0000 & 0.1763 &  & 0.0000 & 0.3026 & 0.0000 \\ 
   &  &  &  &  & wilcox-test & 1.0000 & 0.0000 & 0.0000 & 0.1287 &  & 0.0000 & 0.3376 & 0.0000 \\ 
   \midrule
12 & \cmark & \cmark & \cmark & \xmark & mean & 0.1104 & 0.1064 & 0.1027 & 0.0616 & 0.0506 & 0.0737 & 0.0540 & 0.0751 \\ 
   &  &  &  &  & st.dev. & 0.0039 & 0.0051 & 0.0043 & 0.0020 & 0.0024 & 0.0022 & 0.0021 & 0.0019 \\ 
   &  &  &  &  & t-test & 0.0000 & 0.0000 & 0.0000 & 0.0000 &  & 0.0000 & 0.0000 & 0.0000 \\ 
   &  &  &  &  & wilcox-test & 0.0000 & 0.0000 & 0.0000 & 0.0000 &  & 0.0000 & 0.0000 & 0.0000 \\ 
   \midrule
13 & \cmark & \cmark & \xmark & \cmark & mean & 0.0765 & 0.0595 & 0.0630 & 0.0587 & 0.0632 & 0.0646 & 0.0605 & 0.0646 \\ 
   &  &  &  &  & st.dev. & 0.0050 & 0.0016 & 0.0029 & 0.0016 & 0.0019 & 0.0011 & 0.0016 & 0.0011 \\ 
   &  &  &  &  & t-test & 0.0000 & 1.0000 & 0.7290 & 1.0000 &  & 0.0000 & 1.0000 & 0.0000 \\ 
   &  &  &  &  & wilcox-test & 0.0000 & 1.0000 & 0.8231 & 1.0000 &  & 0.0000 & 1.0000 & 0.0000 \\ 
   \midrule
14 & \cmark & \xmark & \cmark & \cmark & mean & 0.0311 & 0.1273 & 0.1244 & 0.0413 & 0.0408 & 0.0553 & 0.0420 & 0.0626 \\ 
   &  &  &  &  & st.dev. & 0.0090 & 0.0044 & 0.0043 & 0.0019 & 0.0019 & 0.0021 & 0.0018 & 0.0023 \\ 
   &  &  &  &  & t-test & 1.0000 & 0.0000 & 0.0000 & 0.0584 &  & 0.0000 & 0.0000 & 0.0000 \\ 
   &  &  &  &  & wilcox-test & 1.0000 & 0.0000 & 0.0000 & 0.0428 &  & 0.0000 & 0.0000 & 0.0000 \\ 
   \midrule
15 & \xmark & \cmark & \cmark & \cmark & mean & 0.0878 & 0.1016 & 0.0962 & 0.0420 & 0.0374 & 0.0566 & 0.0387 & 0.0593 \\ 
   &  &  &  &  & st.dev. & 0.0031 & 0.0081 & 0.0072 & 0.0019 & 0.0022 & 0.0018 & 0.0020 & 0.0018 \\ 
   &  &  &  &  & t-test & 0.0000 & 0.0000 & 0.0000 & 0.0000 &  & 0.0000 & 0.0000 & 0.0000 \\ 
   &  &  &  &  & wilcox-test & 0.0000 & 0.0000 & 0.0000 & 0.0000 &  & 0.0000 & 0.0000 & 0.0000 \\ 
   \midrule
16 & \cmark & \cmark & \cmark & \cmark & mean & 0.1081 & 0.0985 & 0.0965 & 0.0637 & 0.0543 & 0.0752 & 0.0572 & 0.0768 \\ 
   &  &  &  &  & st.dev. & 0.0039 & 0.0034 & 0.0029 & 0.0020 & 0.0026 & 0.0021 & 0.0022 & 0.0019 \\ 
   &  &  &  &  & t-test & 0.0000 & 0.0000 & 0.0000 & 0.0000 &  & 0.0000 & 0.0000 & 0.0000 \\ 
   &  &  &  &  & wilcox-test & 0.0000 & 0.0000 & 0.0000 & 0.0000 &  & 0.0000 & 0.0000 & 0.0000 \\ 
   \midrule
 \bottomrule
\end{tabular}
\begin{tablenotes}\footnotesize
\item [] \hspace{-0.17cm} \textit{Notes:} Table reports the average measures of the MSE based on 100 simulation replications for the sample size of 200 observations with 3 outcome classes. Columns 1 to 5 specify the DGP identifier and its features, namely 15 additional noise variables (\textit{noise}), nonlinear effects (\textit{nonlin}), multicollinearity among covariates (\textit{multi}), and randomly spaced thresholds (\textit{random}). The sixth column \textit{Statistic} shows the mean and the standard deviation of the accuracy measure for all methods. Additionally, \textit{t-test} and \textit{wilcox-test} contain the p-values of the parametric t-test as well as the nonparametric Wilcoxon test for the equality of means between the results of the \textit{Ordered Forest} and all the other methods.
\end{tablenotes}
\end{threeparttable}
}
\end{table}
\pagebreak
\subsubsection{AMSE: Low Dimension with 6 Classes}
\begin{table}[h!]
\footnotesize
\centering
\caption{Simulation Results: Accuracy Measure = AMSE \& Low Dimension with 6 Classes}\label{tab:allsim8} 
\resizebox{0.94\textwidth}{!}{
\begin{threeparttable}
\begin{tabular}{c|cccc|lrrrrrrrr}
\toprule
\midrule
\multicolumn{6}{c}{\normalsize{\textbf{Simulation Design}}} & \multicolumn{8}{c}{\normalsize{\textbf{Comparison of Methods}}}\\
\midrule
\midrule
 DGP & noise & nonlin & multi & random & Statistic & Ologit & Naive & Ordinal & Cond. & Ordered & Ordered* & Multi & Multi* \\ 
  \midrule
 \midrule
17 & \xmark & \xmark & \xmark & \xmark & mean & 0.0043 & 0.0284 & 0.0283 & 0.0248 & 0.0291 & 0.0324 & 0.0287 & 0.0327 \\ 
   &  &  &  &  & st.dev. & 0.0014 & 0.0012 & 0.0018 & 0.0007 & 0.0010 & 0.0005 & 0.0008 & 0.0005 \\ 
   &  &  &  &  & t-test & 1.0000 & 1.0000 & 0.9998 & 1.0000 &  & 0.0000 & 0.9958 & 0.0000 \\ 
   &  &  &  &  & wilcox-test & 1.0000 & 1.0000 & 1.0000 & 1.0000 &  & 0.0000 & 0.9953 & 0.0000 \\ 
   \midrule
18 & \cmark & \xmark & \xmark & \xmark & mean & 0.0083 & 0.0294 & 0.0292 & 0.0272 & 0.0311 & 0.0337 & 0.0310 & 0.0341 \\ 
   &  &  &  &  & st.dev. & 0.0021 & 0.0007 & 0.0010 & 0.0005 & 0.0006 & 0.0003 & 0.0005 & 0.0003 \\ 
   &  &  &  &  & t-test & 1.0000 & 1.0000 & 1.0000 & 1.0000 &  & 0.0000 & 0.9791 & 0.0000 \\ 
   &  &  &  &  & wilcox-test & 1.0000 & 1.0000 & 1.0000 & 1.0000 &  & 0.0000 & 0.9805 & 0.0000 \\ 
   \midrule
19 & \xmark & \cmark & \xmark & \xmark & mean & 0.0245 & 0.0216 & 0.0222 & 0.0207 & 0.0257 & 0.0237 & 0.0240 & 0.0237 \\ 
   &  &  &  &  & st.dev. & 0.0007 & 0.0009 & 0.0009 & 0.0006 & 0.0009 & 0.0004 & 0.0008 & 0.0004 \\ 
   &  &  &  &  & t-test & 1.0000 & 1.0000 & 1.0000 & 1.0000 &  & 1.0000 & 1.0000 & 1.0000 \\ 
   &  &  &  &  & wilcox-test & 1.0000 & 1.0000 & 1.0000 & 1.0000 &  & 1.0000 & 1.0000 & 1.0000 \\ 
   \midrule
20 & \xmark & \xmark & \cmark & \xmark & mean & 0.0065 & 0.0600 & 0.0568 & 0.0238 & 0.0259 & 0.0299 & 0.0263 & 0.0356 \\ 
   &  &  &  &  & st.dev. & 0.0017 & 0.0019 & 0.0023 & 0.0008 & 0.0009 & 0.0007 & 0.0008 & 0.0007 \\ 
   &  &  &  &  & t-test & 1.0000 & 0.0000 & 0.0000 & 1.0000 &  & 0.0000 & 0.0006 & 0.0000 \\ 
   &  &  &  &  & wilcox-test & 1.0000 & 0.0000 & 0.0000 & 1.0000 &  & 0.0000 & 0.0002 & 0.0000 \\ 
   \midrule
21 & \xmark & \xmark & \xmark & \cmark & mean & 0.0043 & 0.0283 & 0.0282 & 0.0248 & 0.0291 & 0.0324 & 0.0289 & 0.0327 \\ 
   &  &  &  &  & st.dev. & 0.0014 & 0.0013 & 0.0015 & 0.0007 & 0.0009 & 0.0005 & 0.0007 & 0.0005 \\ 
   &  &  &  &  & t-test & 1.0000 & 1.0000 & 1.0000 & 1.0000 &  & 0.0000 & 0.9689 & 0.0000 \\ 
   &  &  &  &  & wilcox-test & 1.0000 & 1.0000 & 1.0000 & 1.0000 &  & 0.0000 & 0.9661 & 0.0000 \\ 
   \midrule
22 & \cmark & \cmark & \xmark & \xmark & mean & 0.0273 & 0.0223 & 0.0228 & 0.0220 & 0.0263 & 0.0242 & 0.0249 & 0.0243 \\ 
   &  &  &  &  & st.dev. & 0.0015 & 0.0007 & 0.0008 & 0.0005 & 0.0007 & 0.0004 & 0.0006 & 0.0003 \\ 
   &  &  &  &  & t-test & 0.0000 & 1.0000 & 1.0000 & 1.0000 &  & 1.0000 & 1.0000 & 1.0000 \\ 
   &  &  &  &  & wilcox-test & 0.0000 & 1.0000 & 1.0000 & 1.0000 &  & 1.0000 & 1.0000 & 1.0000 \\ 
   \midrule
23 & \cmark & \xmark & \cmark & \xmark & mean & 0.0114 & 0.0607 & 0.0580 & 0.0258 & 0.0266 & 0.0319 & 0.0305 & 0.0396 \\ 
   &  &  &  &  & st.dev. & 0.0030 & 0.0014 & 0.0017 & 0.0007 & 0.0007 & 0.0008 & 0.0008 & 0.0007 \\ 
   &  &  &  &  & t-test & 1.0000 & 0.0000 & 0.0000 & 1.0000 &  & 0.0000 & 0.0000 & 0.0000 \\ 
   &  &  &  &  & wilcox-test & 1.0000 & 0.0000 & 0.0000 & 1.0000 &  & 0.0000 & 0.0000 & 0.0000 \\ 
   \midrule
24 & \cmark & \xmark & \xmark & \cmark & mean & 0.0088 & 0.0306 & 0.0296 & 0.0274 & 0.0306 & 0.0346 & 0.0310 & 0.0350 \\ 
   &  &  &  &  & st.dev. & 0.0023 & 0.0010 & 0.0011 & 0.0005 & 0.0006 & 0.0004 & 0.0006 & 0.0004 \\ 
   &  &  &  &  & t-test & 1.0000 & 0.6721 & 1.0000 & 1.0000 &  & 0.0000 & 0.0000 & 0.0000 \\ 
   &  &  &  &  & wilcox-test & 1.0000 & 0.3992 & 1.0000 & 1.0000 &  & 0.0000 & 0.0000 & 0.0000 \\ 
   \midrule
25 & \xmark & \cmark & \cmark & \xmark & mean & 0.0374 & 0.0396 & 0.0377 & 0.0234 & 0.0256 & 0.0292 & 0.0254 & 0.0315 \\ 
   &  &  &  &  & st.dev. & 0.0007 & 0.0016 & 0.0013 & 0.0008 & 0.0011 & 0.0008 & 0.0009 & 0.0007 \\ 
   &  &  &  &  & t-test & 0.0000 & 0.0000 & 0.0000 & 1.0000 &  & 0.0000 & 0.9637 & 0.0000 \\ 
   &  &  &  &  & wilcox-test & 0.0000 & 0.0000 & 0.0000 & 1.0000 &  & 0.0000 & 0.9567 & 0.0000 \\ 
   \midrule
26 & \xmark & \cmark & \xmark & \cmark & mean & 0.0245 & 0.0215 & 0.0224 & 0.0207 & 0.0256 & 0.0236 & 0.0241 & 0.0237 \\ 
   &  &  &  &  & st.dev. & 0.0008 & 0.0007 & 0.0009 & 0.0006 & 0.0008 & 0.0004 & 0.0007 & 0.0005 \\ 
   &  &  &  &  & t-test & 1.0000 & 1.0000 & 1.0000 & 1.0000 &  & 1.0000 & 1.0000 & 1.0000 \\ 
   &  &  &  &  & wilcox-test & 1.0000 & 1.0000 & 1.0000 & 1.0000 &  & 1.0000 & 1.0000 & 1.0000 \\ 
   \midrule
27 & \xmark & \xmark & \cmark & \cmark & mean & 0.0060 & 0.0587 & 0.0560 & 0.0236 & 0.0254 & 0.0297 & 0.0262 & 0.0355 \\ 
   &  &  &  &  & st.dev. & 0.0015 & 0.0017 & 0.0019 & 0.0007 & 0.0009 & 0.0007 & 0.0009 & 0.0007 \\ 
   &  &  &  &  & t-test & 1.0000 & 0.0000 & 0.0000 & 1.0000 &  & 0.0000 & 0.0000 & 0.0000 \\ 
   &  &  &  &  & wilcox-test & 1.0000 & 0.0000 & 0.0000 & 1.0000 &  & 0.0000 & 0.0000 & 0.0000 \\ 
   \midrule
28 & \cmark & \cmark & \cmark & \xmark & mean & 0.0416 & 0.0396 & 0.0384 & 0.0271 & 0.0272 & 0.0312 & 0.0280 & 0.0338 \\ 
   &  &  &  &  & st.dev. & 0.0014 & 0.0012 & 0.0009 & 0.0006 & 0.0009 & 0.0008 & 0.0008 & 0.0006 \\ 
   &  &  &  &  & t-test & 0.0000 & 0.0000 & 0.0000 & 0.8880 &  & 0.0000 & 0.0000 & 0.0000 \\ 
   &  &  &  &  & wilcox-test & 0.0000 & 0.0000 & 0.0000 & 0.8212 &  & 0.0000 & 0.0000 & 0.0000 \\ 
   \midrule
29 & \cmark & \cmark & \xmark & \cmark & mean & 0.0292 & 0.0239 & 0.0240 & 0.0231 & 0.0268 & 0.0255 & 0.0261 & 0.0256 \\ 
   &  &  &  &  & st.dev. & 0.0016 & 0.0009 & 0.0008 & 0.0005 & 0.0007 & 0.0003 & 0.0005 & 0.0003 \\ 
   &  &  &  &  & t-test & 0.0000 & 1.0000 & 1.0000 & 1.0000 &  & 1.0000 & 1.0000 & 1.0000 \\ 
   &  &  &  &  & wilcox-test & 0.0000 & 1.0000 & 1.0000 & 1.0000 &  & 1.0000 & 1.0000 & 1.0000 \\ 
   \midrule
30 & \cmark & \xmark & \cmark & \cmark & mean & 0.0115 & 0.0618 & 0.0580 & 0.0242 & 0.0246 & 0.0306 & 0.0285 & 0.0375 \\ 
   &  &  &  &  & st.dev. & 0.0029 & 0.0018 & 0.0020 & 0.0006 & 0.0006 & 0.0007 & 0.0008 & 0.0006 \\ 
   &  &  &  &  & t-test & 1.0000 & 0.0000 & 0.0000 & 0.9999 &  & 0.0000 & 0.0000 & 0.0000 \\ 
   &  &  &  &  & wilcox-test & 1.0000 & 0.0000 & 0.0000 & 0.9999 &  & 0.0000 & 0.0000 & 0.0000 \\ 
   \midrule
31 & \xmark & \cmark & \cmark & \cmark & mean & 0.0378 & 0.0394 & 0.0375 & 0.0236 & 0.0256 & 0.0295 & 0.0256 & 0.0317 \\ 
   &  &  &  &  & st.dev. & 0.0008 & 0.0014 & 0.0013 & 0.0007 & 0.0009 & 0.0007 & 0.0008 & 0.0006 \\ 
   &  &  &  &  & t-test & 0.0000 & 0.0000 & 0.0000 & 1.0000 &  & 0.0000 & 0.6494 & 0.0000 \\ 
   &  &  &  &  & wilcox-test & 0.0000 & 0.0000 & 0.0000 & 1.0000 &  & 0.0000 & 0.6416 & 0.0000 \\ 
   \midrule
32 & \cmark & \cmark & \cmark & \cmark & mean & 0.0433 & 0.0438 & 0.0413 & 0.0270 & 0.0260 & 0.0314 & 0.0274 & 0.0339 \\ 
   &  &  &  &  & st.dev. & 0.0014 & 0.0017 & 0.0014 & 0.0008 & 0.0011 & 0.0009 & 0.0010 & 0.0008 \\ 
   &  &  &  &  & t-test & 0.0000 & 0.0000 & 0.0000 & 0.0000 &  & 0.0000 & 0.0000 & 0.0000 \\ 
   &  &  &  &  & wilcox-test & 0.0000 & 0.0000 & 0.0000 & 0.0000 &  & 0.0000 & 0.0000 & 0.0000 \\ 
   \midrule
 \bottomrule
\end{tabular}
\begin{tablenotes}\footnotesize
\item [] \hspace{-0.17cm} \textit{Notes:} Table reports the average measures of the MSE based on 100 simulation replications for the sample size of 200 observations with 6 outcome classes. Columns 1 to 5 specify the DGP identifier and its features, namely 15 additional noise variables (\textit{noise}), nonlinear effects (\textit{nonlin}), multicollinearity among covariates (\textit{multi}), and randomly spaced thresholds (\textit{random}). The sixth column \textit{Statistic} shows the mean and the standard deviation of the accuracy measure for all methods. Additionally, \textit{t-test} and \textit{wilcox-test} contain the p-values of the parametric t-test as well as the nonparametric Wilcoxon test for the equality of means between the results of the \textit{Ordered Forest} and all the other methods.
\end{tablenotes}
\end{threeparttable}
}
\end{table}
\pagebreak
\subsubsection{AMSE: Low Dimension with 9 Classes}
\begin{table}[h!]
\footnotesize
\centering
\caption{Simulation Results: Accuracy Measure = AMSE \& Low Dimension with 9 Classes}\label{tab:allsim9} 
\resizebox{0.94\textwidth}{!}{
\begin{threeparttable}
\begin{tabular}{c|cccc|lrrrrrrrr}
\toprule
\midrule
\multicolumn{6}{c}{\normalsize{\textbf{Simulation Design}}} & \multicolumn{8}{c}{\normalsize{\textbf{Comparison of Methods}}}\\
\midrule
\midrule
 DGP & noise & nonlin & multi & random & Statistic & Ologit & Naive & Ordinal & Cond. & Ordered & Ordered* & Multi & Multi* \\ 
  \midrule
 \midrule
33 & \xmark & \xmark & \xmark & \xmark & mean & 0.0025 & 0.0150 & 0.0149 & 0.0134 & 0.0170 & 0.0170 & 0.0168 & 0.0172 \\ 
   &  &  &  &  & st.dev. & 0.0008 & 0.0005 & 0.0007 & 0.0004 & 0.0006 & 0.0003 & 0.0005 & 0.0002 \\ 
   &  &  &  &  & t-test & 1.0000 & 1.0000 & 1.0000 & 1.0000 &  & 0.5492 & 0.9993 & 0.0040 \\ 
   &  &  &  &  & wilcox-test & 1.0000 & 1.0000 & 1.0000 & 1.0000 &  & 0.3269 & 0.9985 & 0.0003 \\ 
   \midrule
34 & \cmark & \xmark & \xmark & \xmark & mean & 0.0046 & 0.0155 & 0.0154 & 0.0144 & 0.0176 & 0.0175 & 0.0175 & 0.0178 \\ 
   &  &  &  &  & st.dev. & 0.0011 & 0.0004 & 0.0004 & 0.0003 & 0.0004 & 0.0002 & 0.0003 & 0.0002 \\ 
   &  &  &  &  & t-test & 1.0000 & 1.0000 & 1.0000 & 1.0000 &  & 0.9697 & 0.9696 & 0.0011 \\ 
   &  &  &  &  & wilcox-test & 1.0000 & 1.0000 & 1.0000 & 1.0000 &  & 0.9359 & 0.9544 & 0.0003 \\ 
   \midrule
35 & \xmark & \cmark & \xmark & \xmark & mean & 0.0123 & 0.0110 & 0.0114 & 0.0107 & 0.0147 & 0.0121 & 0.0137 & 0.0121 \\ 
   &  &  &  &  & st.dev. & 0.0004 & 0.0004 & 0.0005 & 0.0004 & 0.0006 & 0.0003 & 0.0005 & 0.0003 \\ 
   &  &  &  &  & t-test & 1.0000 & 1.0000 & 1.0000 & 1.0000 &  & 1.0000 & 1.0000 & 1.0000 \\ 
   &  &  &  &  & wilcox-test & 1.0000 & 1.0000 & 1.0000 & 1.0000 &  & 1.0000 & 1.0000 & 1.0000 \\ 
   \midrule
36 & \xmark & \xmark & \cmark & \xmark & mean & 0.0044 & 0.0333 & 0.0321 & 0.0148 & 0.0168 & 0.0185 & 0.0175 & 0.0222 \\ 
   &  &  &  &  & st.dev. & 0.0010 & 0.0008 & 0.0009 & 0.0004 & 0.0005 & 0.0005 & 0.0005 & 0.0003 \\ 
   &  &  &  &  & t-test & 1.0000 & 0.0000 & 0.0000 & 1.0000 &  & 0.0000 & 0.0000 & 0.0000 \\ 
   &  &  &  &  & wilcox-test & 1.0000 & 0.0000 & 0.0000 & 1.0000 &  & 0.0000 & 0.0000 & 0.0000 \\ 
   \midrule
37 & \xmark & \xmark & \xmark & \cmark & mean & 0.0026 & 0.0152 & 0.0154 & 0.0136 & 0.0173 & 0.0172 & 0.0170 & 0.0175 \\ 
   &  &  &  &  & st.dev. & 0.0009 & 0.0005 & 0.0006 & 0.0003 & 0.0006 & 0.0002 & 0.0004 & 0.0002 \\ 
   &  &  &  &  & t-test & 1.0000 & 1.0000 & 1.0000 & 1.0000 &  & 0.8591 & 1.0000 & 0.0034 \\ 
   &  &  &  &  & wilcox-test & 1.0000 & 1.0000 & 1.0000 & 1.0000 &  & 0.7952 & 1.0000 & 0.0032 \\ 
   \midrule
38 & \cmark & \cmark & \xmark & \xmark & mean & 0.0136 & 0.0112 & 0.0115 & 0.0111 & 0.0144 & 0.0122 & 0.0137 & 0.0122 \\ 
   &  &  &  &  & st.dev. & 0.0006 & 0.0003 & 0.0004 & 0.0003 & 0.0004 & 0.0002 & 0.0003 & 0.0002 \\ 
   &  &  &  &  & t-test & 1.0000 & 1.0000 & 1.0000 & 1.0000 &  & 1.0000 & 1.0000 & 1.0000 \\ 
   &  &  &  &  & wilcox-test & 1.0000 & 1.0000 & 1.0000 & 1.0000 &  & 1.0000 & 1.0000 & 1.0000 \\ 
   \midrule
39 & \cmark & \xmark & \cmark & \xmark & mean & 0.0066 & 0.0335 & 0.0323 & 0.0159 & 0.0167 & 0.0192 & 0.0200 & 0.0242 \\ 
   &  &  &  &  & st.dev. & 0.0012 & 0.0006 & 0.0007 & 0.0005 & 0.0005 & 0.0006 & 0.0005 & 0.0004 \\ 
   &  &  &  &  & t-test & 1.0000 & 0.0000 & 0.0000 & 1.0000 &  & 0.0000 & 0.0000 & 0.0000 \\ 
   &  &  &  &  & wilcox-test & 1.0000 & 0.0000 & 0.0000 & 1.0000 &  & 0.0000 & 0.0000 & 0.0000 \\ 
   \midrule
40 & \cmark & \xmark & \xmark & \cmark & mean & 0.0046 & 0.0152 & 0.0152 & 0.0142 & 0.0175 & 0.0172 & 0.0173 & 0.0174 \\ 
   &  &  &  &  & st.dev. & 0.0011 & 0.0004 & 0.0005 & 0.0003 & 0.0004 & 0.0002 & 0.0003 & 0.0002 \\ 
   &  &  &  &  & t-test & 1.0000 & 1.0000 & 1.0000 & 1.0000 &  & 1.0000 & 0.9998 & 0.9655 \\ 
   &  &  &  &  & wilcox-test & 1.0000 & 1.0000 & 1.0000 & 1.0000 &  & 1.0000 & 0.9999 & 0.9525 \\ 
   \midrule
41 & \xmark & \cmark & \cmark & \xmark & mean & 0.0190 & 0.0198 & 0.0192 & 0.0127 & 0.0158 & 0.0156 & 0.0152 & 0.0170 \\ 
   &  &  &  &  & st.dev. & 0.0004 & 0.0007 & 0.0007 & 0.0004 & 0.0006 & 0.0004 & 0.0005 & 0.0003 \\ 
   &  &  &  &  & t-test & 0.0000 & 0.0000 & 0.0000 & 1.0000 &  & 0.9652 & 1.0000 & 0.0000 \\ 
   &  &  &  &  & wilcox-test & 0.0000 & 0.0000 & 0.0000 & 1.0000 &  & 0.9503 & 1.0000 & 0.0000 \\ 
   \midrule
42 & \xmark & \cmark & \xmark & \cmark & mean & 0.0125 & 0.0112 & 0.0117 & 0.0108 & 0.0147 & 0.0122 & 0.0138 & 0.0122 \\ 
   &  &  &  &  & st.dev. & 0.0004 & 0.0004 & 0.0007 & 0.0003 & 0.0005 & 0.0002 & 0.0005 & 0.0002 \\ 
   &  &  &  &  & t-test & 1.0000 & 1.0000 & 1.0000 & 1.0000 &  & 1.0000 & 1.0000 & 1.0000 \\ 
   &  &  &  &  & wilcox-test & 1.0000 & 1.0000 & 1.0000 & 1.0000 &  & 1.0000 & 1.0000 & 1.0000 \\ 
   \midrule
43 & \xmark & \xmark & \cmark & \cmark & mean & 0.0043 & 0.0335 & 0.0324 & 0.0149 & 0.0167 & 0.0187 & 0.0176 & 0.0225 \\ 
   &  &  &  &  & st.dev. & 0.0013 & 0.0007 & 0.0009 & 0.0005 & 0.0005 & 0.0005 & 0.0005 & 0.0004 \\ 
   &  &  &  &  & t-test & 1.0000 & 0.0000 & 0.0000 & 1.0000 &  & 0.0000 & 0.0000 & 0.0000 \\ 
   &  &  &  &  & wilcox-test & 1.0000 & 0.0000 & 0.0000 & 1.0000 &  & 0.0000 & 0.0000 & 0.0000 \\ 
   \midrule
44 & \cmark & \cmark & \cmark & \xmark & mean & 0.0208 & 0.0198 & 0.0193 & 0.0143 & 0.0159 & 0.0164 & 0.0162 & 0.0181 \\ 
   &  &  &  &  & st.dev. & 0.0007 & 0.0006 & 0.0005 & 0.0003 & 0.0006 & 0.0004 & 0.0005 & 0.0003 \\ 
   &  &  &  &  & t-test & 0.0000 & 0.0000 & 0.0000 & 1.0000 &  & 0.0000 & 0.0001 & 0.0000 \\ 
   &  &  &  &  & wilcox-test & 0.0000 & 0.0000 & 0.0000 & 1.0000 &  & 0.0000 & 0.0000 & 0.0000 \\ 
   \midrule
45 & \cmark & \cmark & \xmark & \cmark & mean & 0.0130 & 0.0110 & 0.0113 & 0.0108 & 0.0142 & 0.0118 & 0.0134 & 0.0118 \\ 
   &  &  &  &  & st.dev. & 0.0006 & 0.0003 & 0.0003 & 0.0003 & 0.0004 & 0.0003 & 0.0003 & 0.0002 \\ 
   &  &  &  &  & t-test & 1.0000 & 1.0000 & 1.0000 & 1.0000 &  & 1.0000 & 1.0000 & 1.0000 \\ 
   &  &  &  &  & wilcox-test & 1.0000 & 1.0000 & 1.0000 & 1.0000 &  & 1.0000 & 1.0000 & 1.0000 \\ 
   \midrule
46 & \cmark & \xmark & \cmark & \cmark & mean & 0.0070 & 0.0335 & 0.0325 & 0.0166 & 0.0173 & 0.0200 & 0.0204 & 0.0250 \\ 
   &  &  &  &  & st.dev. & 0.0016 & 0.0005 & 0.0007 & 0.0004 & 0.0005 & 0.0005 & 0.0005 & 0.0004 \\ 
   &  &  &  &  & t-test & 1.0000 & 0.0000 & 0.0000 & 1.0000 &  & 0.0000 & 0.0000 & 0.0000 \\ 
   &  &  &  &  & wilcox-test & 1.0000 & 0.0000 & 0.0000 & 1.0000 &  & 0.0000 & 0.0000 & 0.0000 \\ 
   \midrule
47 & \xmark & \cmark & \cmark & \cmark & mean & 0.0192 & 0.0203 & 0.0198 & 0.0130 & 0.0159 & 0.0158 & 0.0153 & 0.0173 \\ 
   &  &  &  &  & st.dev. & 0.0004 & 0.0007 & 0.0007 & 0.0004 & 0.0006 & 0.0004 & 0.0005 & 0.0003 \\ 
   &  &  &  &  & t-test & 0.0000 & 0.0000 & 0.0000 & 1.0000 &  & 0.6681 & 1.0000 & 0.0000 \\ 
   &  &  &  &  & wilcox-test & 0.0000 & 0.0000 & 0.0000 & 1.0000 &  & 0.5336 & 1.0000 & 0.0000 \\ 
   \midrule
48 & \cmark & \cmark & \cmark & \cmark & mean & 0.0203 & 0.0194 & 0.0190 & 0.0142 & 0.0159 & 0.0161 & 0.0162 & 0.0179 \\ 
   &  &  &  &  & st.dev. & 0.0006 & 0.0006 & 0.0005 & 0.0003 & 0.0005 & 0.0003 & 0.0004 & 0.0003 \\ 
   &  &  &  &  & t-test & 0.0000 & 0.0000 & 0.0000 & 1.0000 &  & 0.0006 & 0.0000 & 0.0000 \\ 
   &  &  &  &  & wilcox-test & 0.0000 & 0.0000 & 0.0000 & 1.0000 &  & 0.0004 & 0.0000 & 0.0000 \\ 
   \midrule
 \bottomrule
\end{tabular}
\begin{tablenotes}\footnotesize
\item [] \hspace{-0.17cm} \textit{Notes:} Table reports the average measures of the MSE based on 100 simulation replications for the sample size of 200 observations with 9 outcome classes. Columns 1 to 5 specify the DGP identifier and its features, namely 15 additional noise variables (\textit{noise}), nonlinear effects (\textit{nonlin}), multicollinearity among covariates (\textit{multi}), and randomly spaced thresholds (\textit{random}). The sixth column \textit{Statistic} shows the mean and the standard deviation of the accuracy measure for all methods. Additionally, \textit{t-test} and \textit{wilcox-test} contain the p-values of the parametric t-test as well as the nonparametric Wilcoxon test for the equality of means between the results of the \textit{Ordered Forest} and all the other methods.
\end{tablenotes}
\end{threeparttable}
}
\end{table}
\pagebreak
\subsubsection{AMSE: High Dimension with 3 Classes}
\begin{table}[h!]
\footnotesize
\centering
\caption{Simulation Results: Accuracy Measure = AMSE \& High Dimension with 3 Classes}\label{tab:allsim10} 
\resizebox{0.94\textwidth}{!}{
\begin{threeparttable}
\begin{tabular}{c|cccc|lrrrrrrrr}
\toprule
\midrule
\multicolumn{6}{c}{\normalsize{\textbf{Simulation Design}}} & \multicolumn{7}{c}{\normalsize{\textbf{Comparison of Methods}}}\\
\midrule
\midrule
 DGP & noise & nonlin & multi & random & Statistic & Naive & Ordinal & Cond. & Ordered & Ordered* & Multi & Multi* \\ 
  \midrule
 \midrule
49 & \cmark & \xmark & \xmark & \xmark & mean & 0.0923 & 0.0931 & 0.0908 & 0.0930 & 0.0952 & 0.0926 & 0.0952 \\ 
   &  &  &  &  & st.dev. & 0.0008 & 0.0013 & 0.0009 & 0.0009 & 0.0007 & 0.0007 & 0.0007 \\ 
   &  &  &  &  & t-test & 1.0000 & 0.2408 & 1.0000 &  & 0.0000 & 0.9980 & 0.0000 \\ 
   &  &  &  &  & wilcox-test & 1.0000 & 0.5433 & 1.0000 &  & 0.0000 & 0.9977 & 0.0000 \\ 
   \midrule
50 & \cmark & \cmark & \xmark & \xmark & mean & 0.0692 & 0.0698 & 0.0696 & 0.0702 & 0.0699 & 0.0696 & 0.0699 \\ 
   &  &  &  &  & st.dev. & 0.0009 & 0.0013 & 0.0010 & 0.0009 & 0.0009 & 0.0008 & 0.0008 \\ 
   &  &  &  &  & t-test & 1.0000 & 0.9907 & 0.9999 &  & 0.9649 & 1.0000 & 0.9852 \\ 
   &  &  &  &  & wilcox-test & 1.0000 & 1.0000 & 1.0000 &  & 0.9887 & 1.0000 & 0.9944 \\ 
   \midrule
51 & \cmark & \xmark & \cmark & \xmark & mean & 0.1385 & 0.1379 & 0.0864 & 0.0752 & 0.1008 & 0.0881 & 0.1087 \\ 
   &  &  &  &  & st.dev. & 0.0009 & 0.0010 & 0.0019 & 0.0021 & 0.0023 & 0.0019 & 0.0018 \\ 
   &  &  &  &  & t-test & 0.0000 & 0.0000 & 0.0000 &  & 0.0000 & 0.0000 & 0.0000 \\ 
   &  &  &  &  & wilcox-test & 0.0000 & 0.0000 & 0.0000 &  & 0.0000 & 0.0000 & 0.0000 \\ 
   \midrule
52 & \cmark & \xmark & \xmark & \cmark & mean & 0.0906 & 0.0904 & 0.0884 & 0.0902 & 0.0931 & 0.0902 & 0.0931 \\ 
   &  &  &  &  & st.dev. & 0.0011 & 0.0013 & 0.0008 & 0.0008 & 0.0007 & 0.0008 & 0.0007 \\ 
   &  &  &  &  & t-test & 0.0006 & 0.0794 & 1.0000 &  & 0.0000 & 0.3296 & 0.0000 \\ 
   &  &  &  &  & wilcox-test & 0.0010 & 0.1853 & 1.0000 &  & 0.0000 & 0.2606 & 0.0000 \\ 
   \midrule
53 & \cmark & \cmark & \cmark & \xmark & mean & 0.1079 & 0.1083 & 0.0910 & 0.0888 & 0.1010 & 0.0892 & 0.1013 \\ 
   &  &  &  &  & st.dev. & 0.0009 & 0.0011 & 0.0025 & 0.0020 & 0.0019 & 0.0019 & 0.0019 \\ 
   &  &  &  &  & t-test & 0.0000 & 0.0000 & 0.0000 &  & 0.0000 & 0.0936 & 0.0000 \\ 
   &  &  &  &  & wilcox-test & 0.0000 & 0.0000 & 0.0000 &  & 0.0000 & 0.0745 & 0.0000 \\ 
   \midrule
54 & \cmark & \cmark & \xmark & \cmark & mean & 0.0706 & 0.0703 & 0.0703 & 0.0705 & 0.0706 & 0.0704 & 0.0706 \\ 
   &  &  &  &  & st.dev. & 0.0010 & 0.0011 & 0.0010 & 0.0009 & 0.0009 & 0.0008 & 0.0009 \\ 
   &  &  &  &  & t-test & 0.1479 & 0.9409 & 0.8941 &  & 0.1495 & 0.7655 & 0.1796 \\ 
   &  &  &  &  & wilcox-test & 0.1712 & 0.9972 & 0.9496 &  & 0.0718 & 0.8048 & 0.1178 \\ 
   \midrule
55 & \cmark & \xmark & \cmark & \cmark & mean & 0.1291 & 0.1276 & 0.0725 & 0.0678 & 0.0914 & 0.0758 & 0.0954 \\ 
   &  &  &  &  & st.dev. & 0.0016 & 0.0010 & 0.0020 & 0.0021 & 0.0021 & 0.0020 & 0.0020 \\ 
   &  &  &  &  & t-test & 0.0000 & 0.0000 & 0.0000 &  & 0.0000 & 0.0000 & 0.0000 \\ 
   &  &  &  &  & wilcox-test & 0.0000 & 0.0000 & 0.0000 &  & 0.0000 & 0.0000 & 0.0000 \\ 
   \midrule
56 & \cmark & \cmark & \cmark & \cmark & mean & 0.1081 & 0.1079 & 0.0863 & 0.0828 & 0.0970 & 0.0834 & 0.0971 \\ 
   &  &  &  &  & st.dev. & 0.0012 & 0.0009 & 0.0028 & 0.0019 & 0.0021 & 0.0020 & 0.0021 \\ 
   &  &  &  &  & t-test & 0.0000 & 0.0000 & 0.0000 &  & 0.0000 & 0.0264 & 0.0000 \\ 
   &  &  &  &  & wilcox-test & 0.0000 & 0.0000 & 0.0000 &  & 0.0000 & 0.0364 & 0.0000 \\ 
   \midrule
 \bottomrule
\end{tabular}
\begin{tablenotes}\footnotesize
\item [] \hspace{-0.17cm} \textit{Notes:} Table reports the average measures of the MSE based on 100 simulation replications for the sample size of 200 observations with 3 outcome classes. Columns 1 to 5 specify the DGP identifier and its features, namely 1000 additional noise variables (\textit{noise}), nonlinear effects (\textit{nonlin}), multicollinearity among covariates (\textit{multi}), and randomly spaced thresholds (\textit{random}). The sixth column \textit{Statistic} shows the mean and the standard deviation of the accuracy measure for all methods. Additionally, \textit{t-test} and \textit{wilcox-test} contain the p-values of the parametric t-test as well as the nonparametric Wilcoxon test for the equality of means between the results of the \textit{Ordered Forest} and all the other methods.
\end{tablenotes}
\end{threeparttable}
}
\end{table}
\pagebreak
\subsubsection{AMSE: High Dimension with 6 Classes}
\begin{table}[h!]
\footnotesize
\centering
\caption{Simulation Results: Accuracy Measure = AMSE \& High Dimension with 6 Classes}\label{tab:allsim11}
\resizebox{0.94\textwidth}{!}{
\begin{threeparttable}
\begin{tabular}{c|cccc|lrrrrrrrr}
\toprule
\midrule
\multicolumn{6}{c}{\normalsize{\textbf{Simulation Design}}} & \multicolumn{7}{c}{\normalsize{\textbf{Comparison of Methods}}}\\
\midrule
\midrule
 DGP & noise & nonlin & multi & random & Statistic & Naive & Ordinal & Cond. & Ordered & Ordered* & Multi & Multi* \\ 
  \midrule
 \midrule
57 & \cmark & \xmark & \xmark & \xmark & mean & 0.0352 & 0.0352 & 0.0347 & 0.0361 & 0.0361 & 0.0360 & 0.0361 \\ 
   &  &  &  &  & st.dev. & 0.0003 & 0.0004 & 0.0004 & 0.0004 & 0.0004 & 0.0003 & 0.0004 \\ 
   &  &  &  &  & t-test & 1.0000 & 1.0000 & 1.0000 &  & 0.8112 & 0.9994 & 0.6394 \\ 
   &  &  &  &  & wilcox-test & 1.0000 & 1.0000 & 1.0000 &  & 0.8788 & 0.9989 & 0.6579 \\ 
   \midrule
58 & \cmark & \cmark & \xmark & \xmark & mean & 0.0246 & 0.0246 & 0.0246 & 0.0257 & 0.0248 & 0.0252 & 0.0248 \\ 
   &  &  &  &  & st.dev. & 0.0003 & 0.0003 & 0.0003 & 0.0004 & 0.0003 & 0.0002 & 0.0003 \\ 
   &  &  &  &  & t-test & 1.0000 & 1.0000 & 1.0000 &  & 1.0000 & 1.0000 & 1.0000 \\ 
   &  &  &  &  & wilcox-test & 1.0000 & 1.0000 & 1.0000 &  & 1.0000 & 1.0000 & 1.0000 \\ 
   \midrule
59 & \cmark & \xmark & \cmark & \xmark & mean & 0.0622 & 0.0617 & 0.0459 & 0.0383 & 0.0494 & 0.0479 & 0.0553 \\ 
   &  &  &  &  & st.dev. & 0.0003 & 0.0003 & 0.0005 & 0.0007 & 0.0007 & 0.0006 & 0.0006 \\ 
   &  &  &  &  & t-test & 0.0000 & 0.0000 & 0.0000 &  & 0.0000 & 0.0000 & 0.0000 \\ 
   &  &  &  &  & wilcox-test & 0.0000 & 0.0000 & 0.0000 &  & 0.0000 & 0.0000 & 0.0000 \\ 
   \midrule
60 & \cmark & \xmark & \xmark & \cmark & mean & 0.0339 & 0.0341 & 0.0335 & 0.0350 & 0.0347 & 0.0348 & 0.0348 \\ 
   &  &  &  &  & st.dev. & 0.0003 & 0.0004 & 0.0004 & 0.0004 & 0.0004 & 0.0003 & 0.0004 \\ 
   &  &  &  &  & t-test & 1.0000 & 1.0000 & 1.0000 &  & 1.0000 & 0.9993 & 0.9999 \\ 
   &  &  &  &  & wilcox-test & 1.0000 & 1.0000 & 1.0000 &  & 1.0000 & 0.9995 & 1.0000 \\ 
   \midrule
61 & \cmark & \cmark & \cmark & \xmark & mean & 0.0397 & 0.0397 & 0.0351 & 0.0358 & 0.0383 & 0.0380 & 0.0399 \\ 
   &  &  &  &  & st.dev. & 0.0004 & 0.0004 & 0.0007 & 0.0006 & 0.0005 & 0.0006 & 0.0004 \\ 
   &  &  &  &  & t-test & 0.0000 & 0.0000 & 1.0000 &  & 0.0000 & 0.0000 & 0.0000 \\ 
   &  &  &  &  & wilcox-test & 0.0000 & 0.0000 & 1.0000 &  & 0.0000 & 0.0000 & 0.0000 \\ 
   \midrule
62 & \cmark & \cmark & \xmark & \cmark & mean & 0.0229 & 0.0231 & 0.0229 & 0.0241 & 0.0231 & 0.0235 & 0.0231 \\ 
   &  &  &  &  & st.dev. & 0.0004 & 0.0005 & 0.0005 & 0.0005 & 0.0005 & 0.0004 & 0.0005 \\ 
   &  &  &  &  & t-test & 1.0000 & 1.0000 & 1.0000 &  & 1.0000 & 1.0000 & 1.0000 \\ 
   &  &  &  &  & wilcox-test & 1.0000 & 1.0000 & 1.0000 &  & 1.0000 & 1.0000 & 1.0000 \\ 
   \midrule
63 & \cmark & \xmark & \cmark & \cmark & mean & 0.0628 & 0.0629 & 0.0481 & 0.0405 & 0.0512 & 0.0506 & 0.0583 \\ 
   &  &  &  &  & st.dev. & 0.0003 & 0.0004 & 0.0005 & 0.0008 & 0.0007 & 0.0006 & 0.0005 \\ 
   &  &  &  &  & t-test & 0.0000 & 0.0000 & 0.0000 &  & 0.0000 & 0.0000 & 0.0000 \\ 
   &  &  &  &  & wilcox-test & 0.0000 & 0.0000 & 0.0000 &  & 0.0000 & 0.0000 & 0.0000 \\ 
   \midrule
64 & \cmark & \cmark & \cmark & \cmark & mean & 0.0383 & 0.0386 & 0.0343 & 0.0350 & 0.0367 & 0.0378 & 0.0387 \\ 
   &  &  &  &  & st.dev. & 0.0003 & 0.0004 & 0.0006 & 0.0005 & 0.0005 & 0.0005 & 0.0004 \\ 
   &  &  &  &  & t-test & 0.0000 & 0.0000 & 1.0000 &  & 0.0000 & 0.0000 & 0.0000 \\ 
   &  &  &  &  & wilcox-test & 0.0000 & 0.0000 & 1.0000 &  & 0.0000 & 0.0000 & 0.0000 \\ 
   \midrule
 \bottomrule
\end{tabular}
\begin{tablenotes}\footnotesize
\item [] \hspace{-0.17cm} \textit{Notes:} Table reports the average measures of the MSE based on 100 simulation replications for the sample size of 200 observations with 6 outcome classes. Columns 1 to 5 specify the DGP identifier and its features, namely 1000 additional noise variables (\textit{noise}), nonlinear effects (\textit{nonlin}), multicollinearity among covariates (\textit{multi}), and randomly spaced thresholds (\textit{random}). The sixth column \textit{Statistic} shows the mean and the standard deviation of the accuracy measure for all methods. Additionally, \textit{t-test} and \textit{wilcox-test} contain the p-values of the parametric t-test as well as the nonparametric Wilcoxon test for the equality of means between the results of the \textit{Ordered Forest} and all the other methods.
\end{tablenotes}
\end{threeparttable}
}
\end{table}
\pagebreak
\subsubsection{AMSE: High Dimension with 9 Classes}
\begin{table}[h!]
\footnotesize
\centering
\caption{Simulation Results: Accuracy Measure = AMSE \& High Dimension with 9 Classes}\label{tab:allsim12} 
\resizebox{0.94\textwidth}{!}{
\begin{threeparttable}
\begin{tabular}{c|cccc|lrrrrrrrr}
\toprule
\midrule
\multicolumn{6}{c}{\normalsize{\textbf{Simulation Design}}} & \multicolumn{7}{c}{\normalsize{\textbf{Comparison of Methods}}}\\
\midrule
\midrule
 DGP & noise & nonlin & multi & random & Statistic & Naive & Ordinal & Cond. & Ordered & Ordered* & Multi & Multi* \\ 
  \midrule
 \midrule
65 & \cmark & \xmark & \xmark & \xmark & mean & 0.0180 & 0.0181 & 0.0178 & 0.0189 & 0.0185 & 0.0188 & 0.0185 \\ 
   &  &  &  &  & st.dev. & 0.0002 & 0.0002 & 0.0002 & 0.0002 & 0.0002 & 0.0002 & 0.0002 \\ 
   &  &  &  &  & t-test & 1.0000 & 1.0000 & 1.0000 &  & 1.0000 & 1.0000 & 1.0000 \\ 
   &  &  &  &  & wilcox-test & 1.0000 & 1.0000 & 1.0000 &  & 1.0000 & 1.0000 & 1.0000 \\ 
   \midrule
66 & \cmark & \cmark & \xmark & \xmark & mean & 0.0123 & 0.0123 & 0.0123 & 0.0133 & 0.0124 & 0.0129 & 0.0124 \\ 
   &  &  &  &  & st.dev. & 0.0002 & 0.0002 & 0.0002 & 0.0002 & 0.0002 & 0.0002 & 0.0002 \\ 
   &  &  &  &  & t-test & 1.0000 & 1.0000 & 1.0000 &  & 1.0000 & 1.0000 & 1.0000 \\ 
   &  &  &  &  & wilcox-test & 1.0000 & 1.0000 & 1.0000 &  & 1.0000 & 1.0000 & 1.0000 \\ 
   \midrule
67 & \cmark & \xmark & \cmark & \xmark & mean & 0.0339 & 0.0337 & 0.0263 & 0.0224 & 0.0281 & 0.0284 & 0.0316 \\ 
   &  &  &  &  & st.dev. & 0.0002 & 0.0002 & 0.0003 & 0.0005 & 0.0004 & 0.0003 & 0.0003 \\ 
   &  &  &  &  & t-test & 0.0000 & 0.0000 & 0.0000 &  & 0.0000 & 0.0000 & 0.0000 \\ 
   &  &  &  &  & wilcox-test & 0.0000 & 0.0000 & 0.0000 &  & 0.0000 & 0.0000 & 0.0000 \\ 
   \midrule
68 & \cmark & \xmark & \xmark & \cmark & mean & 0.0181 & 0.0181 & 0.0179 & 0.0190 & 0.0186 & 0.0188 & 0.0186 \\ 
   &  &  &  &  & st.dev. & 0.0002 & 0.0002 & 0.0003 & 0.0003 & 0.0003 & 0.0002 & 0.0003 \\ 
   &  &  &  &  & t-test & 1.0000 & 1.0000 & 1.0000 &  & 1.0000 & 1.0000 & 1.0000 \\ 
   &  &  &  &  & wilcox-test & 1.0000 & 1.0000 & 1.0000 &  & 1.0000 & 1.0000 & 1.0000 \\ 
   \midrule
69 & \cmark & \cmark & \cmark & \xmark & mean & 0.0198 & 0.0199 & 0.0178 & 0.0187 & 0.0193 & 0.0201 & 0.0201 \\ 
   &  &  &  &  & st.dev. & 0.0002 & 0.0002 & 0.0003 & 0.0003 & 0.0003 & 0.0002 & 0.0002 \\ 
   &  &  &  &  & t-test & 0.0000 & 0.0000 & 1.0000 &  & 0.0000 & 0.0000 & 0.0000 \\ 
   &  &  &  &  & wilcox-test & 0.0000 & 0.0000 & 1.0000 &  & 0.0000 & 0.0000 & 0.0000 \\ 
   \midrule
70 & \cmark & \cmark & \xmark & \cmark & mean & 0.0124 & 0.0124 & 0.0124 & 0.0133 & 0.0125 & 0.0130 & 0.0125 \\ 
   &  &  &  &  & st.dev. & 0.0002 & 0.0002 & 0.0002 & 0.0003 & 0.0002 & 0.0002 & 0.0002 \\ 
   &  &  &  &  & t-test & 1.0000 & 1.0000 & 1.0000 &  & 1.0000 & 1.0000 & 1.0000 \\ 
   &  &  &  &  & wilcox-test & 1.0000 & 1.0000 & 1.0000 &  & 1.0000 & 1.0000 & 1.0000 \\ 
   \midrule
71 & \cmark & \xmark & \cmark & \cmark & mean & 0.0338 & 0.0337 & 0.0262 & 0.0225 & 0.0281 & 0.0285 & 0.0315 \\ 
   &  &  &  &  & st.dev. & 0.0002 & 0.0002 & 0.0004 & 0.0005 & 0.0005 & 0.0003 & 0.0004 \\ 
   &  &  &  &  & t-test & 0.0000 & 0.0000 & 0.0000 &  & 0.0000 & 0.0000 & 0.0000 \\ 
   &  &  &  &  & wilcox-test & 0.0000 & 0.0000 & 0.0000 &  & 0.0000 & 0.0000 & 0.0000 \\ 
   \midrule
72 & \cmark & \cmark & \cmark & \cmark & mean & 0.0200 & 0.0200 & 0.0178 & 0.0187 & 0.0193 & 0.0201 & 0.0202 \\ 
   &  &  &  &  & st.dev. & 0.0002 & 0.0002 & 0.0003 & 0.0003 & 0.0003 & 0.0003 & 0.0002 \\ 
   &  &  &  &  & t-test & 0.0000 & 0.0000 & 1.0000 &  & 0.0000 & 0.0000 & 0.0000 \\ 
   &  &  &  &  & wilcox-test & 0.0000 & 0.0000 & 1.0000 &  & 0.0000 & 0.0000 & 0.0000 \\ 
   \midrule
 \bottomrule
\end{tabular}
\begin{tablenotes}\footnotesize
\item [] \hspace{-0.17cm} \textit{Notes:} Table reports the average measures of the MSE based on 100 simulation replications for the sample size of 200 observations with 9 outcome classes. Columns 1 to 5 specify the DGP identifier and its features, namely 1000 additional noise variables (\textit{noise}), nonlinear effects (\textit{nonlin}), multicollinearity among covariates (\textit{multi}), and randomly spaced thresholds (\textit{random}). The sixth column \textit{Statistic} shows the mean and the standard deviation of the accuracy measure for all methods. Additionally, \textit{t-test} and \textit{wilcox-test} contain the p-values of the parametric t-test as well as the nonparametric Wilcoxon test for the equality of means between the results of the \textit{Ordered Forest} and all the other methods.
\end{tablenotes}
\end{threeparttable}
}
\end{table}
\pagebreak

\subsection{Empirical Results}\label{Appendix:empsim}
In this section we present more detailed and supplementary results regarding the empirical results (Section \ref{sec:boxplotsemp}) discussed in the main text. In the following the descriptive statistics for the considered datasets and the results for the prediction accuracy are summarized.

\subsubsection{Descriptive Statistics}\label{Appendix:desc}

\begin{table}[h!]
\centering
\caption{Descriptive Statistics: mammography dataset} 
\scalebox{0.9}{
\begin{tabular}{llrrrrr}
\toprule
\multicolumn{7}{c}{\textbf{Mammography Dataset}}\\
  \midrule
variable & type & mean & sd & median & min & max \\ 
  \midrule
SYMPT* & Categorical & 2.97 & 0.95 & 3.00 & 1.00 & 4.00 \\ 
  PB & Numeric & 7.56 & 2.10 & 7.00 & 5.00 & 17.00 \\ 
  HIST* & Categorical & 1.11 & 0.31 & 1.00 & 1.00 & 2.00 \\ 
  BSE* & Categorical & 1.87 & 0.34 & 2.00 & 1.00 & 2.00 \\ 
  DECT* & Categorical & 2.66 & 0.56 & 3.00 & 1.00 & 3.00 \\ 
  y* & Categorical & 1.61 & 0.77 & 1.00 & 1.00 & 3.00 \\ 
   \bottomrule
\end{tabular}
}
\end{table}

\begin{table}[h!]
\centering
\caption{Descriptive Statistics: nhanes dataset} 
\scalebox{0.9}{
\begin{tabular}{llrrrrr}
\toprule
\multicolumn{7}{c}{\textbf{Nhanes Dataset}}\\
  \midrule
variable & type & mean & sd & median & min & max \\ 
  \midrule
sex* & Categorical & 1.51 & 0.50 & 2.00 & 1.00 & 2.00 \\ 
  race* & Categorical & 2.87 & 1.00 & 3.00 & 1.00 & 5.00 \\ 
  country\_of\_birth* & Categorical & 1.34 & 0.79 & 1.00 & 1.00 & 4.00 \\ 
  education* & Categorical & 3.37 & 1.24 & 3.00 & 1.00 & 5.00 \\ 
  marital\_status* & Categorical & 2.31 & 1.74 & 1.00 & 1.00 & 6.00 \\ 
  waistcircum & Numeric & 100.37 & 16.37 & 99.40 & 61.60 & 176.70 \\ 
  Cholesterol & Numeric & 196.89 & 41.59 & 193.00 & 97.00 & 432.00 \\ 
  WBCcount & Numeric & 7.30 & 2.88 & 6.90 & 1.60 & 83.20 \\ 
  AcuteIllness* & Categorical & 1.25 & 0.43 & 1.00 & 1.00 & 2.00 \\ 
  depression* & Categorical & 1.39 & 0.76 & 1.00 & 1.00 & 4.00 \\ 
  ToothCond* & Categorical & 3.05 & 1.24 & 3.00 & 1.00 & 5.00 \\ 
  sleepTrouble* & Categorical & 2.28 & 1.28 & 2.00 & 1.00 & 5.00 \\ 
  wakeUp* & Categorical & 2.41 & 1.30 & 2.00 & 1.00 & 5.00 \\ 
  cig* & Categorical & 1.51 & 0.50 & 2.00 & 1.00 & 2.00 \\ 
  diabetes* & Categorical & 1.14 & 0.34 & 1.00 & 1.00 & 2.00 \\ 
  asthma* & Categorical & 1.15 & 0.36 & 1.00 & 1.00 & 2.00 \\ 
  heartFailure* & Categorical & 1.03 & 0.16 & 1.00 & 1.00 & 2.00 \\ 
  stroke* & Categorical & 1.03 & 0.18 & 1.00 & 1.00 & 2.00 \\ 
  chronicBronchitis* & Categorical & 1.07 & 0.26 & 1.00 & 1.00 & 2.00 \\ 
  alcohol & Numeric & 3.93 & 20.18 & 2.00 & 0.00 & 365.00 \\ 
  heavyDrinker* & Categorical & 1.17 & 0.37 & 1.00 & 1.00 & 2.00 \\ 
  medicalPlaceToGo* & Categorical & 1.92 & 0.67 & 2.00 & 1.00 & 5.00 \\ 
  BPsys & Numeric & 124.44 & 18.62 & 122.00 & 78.00 & 230.00 \\ 
  BPdias & Numeric & 71.18 & 11.84 & 72.00 & 10.00 & 114.00 \\ 
  age & Numeric & 49.96 & 16.68 & 50.00 & 20.00 & 80.00 \\ 
  BMI & Numeric & 29.33 & 6.66 & 28.32 & 14.20 & 73.43 \\ 
  y* & Categorical & 2.77 & 1.00 & 3.00 & 1.00 & 5.00 \\ 
   \bottomrule
\end{tabular}
}
\end{table}

\begin{table}[h!]
\centering
\caption{Descriptive Statistics: supportstudy dataset} 
\scalebox{0.9}{
\begin{tabular}{llrrrrr}
\toprule
\multicolumn{7}{c}{\textbf{Supportstudy Dataset}}\\
  \midrule
variable & type & mean & sd & median & min & max \\ 
  \midrule
age & Numeric & 62.80 & 16.27 & 65.29 & 20.30 & 100.13 \\ 
  sex* & Categorical & 1.54 & 0.50 & 2.00 & 1.00 & 2.00 \\ 
  dzgroup* & Categorical & 3.23 & 2.48 & 2.00 & 1.00 & 8.00 \\ 
  num.co & Numeric & 1.90 & 1.34 & 2.00 & 0.00 & 7.00 \\ 
  scoma & Numeric & 12.45 & 25.29 & 0.00 & 0.00 & 100.00 \\ 
  charges & Numeric & 59307.91 & 86620.70 & 28416.50 & 1635.75 & 740010.00 \\ 
  avtisst & Numeric & 23.53 & 13.60 & 20.00 & 1.67 & 64.00 \\ 
  race* & Categorical & 1.36 & 0.88 & 1.00 & 1.00 & 5.00 \\ 
  meanbp & Numeric & 84.52 & 27.64 & 77.00 & 0.00 & 180.00 \\ 
  wblc & Numeric & 12.62 & 9.31 & 10.50 & 0.05 & 100.00 \\ 
  hrt & Numeric & 98.59 & 32.93 & 102.50 & 0.00 & 300.00 \\ 
  resp & Numeric & 23.60 & 9.54 & 24.00 & 0.00 & 64.00 \\ 
  temp & Numeric & 37.08 & 1.25 & 36.70 & 32.50 & 41.20 \\ 
  crea & Numeric & 1.80 & 1.74 & 1.20 & 0.30 & 11.80 \\ 
  sod & Numeric & 137.64 & 6.34 & 137.00 & 118.00 & 175.00 \\ 
  y* & Categorical & 2.90 & 1.81 & 2.00 & 1.00 & 5.00 \\ 
   \bottomrule
\end{tabular}
}
\end{table}

\begin{table}[h!]
\centering
\caption{Descriptive Statistics: vlbw dataset} 
\scalebox{0.9}{
\begin{tabular}{llrrrrr}
\toprule
\multicolumn{7}{c}{\textbf{Vlbw Dataset}}\\
  \midrule
variable & type & mean & sd & median & min & max \\ 
  \midrule
race* & Categorical & 1.57 & 0.50 & 2.00 & 1.00 & 2.00 \\ 
  bwt & Numeric & 1094.89 & 260.44 & 1140.00 & 430.00 & 1500.00 \\ 
  inout* & Categorical & 1.03 & 0.16 & 1.00 & 1.00 & 2.00 \\ 
  twn* & Categorical & 1.24 & 0.43 & 1.00 & 1.00 & 2.00 \\ 
  lol & Numeric & 7.73 & 19.47 & 3.00 & 0.00 & 192.00 \\ 
  magsulf* & Categorical & 1.18 & 0.39 & 1.00 & 1.00 & 2.00 \\ 
  meth* & Categorical & 1.44 & 0.50 & 1.00 & 1.00 & 2.00 \\ 
  toc* & Categorical & 1.24 & 0.43 & 1.00 & 1.00 & 2.00 \\ 
  delivery* & Categorical & 1.41 & 0.49 & 1.00 & 1.00 & 2.00 \\ 
  sex* & Categorical & 1.50 & 0.50 & 1.00 & 1.00 & 2.00 \\ 
  y* & Categorical & 5.09 & 2.58 & 6.00 & 1.00 & 9.00 \\ 
   \bottomrule
\end{tabular}
}
\end{table}

\begin{table}[h!]
\centering
\caption{Descriptive Statistics: winequality dataset}
\scalebox{0.9}{
\begin{tabular}{llrrrrr}
\toprule
\multicolumn{7}{c}{\textbf{Winequality Dataset}}\\
  \midrule
variable & type & mean & sd & median & min & max \\ 
  \midrule
fixed.acidity & Numeric & 6.85 & 0.84 & 6.80 & 3.80 & 14.20 \\ 
  volatile.acidity & Numeric & 0.28 & 0.10 & 0.26 & 0.08 & 1.10 \\ 
  citric.acid & Numeric & 0.33 & 0.12 & 0.32 & 0.00 & 1.66 \\ 
  residual.sugar & Numeric & 6.39 & 5.07 & 5.20 & 0.60 & 65.80 \\ 
  chlorides & Numeric & 0.05 & 0.02 & 0.04 & 0.01 & 0.35 \\ 
  free.sulfur.dioxide & Numeric & 35.31 & 17.01 & 34.00 & 2.00 & 289.00 \\ 
  total.sulfur.dioxide & Numeric & 138.38 & 42.51 & 134.00 & 9.00 & 440.00 \\ 
  density & Numeric & 0.99 & 0.00 & 0.99 & 0.99 & 1.04 \\ 
  pH & Numeric & 3.19 & 0.15 & 3.18 & 2.72 & 3.82 \\ 
  sulphates & Numeric & 0.49 & 0.11 & 0.47 & 0.22 & 1.08 \\ 
  alcohol & Numeric & 10.51 & 1.23 & 10.40 & 8.00 & 14.20 \\ 
  y* & Categorical & 3.87 & 0.88 & 4.00 & 1.00 & 6.00 \\ 
   \bottomrule
\end{tabular}
} 
\end{table}

\pagebreak
\clearpage

\subsubsection{Prediction Accuracy}\label{Appendix:emptables}
Tables \ref{emp:RPS} and \ref{emp:MSE} summarize in detail the results of the prediction accuracy exercise using real datasets for the ARPS and the AMSE, respectively. The first column \textit{Data} specifies the dataset, the second column \textit{Class} defines the number of outcome classes of the dependent variable and the third column \textit{Size} indicates the number of observations. Similarly to the simulation results, the column \textit{Statistic} contains summary statistics and statistical tests results for the equality of means between the results of the \textit{Ordered Forest} and all the other methods.

\begin{table}[h!]
\footnotesize
\centering
\caption{Empirical Results: Accuracy Measure = ARPS}
\label{emp:RPS}
\resizebox{0.92\textwidth}{!}{
\begin{threeparttable}
\begin{tabular}{lcllrrrrrrrr}
\toprule
\midrule
\multicolumn{4}{c}{\normalsize{\textbf{Dataset Summary}}} & \multicolumn{8}{c}{\normalsize{\textbf{Comparison of Methods}}}\\
\midrule
\midrule
Data & Class & Size & Statistic & Ologit & Naive & Ordinal & Cond. & Ordered & Ordered* & Multi & Multi* \\ 
  \midrule
  \midrule
mammography & 3 & 412 & mean & 0.1776 & 0.2251 & 0.2089 & 0.1767 & 0.1823 & 0.1766 & 0.1826 & 0.1767 \\ 
   &  &  & st.dev. & 0.0010 & 0.0027 & 0.0021 & 0.0013 & 0.0018 & 0.0008 & 0.0019 & 0.0007 \\ 
   &  &  & t-test & 1.0000 & 0.0000 & 0.0000 & 1.0000 &  & 1.0000 & 0.3999 & 1.0000 \\ 
   &  &  & wilcox-test & 1.0000 & 0.0000 & 0.0000 & 1.0000 &  & 1.0000 & 0.3153 & 1.0000 \\ 
   \midrule
nhanes & 5 & 1914 & mean & 0.1088 & 0.1089 & 0.1100 & 0.1085 & 0.1103 & 0.1137 & 0.1104 & 0.1159 \\ 
   &  &  & st.dev. & 0.0004 & 0.0003 & 0.0004 & 0.0001 & 0.0002 & 0.0001 & 0.0002 & 0.0001 \\ 
   &  &  & t-test & 1.0000 & 1.0000 & 0.9839 & 1.0000 &  & 0.0000 & 0.2106 & 0.0000 \\ 
   &  &  & wilcox-test & 1.0000 & 1.0000 & 0.9738 & 1.0000 &  & 0.0000 & 0.2179 & 0.0000 \\ 
   \midrule
supportstudy & 5 & 798 & mean & 0.1872 & 0.1849 & 0.1834 & 0.1800 & 0.1823 & 0.1931 & 0.1857 & 0.1944 \\ 
   &  &  & st.dev. & 0.0011 & 0.0010 & 0.0009 & 0.0008 & 0.0008 & 0.0003 & 0.0007 & 0.0004 \\ 
   &  &  & t-test & 0.0000 & 0.0000 & 0.0052 & 1.0000 &  & 0.0000 & 0.0000 & 0.0000 \\ 
   &  &  & wilcox-test & 0.0000 & 0.0000 & 0.0073 & 1.0000 &  & 0.0000 & 0.0000 & 0.0000 \\ 
   \midrule
vlbw & 9 & 218 & mean & 0.1595 & 0.1713 & 0.1724 & 0.1603 & 0.1686 & 0.1623 & 0.1685 & 0.1642 \\ 
   &  &  & st.dev. & 0.0011 & 0.0026 & 0.0030 & 0.0014 & 0.0021 & 0.0005 & 0.0020 & 0.0003 \\ 
   &  &  & t-test & 1.0000 & 0.0100 & 0.0023 & 1.0000 &  & 1.0000 & 0.5143 & 1.0000 \\ 
   &  &  & wilcox-test & 1.0000 & 0.0116 & 0.0010 & 1.0000 &  & 1.0000 & 0.5733 & 1.0000 \\ 
   \midrule
winequality & 6 & 4893 & mean & 0.0756 & 0.0501 & 0.0503 & 0.0596 & 0.0507 & 0.0673 & 0.0504 & 0.0683 \\ 
   &  &  & st.dev. & 0.0000 & 0.0003 & 0.0002 & 0.0001 & 0.0002 & 0.0001 & 0.0002 & 0.0000 \\ 
   &  &  & t-test & 0.0000 & 1.0000 & 0.9992 & 0.0000 &  & 0.0000 & 0.9971 & 0.0000 \\ 
   &  &  & wilcox-test & 0.0000 & 0.9999 & 0.9986 & 0.0000 &  & 0.0000 & 0.9966 & 0.0000 \\ 
   \midrule
   \bottomrule
\end{tabular}
\begin{tablenotes}\footnotesize
\item [] \hspace{-0.17cm} \textit{Notes:} Table reports the average measures of the RPS based on 10 repetitions of 10-fold cross-validation. The fourth column \textit{Statistic} shows the mean and the standard deviation of the accuracy measure for all methods. Additionally, \textit{t-test} and \textit{wilcox-test} contain the p-values of the parametric t-test as well as the nonparametric Wilcoxon test for the equality of means between the results of the \textit{Ordered Forest} and all the other methods.
\end{tablenotes}
\end{threeparttable}
}
\end{table}

\begin{table}[h!]
\footnotesize
\centering
\caption{Empirical Results: Accuracy Measure = AMSE}
\label{emp:MSE}
\resizebox{0.92\textwidth}{!}{
\begin{threeparttable}
\begin{tabular}{lcllrrrrrrrr}
\toprule
\midrule
\multicolumn{4}{c}{\normalsize{\textbf{Dataset Summary}}} & \multicolumn{8}{c}{\normalsize{\textbf{Comparison of Methods}}}\\
\midrule
\midrule
 Data & Class & Size & Statistic & Ologit & Naive & Ordinal & Cond. & Ordered & Ordered* & Multi & Multi* \\ 
  \midrule
  \midrule
mammography & 3 & 412 & mean & 0.1754 & 0.2593 & 0.2222 & 0.1720 & 0.1766 & 0.1726 & 0.1770 & 0.1726 \\ 
   &  &  & st.dev. & 0.0007 & 0.0025 & 0.0031 & 0.0008 & 0.0012 & 0.0004 & 0.0013 & 0.0004 \\ 
   &  &  & t-test & 0.9923 & 0.0000 & 0.0000 & 1.0000 &  & 1.0000 & 0.2467 & 1.0000 \\ 
   &  &  & wilcox-test & 0.9943 & 0.0000 & 0.0000 & 1.0000 &  & 1.0000 & 0.2179 & 1.0000 \\ 
   \midrule
nhanes & 5 & 1914 & mean & 0.1310 & 0.1309 & 0.1332 & 0.1304 & 0.1332 & 0.1329 & 0.1319 & 0.1343 \\ 
   &  &  & st.dev. & 0.0003 & 0.0003 & 0.0003 & 0.0002 & 0.0003 & 0.0001 & 0.0003 & 0.0001 \\ 
   &  &  & t-test & 1.0000 & 1.0000 & 0.7067 & 1.0000 &  & 0.9936 & 1.0000 & 0.0000 \\ 
   &  &  & wilcox-test & 1.0000 & 1.0000 & 0.6579 & 1.0000 &  & 0.9955 & 1.0000 & 0.0000 \\ 
   \midrule
supportstudy & 5 & 798 & mean & 0.1124 & 0.1110 & 0.1094 & 0.1078 & 0.1088 & 0.1129 & 0.1101 & 0.1135 \\ 
   &  &  & st.dev. & 0.0005 & 0.0004 & 0.0004 & 0.0004 & 0.0004 & 0.0002 & 0.0003 & 0.0002 \\ 
   &  &  & t-test & 0.0000 & 0.0000 & 0.0020 & 1.0000 &  & 0.0000 & 0.0000 & 0.0000 \\ 
   &  &  & wilcox-test & 0.0000 & 0.0000 & 0.0008 & 0.9999 &  & 0.0000 & 0.0000 & 0.0000 \\ 
   \midrule
vlbw & 9 & 218 & mean & 0.0944 & 0.0986 & 0.0990 & 0.0956 & 0.1008 & 0.0958 & 0.1006 & 0.0956 \\ 
   &  &  & st.dev. & 0.0002 & 0.0008 & 0.0009 & 0.0004 & 0.0008 & 0.0003 & 0.0009 & 0.0002 \\ 
   &  &  & t-test & 1.0000 & 1.0000 & 0.9999 & 1.0000 &  & 1.0000 & 0.7224 & 1.0000 \\ 
   &  &  & wilcox-test & 1.0000 & 1.0000 & 0.9999 & 1.0000 &  & 1.0000 & 0.7821 & 1.0000 \\ 
   \midrule
winequality & 6 & 4893 & mean & 0.1001 & 0.0692 & 0.0698 & 0.0831 & 0.0702 & 0.0906 & 0.0693 & 0.0913 \\ 
   &  &  & st.dev. & 0.0000 & 0.0003 & 0.0003 & 0.0001 & 0.0003 & 0.0001 & 0.0003 & 0.0001 \\ 
   &  &  & t-test & 0.0000 & 1.0000 & 0.9960 & 0.0000 &  & 0.0000 & 1.0000 & 0.0000 \\ 
   &  &  & wilcox-test & 0.0000 & 1.0000 & 0.9974 & 0.0000 &  & 0.0000 & 1.0000 & 0.0000 \\ 
   \midrule
   \bottomrule
\end{tabular}
\begin{tablenotes}\footnotesize
\item [] \hspace{-0.17cm} \textit{Notes:} Table reports the average measures of the MSE based on 10 repetitions of 10-fold cross-validation. The fourth column \textit{Statistic} shows the mean and the standard deviation of the accuracy measure for all methods. Additionally, \textit{t-test} and \textit{wilcox-test} contain the p-values of the parametric t-test as well as the nonparametric Wilcoxon test for the equality of means between the results of the \textit{Ordered Forest} and all the other methods.
\end{tablenotes}
\end{threeparttable}
}
\end{table}

\pagebreak

\subsection{Software Implementation}\label{Appendix:tuning}
The Monte Carlo study has been conducted using the \textsf{R} statistical software \parencite{rstats} in version 3.5.2 (Eggshell Igloo) and the respective packages implementing the estimators used. With regards to the forest-based estimators the main tuning parameters, namely the number of trees, the number of randomly chosen covariates and the minimum leaf size have been specified according to the values in Table \ref{tab:sim} in the main text.

\begin{table}[h!]
\centering
\caption{Overview of Software Packages and Tuning Parameters}
\label{tab:packages}
\resizebox{0.99\textwidth}{!}{
\begin{tabular}{lcccccccc}
\toprule
\multicolumn{9}{c}{Software Implementation and Tuning Parameters} \\
\midrule
method & Ologit & Naive & Ordinal & Conditional & Ordered & Ordered* & Multi & Multi* \\
package & \textsf{rms} & \textsf{ordinalForest} & \textsf{ordinalForest} & \textsf{party} & \textsf{ranger} & \textsf{grf} & \textsf{ranger} & \textsf{grf} \\
function & \textsf{lrm} & \textsf{ordfor} & \textsf{ordfor} & \textsf{cforest} & \textsf{ranger} & \textsf{regression\_forest} & \textsf{ranger} & \textsf{regression\_forest} \\
\midrule
max. iterations & 25 & -& -& -& -& -& -&-\\
trees & - & 1000 & 1000 & 1000 & 1000 & 1000 & 1000 & 1000 \\
random subset & - & $\sqrt{p}$& $\sqrt{p}$& $\sqrt{p}$& $\sqrt{p}$& $\sqrt{p}$& $\sqrt{p}$& $\sqrt{p}$\\
leaf size & -& 5& 5& 0& 5& 5& 5& 5 \\
$B_{sets}$ & - & 0 & 1000 & -& -& -& -& -\\
$B_{prior}$ & - & 0 & 100 & -& -& -& -& - \\
performance & -& \textsf{equal}& \textsf{equal}& -& -& -& -& - \\
$S_{best}$ & -& 0 & 10& -& -& -& -& -\\
\bottomrule
\end{tabular}
}
\end{table}

In terms of the particular \textsf{R} packages used the ordered logistic regression has been implemented using the \textsf{rms} package (version 5.1-3) written by \textcite{rms}. The respective \textsf{lrm} function for fitting the ordered logit has been used with the default parameters, except setting the maximum number of iterations, \textsf{maxit}=25 as for some of the DGPs the ordered logit has experienced convergence issues. Next, the naive and the ordinal forest have been applied based on the \textsf{ordinalForest} package in version 2.3 \parencite{ordinalForest} with the \textsf{ordfor} function. As described in Appendix \ref{Appendix:hornung} the ordinal forest introduces additional tuning parameters for which we use the default parameters as suggested in the package manual. Further, the conditional forest has been estimated with the package \textsf{party} in version 1.3-1 \parencites{Hothorn2006a, Strobl2007, Strobl2008}. Regarding the choice of the tuning parameters, we rely on the default parameters of the \textsf{cforest} function. A particularity of the conditional forest is, due to the conceptual differences to standard regression forest in terms of the splitting criterion, the choice of the stopping rule. This is controlled by the significance level $\alpha$ (see Appendix \ref{Appendix:hornik} for details). However, in order to grow deep trees we follow the suggestion in the package manual to set \textsf{mincriterion}$=0$, which has been also used in the simulation study conducted in \textcite{Janitza2016}. Lastly, the \textit{Ordered Forest} as well as the multinomial forest algorithms are implemented using the package \textsf{ranger} in version 0.11.1 \parencite{Wright2017} with the default hyperparameters. The honest versions of the above two estimators rely on the \textsf{grf} package in version 0.10.2 \parencite{grf} with the default hyperparameters as well. A detailed overview of packages with the corresponding tuning parameters is provided in Table \ref{tab:packages}.

Furthermore, Tables \ref{tab:abs_time} and \ref{tab:rel_time} compare the absolute and relative computation time of the respective methods. For comparison purposes, we measure the computation time for the four main DGPs presented in Section \ref{sec:boxplots} of the main text, namely the simple DGP in the low- and high-dimensional case as well as the complex DGP in the low- and high-dimensional case, for both the small sample size ($N=200$) and the big sample size ($N=800$) for all considered number of outcome classes. We estimate the model based on the training set and predict the class probabilities for a test set of size $N=10'000$ as in the main simulation. We repeat this procedure 10 times and report the average computation time. The tuning parameters and the software implementations are chosen as defined in Table \ref{tab:sim} in the main text and Table \ref{tab:packages} herein, respectively. All simulations are computed on a 64-Bit Windows machine with 4 cores (1.80GHz) and 16GB RAM storage.\\

\begin{table}[h!]
\centering
\caption{Absolute Computation Time in Seconds}
\label{tab:abs_time}
\resizebox{0.99\textwidth}{!}{
\begin{threeparttable}
\begin{tabular}{clllrrrrrrrr}
\toprule
\midrule
\multicolumn{4}{c}{\normalsize{\textbf{Simulation Design}}} & \multicolumn{8}{c}{\normalsize{\textbf{Comparison of Methods}}}\\
\midrule
\midrule
 Class & Dim. & DGP & Size & Ologit & Naive & Ordinal & Cond. & Ordered & Ordered* & Multi & Multi* \\ 
  \midrule
 \midrule
3 & Low & Simple & 200 & 0.01 & 1.22 & 10.33 & 46.61 & 0.62 & 1.24 & 0.91 & 1.86 \\ 
  3 & Low & Simple & 800 & 0.02 & 1.58 & 40.83 & 150.84 & 1.03 & 1.96 & 1.61 & 2.98 \\ 
  3 & Low & Complex & 200 & 0.02 & 1.19 & 11.93 & 47.43 & 0.63 & 1.26 & 0.98 & 1.92 \\ 
  3 & Low & Complex & 800 & 0.03 & 1.71 & 52.45 & 150.59 & 1.08 & 1.94 & 1.73 & 3.06 \\ 
   \midrule
3 & High & Simple & 200 &  & 3.50 & 61.89 & 64.28 & 4.05 & 5.08 & 6.06 & 7.27 \\ 
  3 & High & Simple & 800 &  & 13.91 & 332.60 & 175.76 & 7.19 & 7.10 & 12.19 & 11.02 \\ 
  3 & High & Complex & 200 &  & 3.46 & 60.25 & 59.98 & 4.02 & 4.96 & 6.02 & 7.10 \\ 
  3 & High & Complex & 800 &  & 13.83 & 325.65 & 173.63 & 6.83 & 6.61 & 11.50 & 10.66 \\ 
   \midrule
 \midrule
6 & Low & Simple & 200 & 0.02 & 1.88 & 12.79 & 46.80 & 1.47 & 3.00 & 1.74 & 3.52 \\ 
  6 & Low & Simple & 800 & 0.03 & 2.28 & 48.98 & 151.58 & 2.45 & 4.75 & 3.10 & 5.82 \\ 
  6 & Low & Complex & 200 & 0.03 & 1.85 & 14.75 & 46.97 & 1.56 & 3.12 & 1.85 & 3.66 \\ 
  6 & Low & Complex & 800 & 0.04 & 2.54 & 64.44 & 151.84 & 2.68 & 4.82 & 3.30 & 6.02 \\ 
   \midrule
6 & High & Simple & 200 &  & 4.21 & 69.80 & 64.14 & 10.24 & 11.74 & 12.01 & 13.63 \\ 
  6 & High & Simple & 800 &  & 15.86 & 386.02 & 176.27 & 19.34 & 17.43 & 26.24 & 19.97 \\ 
  6 & High & Complex & 200 &  & 4.11 & 70.51 & 60.85 & 9.98 & 11.52 & 11.95 & 13.61 \\ 
  6 & High & Complex & 800 &  & 15.85 & 371.69 & 174.17 & 18.11 & 17.18 & 24.43 & 19.52 \\ 
   \midrule
 \midrule
9 & Low & Simple & 200 & 0.03 & 2.32 & 20.53 & 46.70 & 2.27 & 4.71 & 2.44 & 5.03 \\ 
  9 & Low & Simple & 800 & 0.04 & 2.69 & 57.22 & 145.21 & 3.82 & 7.29 & 4.61 & 7.99 \\ 
  9 & Low & Complex & 200 & 0.03 & 2.29 & 22.86 & 47.36 & 2.40 & 4.83 & 2.65 & 5.28 \\ 
  9 & Low & Complex & 800 & 0.05 & 3.07 & 79.15 & 151.36 & 4.27 & 7.75 & 5.81 & 8.68 \\ 
   \midrule
9 & High & Simple & 200 &  & 4.85 & 80.76 & 63.25 & 16.05 & 17.84 & 17.69 & 19.56 \\ 
  9 & High & Simple & 800 &  & 16.91 & 413.74 & 169.91 & 31.34 & 26.91 & 38.95 & 27.38 \\ 
  9 & High & Complex & 200 &  & 4.62 & 78.86 & 57.68 & 15.79 & 17.78 & 17.57 & 19.59 \\ 
  9 & High & Complex & 800 &  & 18.10 & 437.04 & 175.07 & 31.12 & 27.33 & 37.59 & 28.16 \\ 
   \midrule
 \bottomrule
\end{tabular}
\begin{tablenotes}\footnotesize
\item [] \hspace{-0.17cm} \textit{Notes:} Table reports the average absolute computation time in seconds based on 10 simulation replications of training and prediction. The first column denotes the number of outcome classes. Columns 2 and 3 specify the dimension and the DGP, respectively. The fourth column contains the number of observations in the training set. The prediction set consists of 10 000 observations.
\end{tablenotes}
\end{threeparttable}
}
\end{table}

The results reveal the expected pattern for the \textit{Ordered Forest}. The more outcome classes the longer the computation time as by definition of the algorithm more forests have to be estimated. Furthermore, we also observe a longer computation time if the number of observation and/or the number of considered splitting covariates increases which is also an expected behaviour. However, the computation time is not sensitive to the particular DGP which it should not be either. The latter two patterns are true for all considered methods. In comparison to the other forest-based methods, the computational advantage of the \textit{Ordered Forest} becomes apparent. The \textit{Ordered Forest} outperforms the ordinal and the conditional forest in all cases. In some cases the \textit{Ordered Forest} is even more than 100 times faster and even in the closest cases it is more than 3 times faster than the two. In absolute terms this translates to computation time of around 1 second for the \textit{Ordered Forest} and around 50 seconds for the ordinal and around 150 seconds for the conditional forest in the most extreme case. Contrarily, in the closest case, the computation time for the \textit{Ordered Forest} is around 15 seconds, while for the ordinal forest this is around 80 seconds and around 60 seconds for the conditional forest. This points to the additional computation burden of the ordinal and the conditional forest due to the optimization procedure and the permutation tests, respectively. The only exception is the naive forest which does not include the optimization step. Furthermore, we observe a slightly longer computation time for the multinomial forest in comparison to the \textit{Ordered Forest}, which is due to one extra forest being estimated. The honest versions of the two forests take a bit longer in general, but this seems to reverse once bigger samples are considered (in terms of both number of observations as well as number of considered covariates).

\begin{table}[h!]
\centering
\caption{Relative Computation Time}
\label{tab:rel_time}
\resizebox{0.99\textwidth}{!}{
\begin{threeparttable}
\begin{tabular}{clllrrrrrrrr}
\toprule
\midrule
\multicolumn{4}{c}{\normalsize{\textbf{Simulation Design}}} & \multicolumn{8}{c}{\normalsize{\textbf{Comparison of Methods}}}\\
\midrule
\midrule
 Class & Dim. & DGP & Size & Ologit & Naive & Ordinal & Cond. & Ordered & Ordered* & Multi & Multi* \\ 
  \midrule
 \midrule
3 & Low & Simple & 200 & 0.02 & 1.98 & 16.76 & 75.66 & 1 & 2.02 & 1.48 & 3.02 \\ 
  3 & Low & Simple & 800 & 0.02 & 1.53 & 39.68 & 146.59 & 1 & 1.91 & 1.56 & 2.90 \\ 
  3 & Low & Complex & 200 & 0.03 & 1.87 & 18.79 & 74.70 & 1 & 1.99 & 1.55 & 3.03 \\ 
  3 & Low & Complex & 800 & 0.03 & 1.59 & 48.79 & 140.09 & 1 & 1.81 & 1.61 & 2.84 \\ 
   \midrule
3 & High & Simple & 200 &  & 0.86 & 15.27 & 15.86 & 1 & 1.25 & 1.50 & 1.79 \\ 
  3 & High & Simple & 800 &  & 1.94 & 46.28 & 24.46 & 1 & 0.99 & 1.70 & 1.53 \\ 
  3 & High & Complex & 200 &  & 0.86 & 14.99 & 14.92 & 1 & 1.23 & 1.50 & 1.77 \\ 
  3 & High & Complex & 800 &  & 2.02 & 47.68 & 25.42 & 1 & 0.97 & 1.68 & 1.56 \\ 
   \midrule
 \midrule
6 & Low & Simple & 200 & 0.02 & 1.28 & 8.73 & 31.95 & 1 & 2.05 & 1.19 & 2.40 \\ 
  6 & Low & Simple & 800 & 0.01 & 0.93 & 19.95 & 61.74 & 1 & 1.94 & 1.26 & 2.37 \\ 
  6 & Low & Complex & 200 & 0.02 & 1.18 & 9.45 & 30.09 & 1 & 2.00 & 1.19 & 2.34 \\ 
  6 & Low & Complex & 800 & 0.02 & 0.94 & 24.02 & 56.59 & 1 & 1.80 & 1.23 & 2.24 \\ 
   \midrule
6 & High & Simple & 200 &  & 0.41 & 6.81 & 6.26 & 1 & 1.15 & 1.17 & 1.33 \\ 
  6 & High & Simple & 800 &  & 0.82 & 19.96 & 9.11 & 1 & 0.90 & 1.36 & 1.03 \\ 
  6 & High & Complex & 200 &  & 0.41 & 7.07 & 6.10 & 1 & 1.16 & 1.20 & 1.36 \\ 
  6 & High & Complex & 800 &  & 0.88 & 20.52 & 9.62 & 1 & 0.95 & 1.35 & 1.08 \\ 
   \midrule
 \midrule
9 & Low & Simple & 200 & 0.01 & 1.02 & 9.03 & 20.54 & 1 & 2.07 & 1.07 & 2.21 \\ 
  9 & Low & Simple & 800 & 0.01 & 0.70 & 14.98 & 38.01 & 1 & 1.91 & 1.21 & 2.09 \\ 
  9 & Low & Complex & 200 & 0.01 & 0.95 & 9.51 & 19.69 & 1 & 2.01 & 1.10 & 2.19 \\ 
  9 & Low & Complex & 800 & 0.01 & 0.72 & 18.55 & 35.48 & 1 & 1.82 & 1.36 & 2.03 \\ 
   \midrule
9 & High & Simple & 200 &  & 0.30 & 5.03 & 3.94 & 1 & 1.11 & 1.10 & 1.22 \\ 
  9 & High & Simple & 800 &  & 0.54 & 13.20 & 5.42 & 1 & 0.86 & 1.24 & 0.87 \\ 
  9 & High & Complex & 200 &  & 0.29 & 5.00 & 3.65 & 1 & 1.13 & 1.11 & 1.24 \\ 
  9 & High & Complex & 800 &  & 0.58 & 14.04 & 5.63 & 1 & 0.88 & 1.21 & 0.90 \\ 
   \midrule
 \bottomrule
\end{tabular}
\begin{tablenotes}\footnotesize
\item [] \hspace{-0.17cm} \textit{Notes:} Table reports the average relative computation time with regards to the \textit{Ordered Forest} estimator based on 10 simulation replications of training and prediction. The first column denotes the number of outcome classes. Columns 2 and 3 specify the dimension and the DGP, respectively. The fourth column contains the number of observations in the training set. The prediction set consists of 10 000 observations.
\end{tablenotes}
\end{threeparttable}
}
\end{table}

Generally, the sensitivity with regards to the computation time appears to be very different for the considered methods. For the \textit{Ordered Forest} as well as the multinomial forest, including their honest versions, the most important aspect is clearly the number of outcome classes. For the naive and the ordinal forest the number of observations seems to be most decisive and for the conditional forest paradoxically the size of the prediction set is most relevant. Overall, the above result support the theoretical argument of the \textit{Ordered Forest} being computationally advantageous in comparison to the ordinal and the conditional forest.

\pagebreak

\section{Empirical Application}\label{Appendix:applications}

In this appendix we provide the descriptive statistics for the dataset used in the empirical
application of the main text as well as supplementary results containing the estimation of marginal
effects.

\subsection{Descriptive Statistics}\label{Appendix:nhis}

\begin{table}[ht]
\centering
\caption{Descriptive Statistics: NHIS Dataset} 
\begin{tabular}{llrrrrr}
  \toprule
\multicolumn{7}{c}{\textbf{NHIS Dataset}}\\
  \midrule
variable & type & mean & sd & median & min & max \\ 
  \midrule
   Health Status* & Categorical & 3.93 & 0.95 & 4.00 & 1.00 & 5.00 \\ 
Health Insurance* & Categorical & 0.84 & 0.37 & 1.00 & 0.00 & 1.00 \\
Female* & Categorical & 0.50 & 0.50 & 0.50 & 0.00 & 1.00 \\ 
  Non White* & Categorical & 0.20 & 0.40 & 0.00 & 0.00 & 1.00 \\ 
  Age & Numeric & 42.72 & 8.70 & 43.00 & 26.00 & 59.00 \\ 
  Education & Numeric & 13.74 & 2.99 & 14.00 & 0.00 & 18.00 \\ 
  Family Size & Numeric & 3.63 & 1.37 & 4.00 & 2.00 & 18.00 \\ 
  Employed* & Categorical & 0.82 & 0.39 & 1.00 & 0.00 & 1.00 \\ 
  Income* & Categorical & 94178.04 & 56738.46 & 85985.78 & 19282.93 & 167844.53 \\ 
   \bottomrule
\end{tabular}
\end{table}

\begin{table}[ht]
\centering
\caption{Descriptive Statistics by Class: NHIS Dataset} 
\begin{tabular}{lrrrrr}
    \toprule
\multicolumn{6}{c}{\textbf{NHIS Dataset}}\\
\midrule
& \multicolumn{5}{c}{Health Status}\\
\midrule
variable & poor & fair & good & very good & excellent \\ 
  \midrule
  Health Status & 1.14 & 5.66 & 25.14 & 34.92 & 33.13 \\ 
Health Insurance & 79.07 & 71.50 & 77.88 & 87.52 & 87.76 \\
Female & 49.77 & 51.08 & 49.28 & 50.43 & 49.92 \\ 
  Non White & 31.63 & 23.89 & 22.84 & 18.18 & 18.21 \\ 
  Age & 47.65 & 45.37 & 43.75 & 42.73 & 41.30 \\ 
  Education & 12.11 & 12.20 & 12.89 & 13.97 & 14.46 \\ 
  Family Size & 3.33 & 3.68 & 3.68 & 3.59 & 3.64 \\ 
  Employed & 28.84 & 65.57 & 80.99 & 84.35 & 84.21 \\ 
  Income & 53409.03 & 62473.99 & 78957.11 & 99685.45 & 106743.21 \\
  \midrule
  N & 215 & 1063 & 4724 & 6562 & 6226 \\
  share in \% & 1.14 & 5.66 & 25.14 & 34.92 & 33.13 \\
   \bottomrule
\multicolumn{6}{l}{\footnotesize{\textit{Note:} Means of variables for respective outcome class displayed. Shares for dummy var-}}\\ \multicolumn{6}{l}{\footnotesize{iables are indicated in \%.}}
\end{tabular}
\end{table}

\pagebreak

\subsection{Marginal Effects}\label{AppendixME}
In what follows, the results for the marginal effects at mean are presented for the considered NHIS dataset. Similarly as in the main text, the effects are computed for each outcome class of the dependent variable both for the \textit{Ordered Forest} as well as for the ordered logit. The estimations are done in \textsf{R} version 3.6.1 using the \textsf{orf} package \parencite{orf2019} in version 0.1.3 for the \textit{Ordered Forest} and the \textsf{oglmx} package \parencite{Carroll2018} in version 3.0.0.0 for the ordered logit.
\vspace{1.0cm}

\begin{table}[h!]
\centering
\caption{Marginal Effects at Mean: NHIS Dataset}
\resizebox{0.99\textwidth}{!}{
\begin{threeparttable}
\begin{tabular}{l@{\phantom{......}}c@{\phantom{......}}@{ }r@{ }@{ }r@{ }@{ }r@{ }@{ }r@{ }@{ }l@{ }HHH@{ }r@{ }@{ }r@{ }@{ }r@{ }@{ }r@{ }@{ }l@{ }}
\toprule
\multicolumn{2}{c}{\normalsize{\textbf{Dataset}}} & \multicolumn{5}{c}{\normalsize{\textbf{Ordered Forest}}} & \multicolumn{3}{c}{} & \multicolumn{5}{c}{\normalsize{\textbf{Ordered Logit}}}\\
\midrule
 Variable & Class & Effect & Std.Error & t-Value & p-Value &  &  & Variable & Category & Effect & Std.Error & t-Value & p-Value &   \\ 
  \hline
Age & 1 & 0.01 & 0.01 & 0.69 & 48.80 &     &  & Age & 1 & 0.04 & 0.00 & 12.77 & 0.00 & *** \\ 
   & 2 & 0.31 & 0.20 & 1.55 & 12.07 &     &  &  & 2 & 0.18 & 0.01 & 20.08 & 0.00 & *** \\ 
   & 3 & -3.76 & 3.10 & -1.21 & 22.49 &     &  &  & 3 & 0.62 & 0.03 & 22.63 & 0.00 & *** \\ 
   & 4 & -1.31 & 4.67 & -0.28 & 77.93 &     &  &  & 4 & 0.00 & 0.01 & 0.15 & 87.78 &     \\ 
   & 5 & 4.75 & 5.63 & 0.84 & 39.88 &     &  &  & 5 & -0.83 & 0.04 & -23.38 & 0.00 & *** \\ 
   \hline
Education & 1 & 0.00 & 0.00 & 0.00 & 100.00 &     &  & Education & 1 & -0.09 & 0.01 & -11.83 & 0.00 & *** \\ 
   & 2 & 0.00 & 0.00 & 0.00 & 100.00 &     &  &  & 2 & -0.46 & 0.03 & -16.90 & 0.00 & *** \\ 
   & 3 & 0.00 & 0.00 & 0.00 & 100.00 &     &  &  & 3 & -1.60 & 0.09 & -18.19 & 0.00 & *** \\ 
   & 4 & 0.00 & 0.00 & 0.00 & 100.00 &     &  &  & 4 & -0.00 & 0.02 & -0.15 & 87.78 &     \\ 
   & 5 & 0.00 & 0.00 & 0.00 & 100.00 &     &  &  & 5 & 2.16 & 0.12 & 18.63 & 0.00 & *** \\ 
   \hline
Employed & 1 & -2.07 & 0.56 & -3.73 & 0.02 & *** &  & Employed & 1 & -0.35 & 0.05 & -7.22 & 0.00 & *** \\ 
   & 2 & -1.79 & 1.39 & -1.28 & 19.89 &     &  &  & 2 & -1.69 & 0.21 & -8.08 & 0.00 & *** \\ 
   & 3 & -6.76 & 11.87 & -0.57 & 56.88 &     &  &  & 3 & -5.49 & 0.61 & -8.94 & 0.00 & *** \\ 
   & 4 & 4.13 & 9.18 & 0.45 & 65.29 &     &  &  & 4 & 0.57 & 0.15 & 3.86 & 0.01 & *** \\ 
   & 5 & 6.49 & 16.14 & 0.40 & 68.74 &     &  &  & 5 & 6.96 & 0.73 & 9.52 & 0.00 & *** \\ 
   \hline
FamilySize & 1 & 0.08 & 0.09 & 0.95 & 34.06 &     &  & FamilySize & 1 & -0.01 & 0.01 & -0.81 & 42.00 &     \\ 
   & 2 & -5.07 & 4.12 & -1.23 & 21.81 &     &  &  & 2 & -0.04 & 0.05 & -0.81 & 41.95 &     \\ 
   & 3 & 3.81 & 15.41 & 0.25 & 80.45 &     &  &  & 3 & -0.13 & 0.16 & -0.81 & 41.94 &     \\ 
   & 4 & 2.96 & 33.98 & 0.09 & 93.07 &     &  &  & 4 & -0.00 & 0.00 & -0.15 & 87.99 &     \\ 
   & 5 & -1.78 & 39.79 & -0.04 & 96.43 &     &  &  & 5 & 0.18 & 0.22 & 0.81 & 41.94 &     \\ 
   \hline
Female & 1 & -0.01 & 0.01 & -0.58 & 56.51 &     &  & Female & 1 & 0.02 & 0.03 & 0.68 & 49.85 &     \\ 
   & 2 & 0.21 & 0.79 & 0.27 & 78.66 &     &  &  & 2 & 0.09 & 0.13 & 0.68 & 49.81 &     \\ 
   & 3 & -2.61 & 5.01 & -0.52 & 60.28 &     &  &  & 3 & 0.30 & 0.45 & 0.68 & 49.80 &     \\ 
   & 4 & -3.07 & 10.21 & -0.30 & 76.38 &     &  &  & 4 & 0.00 & 0.00 & 0.15 & 88.10 &     \\ 
   & 5 & 5.47 & 11.73 & 0.47 & 64.10 &     &  &  & 5 & -0.41 & 0.60 & -0.68 & 49.80 &     \\ 
   \hline
HealthInsurance & 1 & -0.00 & 0.02 & -0.01 & 98.90 &     &  & HealthInsurance & 1 & -0.09 & 0.04 & -2.17 & 2.97 & **  \\ 
   & 2 & -1.15 & 1.26 & -0.91 & 36.12 &     &  &  & 2 & -0.44 & 0.20 & -2.20 & 2.77 & **  \\ 
   & 3 & 2.96 & 6.17 & 0.48 & 63.14 &     &  &  & 3 & -1.52 & 0.68 & -2.25 & 2.43 & **  \\ 
   & 4 & -9.14 & 16.51 & -0.55 & 57.96 &     &  &  & 4 & 0.05 & 0.05 & 0.98 & 32.83 &     \\ 
   & 5 & 7.34 & 17.48 & 0.42 & 67.48 &     &  &  & 5 & 2.01 & 0.87 & 2.30 & 2.16 & **  \\ 
   \hline
Income & 1 & 0.02 & 0.01 & 1.46 & 14.45 &     &  & Income & 1 & -0.00 & 0.00 & -12.16 & 0.00 & *** \\ 
   & 2 & -0.36 & 0.92 & -0.39 & 69.70 &     &  &  & 2 & -0.00 & 0.00 & -18.01 & 0.00 & *** \\ 
   & 3 & 1.87 & 10.11 & 0.18 & 85.36 &     &  &  & 3 & -0.00 & 0.00 & -19.87 & 0.00 & *** \\ 
   & 4 & -4.32 & 10.98 & -0.39 & 69.37 &     &  &  & 4 & -0.00 & 0.00 & -0.15 & 87.78 &     \\ 
   & 5 & 2.80 & 16.54 & 0.17 & 86.56 &     &  &  & 5 & 0.00 & 0.00 & 20.36 & 0.00 & *** \\ 
   \hline
NonWhite & 1 & 0.03 & 0.03 & 0.92 & 35.56 &     &  & NonWhite & 1 & 0.30 & 0.04 & 6.99 & 0.00 & *** \\ 
   & 2 & 0.96 & 1.59 & 0.60 & 54.76 &     &  &  & 2 & 1.45 & 0.19 & 7.81 & 0.00 & *** \\ 
   & 3 & 1.98 & 6.55 & 0.30 & 76.22 &     &  &  & 3 & 4.76 & 0.56 & 8.48 & 0.00 & *** \\ 
   & 4 & 0.37 & 10.52 & 0.04 & 97.18 &     &  &  & 4 & -0.41 & 0.12 & -3.52 & 0.04 & *** \\ 
   & 5 & -3.34 & 12.61 & -0.27 & 79.10 &     &  &  & 5 & -6.09 & 0.68 & -8.94 & 0.00 & *** \\
   \midrule 
\multicolumn{13}{l}{\footnotesize{Significance levels correspond to: $*** .<0.01$, $** .<0.05$, $* .<0.1$.}}\\
   \bottomrule
\end{tabular}
\begin{tablenotes}\footnotesize
\item [] \hspace{-0.17cm} \textit{Notes:} Table shows the comparison of the marginal effects at mean in \% points between the \textit{Ordered Forest} and the ordered logit. The effects are estimated for all classes, together with the corresponding standard errors, t-values and p-values. The standard errors for the \textit{Ordered Forest} are estimated using the weight-based inference and for the ordered logit are obtained via the delta method.
\end{tablenotes}
\end{threeparttable}
}
\end{table}

\end{document}